\documentclass[final]{article}
\usepackage{anysize,subfigure}
\usepackage[dvips]{graphicx}
\usepackage{amsfonts,amssymb,color}
\usepackage{colortbl}
\usepackage{multirow}
\usepackage[full]{complexity}
\marginsize{1in}{1in}{1.5cm}{1.5cm}

\input{Qcircuit.tex}

\newcommand{\AAA}{${\mathcal A}$}
\newcommand{\T}{${\mathcal T}$}
\newcommand{\N}{${\mathcal N}$}
\newcommand{\CC}{${\mathcal C}$}
\newcommand{\shutup}[1]{}
\newcommand{\rot}{\mathrm{rot}}
\newcommand{\bT}{\mathrm{\bf \scriptscriptstyle {,}}}
\newcommand{\bC}{\mathrm{\bf \scriptscriptstyle {,}}}
\newcommand{\bA}{\mathrm{\bf \scriptscriptstyle {}}}
\newcommand{\tca}[3]{\!\!\!({#1} \bT {#2} \bC {#3}\bA)}
\def\length{16cm}
\newtheorem{theorem}{Theorem}[section]
\newtheorem{example}{Example}[section]

\definecolor{gray}{gray}{0.7}
\newcommand{\qed}{\nobreak \ifvmode \relax \else
      \ifdim\lastskip<1.5em \hskip-\lastskip
      \hskip1.5em plus0em minus0.5em \fi \nobreak
      \vrule height0.75em width0.5em depth0.25em\fi}

\title{\large Constant-Optimized Quantum Circuits for Modular Multiplication and Exponentiation}
\author{Igor L. Markov\footnote{Department of EECS, University of Michigan, Ann Arbor, MI 48109-2121; Email: imarkov@eecs.umich.edu.}
\hspace{1cm}
Mehdi Saeedi\footnote{Computer Engineering Department, Amirkabir University of Technology, Tehran, Iran; Email: msaeedi@aut.ac.ir.}}

\begin{document}
\date{}
\maketitle

\begin{abstract}
Reversible circuits for modular multiplication $Cx\%M$ with $x<M$ arise
as components of modular exponentiation in Shor's quantum number-factoring
algorithm. However, existing generic constructions focus on asymptotic gate count and circuit depth rather than actual values, producing fairly large circuits not optimized for specific $C$ and $M$ values. In this work, we develop such optimizations in a bottom-up fashion, starting with most convenient $C$ values. When zero-initialized ancilla registers are available, we reduce the search for compact circuits to a shortest-path problem.
Some of our modular-multiplication circuits are asymptotically smaller than previous constructions, but worst-case bounds and average sizes remain $\Theta(n^2)$. In the context of modular exponentiation, we offer several constant-factor improvements, as well as an improvement by a constant additive term that is significant for few-qubit circuits arising in ongoing laboratory experiments with Shor's algorithm.
\end{abstract}

\section{Introduction}
\label{sec:intro}

The pursuit of quantum computation~\cite{NielsenC2000} has generated both excitement and controversy,
while producing few compelling empirical demonstrations so far. Adiabatic computing
experiments by DWave Systems were sharply criticized for {\em not} demonstrating quantum entanglement and
not solving hard problem instances that would confound best known problem-specific algorithms on non-quantum
computers. Several academic groups implemented Shor's number-factoring algorithm on several qubits to factor the number 15, recalling that asymptotic worst-case complexity of Shor's algorithm is polynomial while best known number-factoring algorithms for non-quantum computers take more than polynomial time to run, both in theory and in practice. Experiments with photonic quantum gates~\cite{LuBYP2007,Lanyon2007,PolitiMB09}
suggest the presence of entanglement,\footnote{Similar results were shown by simulation for semiconductor nanostructures~\cite{Buscemi2010}.} but leave unclear how entanglement is going to scale in larger systems. Recent ion traps decrease per-gate error rates below the threshold estimate for fault-tolerant quantum computing \cite{Brown2011}, making sophisticated quantum algorithms more practical if appropriate quantum error-correction is used.
Shor's algorithm remains the best candidate for benchmarking quantum algorithms because $(i)$ it solves a practical problem for which optimized non-quantum software is also available, $(ii)$ it has been thoroughly studied, and $(iii)$ it can be implemented with several known circuits.

{Reducing the size of quantum circuits required by Shor's algorithm} \cite{Beckman1996,VanMeter2005} --- the focus of our work --- decreases resource requirements for future quantum computers in a {\em non-linear} way because larger circuits entail heavier overhead for quantum error-correction~\cite{VanMeter2006}. In comparisons to non-quantum number-factoring software, smaller circuits can make quantum computers more competitive. However, quantum simulators \cite{Viamontes2009} can also run faster on smaller circuits. The significance of simulation in benchmarking quantum algorithms is twofold.
{\em First}, simulation can help studying intermediate states generated by a quantum algorithm and estimate the amount of quantum entanglement in these states. {\em Second}, simulators can be viewed as competing non-quantum algorithms. While this aspect of simulation is often dismissed {\em a priori}, an instructive example is given by the Quantum Fourier Transform (QFT). It was recently discovered that QFT can be efficiently simulated {\em when used stand-alone} \cite{Aharonov06,Yoran2007} (but not as part of Shor's algorithm) and thus does not offer a quantum speed-up, despite generating a significant amount of entanglement. This unexpected result was obtained independently by several researchers~\cite{Aharonov06,Yoran2007} by optimizing approximate QFT circuits for a specific simulation technique~\cite{MarkovShi2008}.

\subsection{Shor's algorithm}

Shor's algorithm seeks to factor a given value $M > 0$, which we assume to be semiprime $M = pq$ with unknown factors. The strategy is to consider the functions $f_b(x)=x^b\%M$\label{fn:rem}\footnote{Here and in the remaining text, the percent sign \% denotes the modulo (remainder) operation, as it does in the C and C++ languages.}, potentially with several different $1<b<M$ values and determine their periods in case gcd$(b,M)=1$.
When the period is determined to be even $b^{2\pi}\%M=1$, we have $(b^\pi-1)(b^\pi+1)\%M=0$,
thus either $(b^\pi-1)$ or $(b^\pi+1)$ must share at least one prime factor with $M$. If $b^{\pi}\%M \neq -1$, such a factor can be found using gcd$(b^\pi\pm 1,M)$, otherwise it leads to the trivial factors $1$ and $M$. When the period is determined to be odd, another $b$ value is tried.

\begin{figure}[tb]
\scriptsize
\scalebox{0.8}{
\input{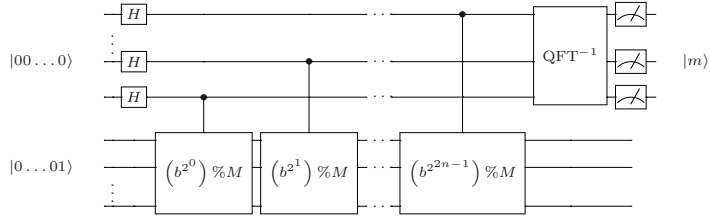}
}
\centering
\vspace{2mm}
\caption{
 \label{fig:shor} An outline of the quantum part of Shor's algorithm. The ``one controlling-qubit trick'' from \cite{Beauregard2003} can significantly reduce the number of qubits required to implement Shor's algorithm (Figure \ref{fig:shor2}).
}
\end{figure}

The period-finding procedure relies on a quantum circuit (Figure \ref{fig:shor}),
instantiated for a given value $1 < b < M$ coprime with $M$. The circuit operates on two 0-initialized quantum registers \cite{NielsenC2000} with
 \begin{itemize}
  \item  a block of parallel Hadamard gates on Register 1,
  \item  a circuit for modular exponentiation (mod-exp) evaluates $f(y)=b^y \% M$ by mapping $|y\rangle |0\rangle \mapsto |y\rangle |f(y)\rangle$, where $y$ is read from Register 1 and $f(y)$ is written to Register 2; Register 1 can be temporarily modified, but must be restored at the end,
  \item a circuit for the Quantum Fourier Transform (QFT) on Register 1,
  \item a block of parallel measurements on Register 1.
 \end{itemize}

\begin{figure}[tb]
\scriptsize
\scalebox{0.8}{
\input{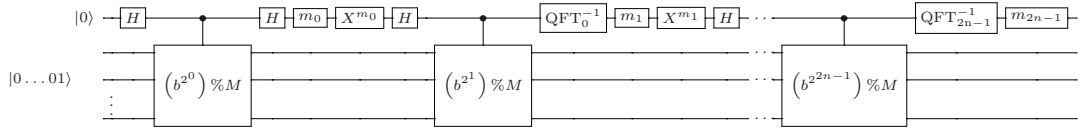}
}
\centering
\vspace{2mm}
\caption{
 \label{fig:shor2} The ``one controlling-qubit trick'' \cite{Beauregard2003} that reduces qubit requirements for Shor's algorithm. In this figure, $m_i$ represents the $i$-th measurement. In addition, $X^{m_i}$ (negation) and QFT$^{-1}_{i}$ gates are conditioned on the result of each measurement.
}
\end{figure}

The first and last blocks cannot be optimized any further. QFT circuits are understood fairly well and are much smaller than circuits for modular exponentiation \cite{NielsenC2000}.
Therefore, our focus is on mod-exp circuits. They typically consist of reversible gates --- NOT (\N), CNOT (\CC) and Toffoli (\T) --- which can be modeled and optimized entirely in terms of Boolean logic \cite{SaeediM2011}. However, in physical implementations, Toffoli gates must be decomposed into smaller gates directly implementable in a given technology \cite{ShendeM2009}. Reversible circuits for modular exponentiation start with an inverter on Register 2 that changes the $|000\cdots0\rangle$ value to
$|000\cdots1\rangle$, and otherwise exhibit the following structure: each
($i$-th) bit of Register 1 enables (controls) a circuit block that multiplies
Register 2 by $C_i=b^{2^i} \% M$ and reduces the result $\%M$. When $b$ and $M$
are known, $C_i$ can be pre-computed without quantum computation. Therefore, we
refer to $C_i x \% M$-blocks below. They are typically implemented using shift and
addition circuits, and a number of relevant quantum adders are known
\cite{Cuccaro2004,Takahashi2010}. The selection of appropriate adder types is
discussed in \cite{VanMeter2005,Fowler2004}.

\begin{figure}[b]
\scriptsize
\input{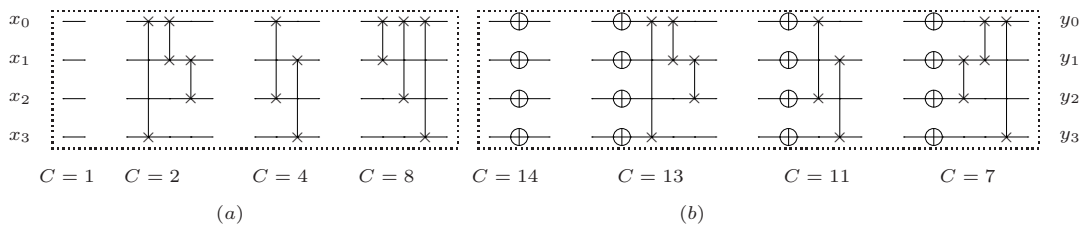}
\centering
\parbox{\length}{
\caption{\label{fig:15}
Circuits for $f(x)=Cx\%M,~M=15$, $\gcd(C, M)$ = 1, (a) $C=2^k$, (b) $C=M-2^k$.
$\oplus$ gates indicate inverters, while the two-bit gates are bit-swaps.
All gates are linear and can thus be simulated on an initial full-superposition state
using the stabilizer formalism \cite{NielsenC2000}.
}
}
\end{figure}

\begin{figure}[tb]
\scriptsize
\scalebox{0.95}{
\input{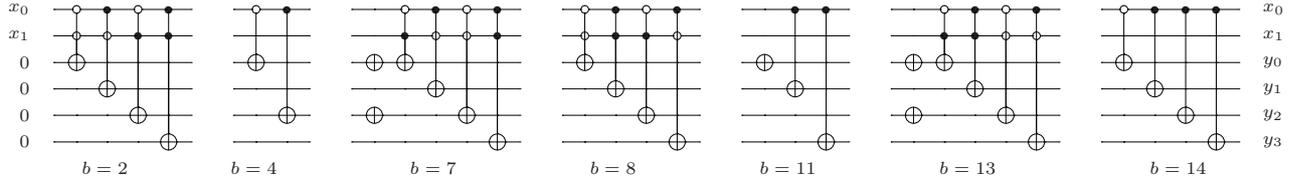}
}
\centering
\parbox{\length}{
\vspace{4mm}
\caption{\label{fig:exp15}
Circuits for $f(x) = b^x\%15$, gcd$(b, 15)$ = 1. These circuits compare favorably with better-known circuits in
terms of controlled-SWAP gates, because each controlled-SWAP is worth three Toffoli gates.
}
}
\end{figure}

 Each controlled modular multiplication is traditionally implemented separately. When dealing with reversible logic and quantum circuits, we note that the coprimality of $C$ and $M$ makes $x\mapsto Cx\%M$ a reversible transformation. The number of coprime $C$ values is $\varphi(M)=(p-1)(q-1)$, where $\varphi(M)$ is the Euler's totient function and gives the size of $(\mathbb{Z}/M\mathbb{Z})^\times$ --- the multiplicative group of integers mod-$M$. For $M=15$, modular multiplication circuits for the eight $C$ coprime values are illustrated in Figure \ref{fig:15}. Figure \ref{fig:exp15} shows circuits for $f(x) = b^x\%15$, gcd$(b, 15)$ = 1.

 When not knowing $p$ and $q$, one should also not assume any knowledge that would
 make it easy to find them.  For example, one should not choose $C$ that satisfies
 $C^{2\pi} = 1 \% M$ with a known (small) $\pi$ because such solutions would
 allow one to factorize $M$ via gcd$(C^\pi\pm1,M)$. Also recall that $(\mathbb{Z}/M\mathbb{Z})^\times$
 is a product of two cyclic groups $\mathbb{Z}/p\mathbb{Z}$ and $\mathbb{Z}/q\mathbb{Z}$,
 and thus $(\mathbb{Z}/M\mathbb{Z})^\times$ admits a generating set with only two elements.
 However, knowing such generators is tantamount to knowing $p$ and $q$.
 When working with specific small $M=pq$,
it is sometimes difficult to avoid using the knowledge of $p$ and $q$, but results obtained
this way do not necessarily scale to large values. The same can be said about results produced
through exhaustive search.

\subsection{Known circuits for modular multiplication by a constant} \label{sec:MM_pre_work}
 We now outline several approaches to modular multiplication by a constant
 and point out their potential inefficiencies. It is commonly agreed that techniques
 that give asymptotically the smallest gate counts (based on Fast Fourier Transforms)
 are not practical for up to several hundred bits, and we do not discuss them here.
 Karatsuba multiplication also does not appear competitive, as far as we can tell.

 \ \\
 {\bf Multiplication by a known constant $C$} (not modular) can be implemented as a sequence of alternating shifts and additions, where one of the addends is always $x$ \cite{BrentZ2010}. When $C$ is even, we factor out a power of two and accumulate a multiplication by a power of two, leaving a smaller odd constant $C'$. For an odd constant $C>1$,
 we subtract one and accumulate a $+x$ operator, leaving a smaller even constant $C''$. This process
 stops at 1 and essentially traverses the binary expansion of $C$ from the least significant bit
 to the most significant bit, resulting in $n_1-1$ additions when the binary expansion of $C$ includes
 $n_1$ non-zero bits. On average, an $n$-bit number has $n/2$ non-zero bits.
 An improvement is possible by also using $-x$ operators. When $C$ is odd, we consider $C\%4$.
 When $C\%4=1$, we proceed as above. When $C\%4=3$, we {\em add} one and accumulate a $-x$ operator.
 This step may temporarily increase an odd constant by one, but always results in constants divisible by four, so the next step will decrease it by more. This algorithm essentially constructs the so-called Canonical Signed
 Digit (CSD) representation \cite{Avizienis1961,BrentZ2010} that prohibits adjacent non-zero bits. Thus, the number of additions and subtractions cannot exceed $n/2$ and averages $n/3$. For example, consider 39=0$b$100111. Rather than expand $39x=2(2(8x+x)+x)+x$ with three additions, we can expand $39x=8(4x+x)-x$ with only two addition/subtraction operations.

\ \\
 {\bf Computing $Cx\%M$ by reversible circuits through binary or CSD expansion of $C$}
    poses several challenges. This technique is based on the operation $2^kx+x$ and thus requires
    a modular adder circuit with two (unknown) arguments, and must also copy the $x$ value to a separate
    register. A single (ancilla) register suffices, but clearing it (reinitializing to 0) after the additions
    requires effort. As we explain below, clearing the ancillae requires another modular-multiplication circuit.
    For constants $C$ with sparse CSD expansion, the ancillae-clearing circuit can be much larger than the multiplication itself,
    as its CSD expansion is unlikely to be sparse. In general, the second circuit
    requires on the order of $n^2$ gates for $n$-bit arguments, and is the same size
    as the first circuit, on average.

   \ \\
 {\bf Computing $Cx\%M$ using binary expansion of $x$}, rather than $C$,\footnote{Given that $x$ is not a constant, its CSD expansion is not easily available and cannot be used in combinational multiplication circuits.}
    entails chaining $x^i$-controlled mod-$M$ additions of constants $(2^iC)\%M$.
    As shown in Section \ref{sec:blocks}, such reversible modular addition-of-a-constant circuits
    can be simplified for each particular constant and require a single (register) argument
    rather than two (cf. previous paragraph).
    Even for the simplest constants, quadratically many gates are required.
    Moreover, $x$ cannot be modified while modular additions are controlled by the bits of $x$.
    Therefore, the additions must be accumulated in a separate register, which again requires clearing the garbage ancillae.

 \ \\
   {\bf Clearing garbage ancillae.}  In reversible circuits, 0-initialized ancillae must be cleared by each circuit block
   (except, possibly, the last), but some blocks produce {\em garbage bits}. For example,
   using traditional implementations of constant-multiplication as a sequence of shifts and adds
   requires creating a copy of the input, e.g., to compute $3x=2x+x$. However, clearing this copy
   (using the result of multiplication) essentially requires a division operation. In the context
   of modular multiplication $Cx\%M$ with gcd$(C,M)=1$, division can be performed as multiplication
   by the modular inverse $C^{-1}\%M$ pre-computed by the extended Euclidean GCD algorithm \cite{BrentZ2010}. We will now show how this approach was developed by Bennett to construct reversible
   modular multiplication circuits that clear their ancillae
   \cite[Section II]{VedralEtAl1995}, \cite[Formulae 4.4-4.6]{Beckman1996}.
   Assume a reversible circuit computing $g(x)=Cx\% M$ using a copy-register:

\begin{small}
\begin{equation}
  U_g : |x\rangle|0\rangle \longmapsto |x\rangle|g(x)\rangle
    \ \ \mathrm{or} \ \
  U^\times_g : |x\rangle|0\rangle \longmapsto |g(x)\rangle|x\rangle
\end{equation}
\end{small}

\noindent
   When gcd$(C,M)=1$, the function $g(x)$ is reversible, and the same construction
   can be applied to $g^{-1}(z)=C^{-1}z\%M$, where $C^{-1}$ is the modular inverse of $C$.
\begin{small}
\begin{equation}
   U_{g^{-1}} : |z\rangle|0\rangle \longmapsto |z\rangle |g^{-1}(z)\rangle
\end{equation}
\end{small}
   when $z=g(x)$, we get
\begin{small}
\begin{equation}
   U_{g^{-1}} : |g(x)\rangle|0\rangle \longmapsto |g(x)\rangle|x\rangle
\end{equation}
\end{small}
   A reversible circuit can be reversed --- by reversing the order of the gates and replacing
   each gate with its inverse, keeping in mind that the gates NOT, CNOT and Toffoli are self-inverse.
\begin{small}
\begin{equation}
   U^{-1}_{g^{-1}} : |g(x)\rangle|x\rangle \longmapsto |g(x)\rangle|0\rangle
\end{equation}
\end{small}
   Applying $U^{-1}_{g^{-1}}$ after $U^\times_g$ replaces $x$ with $g(x)$
   and leaves the copy-register initialized to 0.
\begin{small}
\begin{equation}
    U^{-1}_{g^{-1}} \cdot U^\times_g : |x\rangle|0\rangle \longmapsto |g(x)\rangle|0\rangle
\end{equation}
\end{small}
   Unfortunately, when $C$ has sparse binary or CSD expansion,
   it is unlikely, in general, that so will $C^{-1}\%M$.
   Thus, for constants like 2, 8, 17 and 63, not only we have to implement
   two multiplications rather than one, but the second one may require
   a much larger circuit, and the number of ancillae can be significant.

\subsection{Modular exponentiation circuits}
   A number of mod-exp circuits have been proposed in the literature.
   The traditional approach is to implement each controlled modular multiplication
   separately and chain these operations. Circuits used in laboratory experiments with several
   qubits typically use the following shortcut. Since modular multiplications in mod-exp start
   with the value 1, the number of possible outcomes after the first $k$ multiplications
   is at most $2^k$. Therefore, for $k=1,2$, one can conditionally produce these outcomes
   without performing multiplication. This observation is also useful when many qubits
   are available, but one seeks to decrease the depth of the circuit rather than gate counts.
   In this case, one can establish one register for each conditional multiplication $Cx\%M$
   and use CNOT gates in each register to conditionally replace the initial value 1 with $C$.
   All these operations are done in parallel and followed by a tree of multipliers. At the cost of
   a several-fold increase in gate counts and an asymptotic increase in bitlines (from linear
   to quadratic), circuit depth reduces from linear to logarithmic. As long as bitlines are
   the most valuable and limited resource of quantum computers, this parallel approach remains
   impractical.

\subsection{Paper outline} \label{sec:outline}
Basic circuit blocks for addition, comparison and modular reduction
are introduced in Section \ref{sec:blocks}.
Based on these blocks, we develop
{\em multiplicative blocks} in Section \ref{sec:mult},
such as inversion, division with remainder, and multiplication by constants. In several important cases, we develop linear-sized modular multiplication circuits which were not known before. Whereas traditional circuit-synthesis algorithms \cite{SaeediM2011} operate at the bit level, we introduce word-level algorithms that perform dramatically better.
Section \ref{sec:divrem} proposes a new approach for building $Cx\%M$ circuits
based on modular decomposition of $x$ that can be implemented by compact circuits
in some cases. Section \ref{sec:trans} defines
several circuit operators for producing additional circuits.
Examples are given in Section \ref{sec:mod-mult-ex}.
Section \ref{sec:mod-exp} proposes circuits for modular exponentiation, based on techniques
from earlier sections. Section \ref{sec:modexp-ex} shows examples.

\section{Additive circuit blocks}
\label{sec:blocks}
\label{sec:add}

\begin{figure}[t]
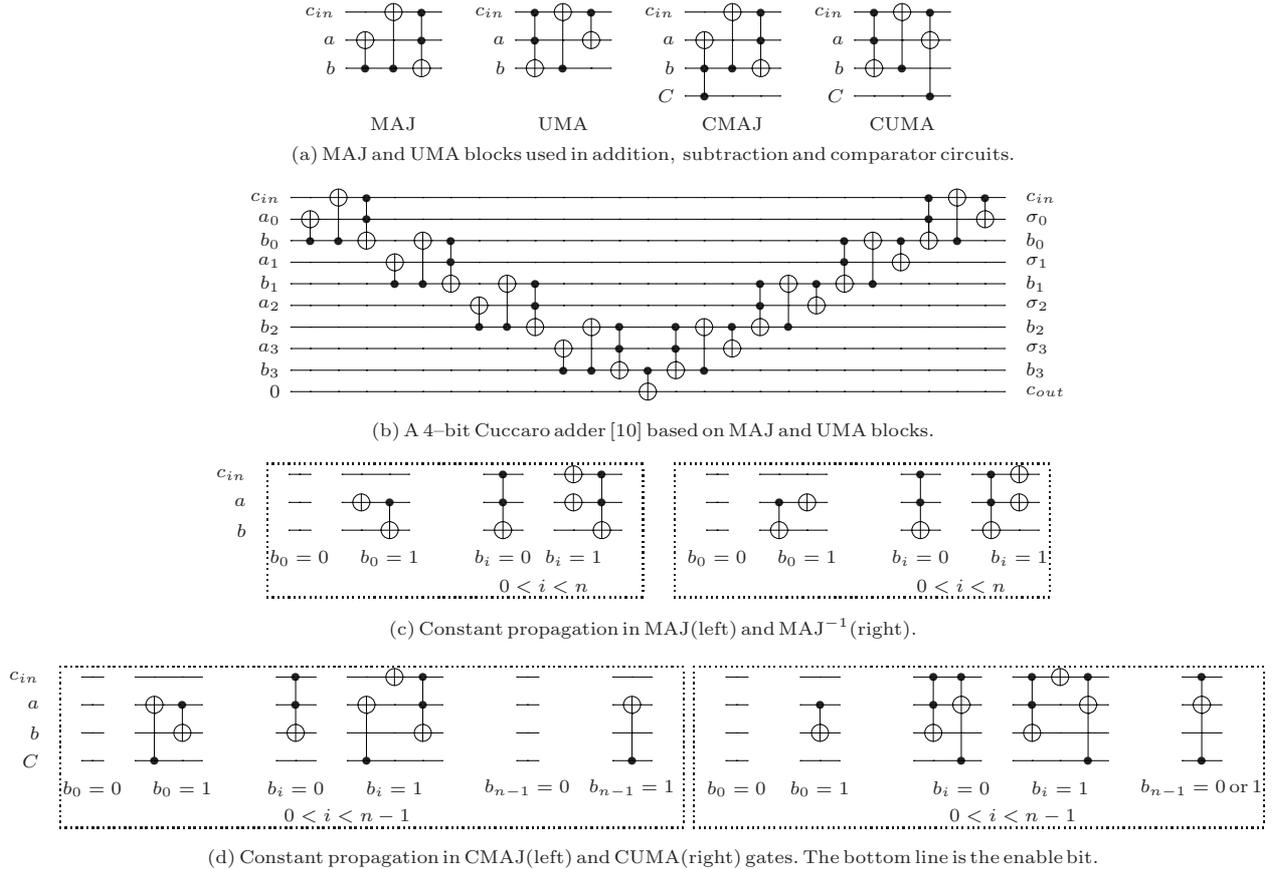

    \begin{center}$
        \scriptsize
        \begin{array}{cc}
            \input{qcircuits/MAJ-UMA}\\\\
            \rm{(a)\,MAJ\,and\,UMA\,blocks\,used\,in\,addition,\,subtraction\,and\,comparator\,circuits.}\\\\
            \input{qcircuits/CuccaroAdder}\\\\
            \rm{(b)\,A\,4 \textendash bit\,Cuccaro\,adder\,\cite{Cuccaro2004}\,based\,on\,MAJ\,and\,UMA\,blocks.}\\\\
            \input{qcircuits/MAJ-simplified-CMP}\\\\
            \rm{(c)\,Constant\,propagation\,in\,MAJ(left)\,and\,MAJ^{-1}(right).}\\\\
\scalebox{1}{
            \input{qcircuits/MAJ-UMA-simplified-ADD}
}\\\\
            \rm{(d)\,Constant\,propagation\,in\,CMAJ(left)\,and\,CUMA(right)\,gates.\,The\,bottom\,line\,is\,the\,enable\, bit.}
        \end{array}$
    \end{center}
    \centering
    \vspace{2mm}
    \parbox{\length}{
    \caption{\label{fig:blocks} Building blocks for addition, subtraction, and comparator circuits;
    $b_i$ ($0 \leq i \leq n-1$) is the $i$-th bit. In (c), $c_{out}$ is discarded. In (c) and (d), $c_{in}=0$ for the first MAJ and CMAJ block. The order of lines in (c) and (d) is the same.}
    }
\end{figure}

  Key arithmetic blocks used by modular multiplication
  are {\em adders}, {\em subtractors} and {\em comparators},
  along with their controlled variants. Such circuit blocks
  are well-known for conventional digital logic, but must be
  adapted to the reversible context so as to avoid explicit
  fanout and minimize the number of ancillae. We introduce such
  reversible blocks below and illustrate several possible circuit
  optimizations. One such optimization deals with the insertion of
  control (enable) signals.

\ \\
{\bf Addition and subtraction.}
  A number of adder circuits developed in the literature
  can be used in our constructions. To this end, Takahashi~\cite{Takahashi2010}
  describes several other adders with different trade-offs
  between circuit size, circuit depth and the required number of ancillae.
  To be specific, we are using linear-sized adders by Cuccaro et al \cite{Cuccaro2004},
  illustrated in Figure \ref{fig:blocks}b, which are the smallest known. They are built using MAJ and UMA blocks shown in Figure \ref{fig:blocks}a. An $n$-bit Cuccaro adder requires $2n$ Toffoli gates and $4n+1$ CNOT gates.
  Subtraction can be evaluated using bitwise negation as $(x-y)=(x'+y)'$ or
  as $(x-y)\% M = (x+ (M-y)) \% M$. The latter formula becomes competitive
  when the minuend $y$ is known and $M-y$ contains more $0$ bits than does $y$.

\begin{figure}[tb]
\scriptsize
\scalebox{0.95}{
\input{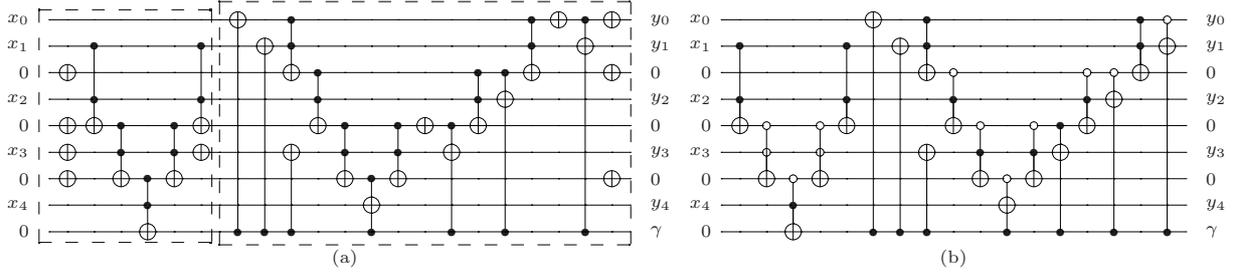}
}
\centering
\parbox{\length}{
    \vspace{2mm}
\caption{
\label{fig:Mod21}
(a) A circuit for modular reduction with $M=21$ (after optimization)
consisting of a comparator and subtractor based on the Cuccaro construction \cite{Cuccaro2004}. For $x>21$, $\gamma=1$ which activates the subtractor block. Using this circuit as a building block may require clearing or avoiding the garbage output bit $\gamma$ as in Figure \ref{fig:2x21}. (b) Further optimization using negative controls, shown with hollow circles.
}
}
\end{figure}

\begin{figure}[tb]
\scriptsize

\input{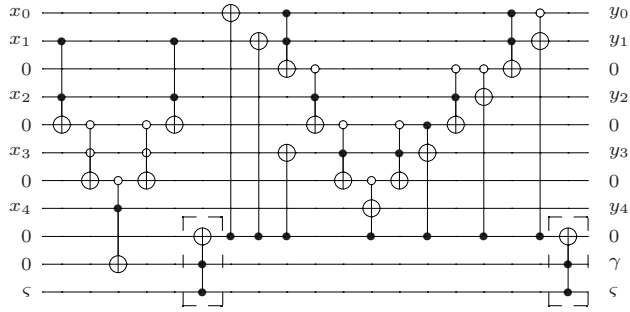}
\centering
\parbox{\length}{
\caption{
\label{fig:controlled_Mod21}
Modular reduction with $M=21$ controlled by $\varsigma$. The last Toffoli gate clears the added ancilla. Compared to Figure \ref{fig:Mod21}b, added are one ancilla and two Toffoli gates incident to $\varsigma$, of which the first enables subtraction and the second clears the ancilla.
}
}
\end{figure}

\ \\
{\bf Controlled addition.} The structure of Cuccaro adders
facilitates controlled addition with a smaller overhead.
The straightforward solution is to {\em enable} such an adder
by adding a control to every gate, requiring Toffoli gates with three
controls that need to be broken down into smaller gates.
A more economical solution is to {\em disable} a Cuccaro
adder by (1) disabling the middle CNOT gate by adding a control,
(2) ensuring that the matching MAJ and UMA gates cancel out.
A close inspection of MAJ and UMA gates suggests that their
Toffoli gates and their middle CNOT gates cancel out.
The outer CNOT gates can be disabled by adding controls,
turning them into Toffoli gates, as illustrated by CMAJ and CUMA
blocks in Figure \ref{fig:blocks}a. Thus, an $n$-bit controlled addition
is possible with $4n+1$ Toffoli gates and $2n$ CNOTs.

\ \\
{\bf Controlled addition of a constant (not modular).}
  A known $n$-bit value with $n_1$ non-zero bits can be set on zero-initialized ancillae
  using $n_1$ inverters. The adder may modify those values temporarily, but restores them
  at the end, which allows one to restore the ancillae lines to zeros for use in subsequent
  circuit blocks. Some of these inverters cancel out in the final circuit.
  When a control input of a gate is known to be 0 or 1, the gate can be simplified or
  removed entirely, as shown in Figure \ref{fig:blocks}c.
  Such optimizations can be performed by a straightforward circuit traversal.
  Not counting some of the above simplifications, such a circuit requires
  no more than $3n-5$ Toffoli gates, $n_1+2$ CNOT gates, $2n_1$ inverters.
  Given a constant $C_0$, one can compute $M-C_0$ and compare possible circuits
  for adding $C_0$ and subtracting $M-C_0$.

\ \\
  {\bf Comparators} $a<b$ are similar to subtractors --- one subtracts $a-b$ and checks $a-b<0$.
  Cuccaro adders can be modified to perform comparison, leaving their data inputs unchanged
  and producing a one-bit result as the most significant carry-bit of subtraction. Therefore,
  after the MAJ blocks, one uses inverse MAJ blocks instead of UMA blocks used in adders.
  When comparing to a known constant, simplifications are possible as in Figure \ref{fig:blocks}d.
  Comparing to a known $n$-bit constant with $n_1$ non-zero bits, such circuits require
  no more than $2n-2$ Toffoli gates, $3$ CNOT gates, and $4n_1$ inverters.

\ \\
  {\bf Modular reduction $x\%M$ for $x\leq 2M$}
  can be performed with one comparator and one conditional subtraction,
  connected serially with at most $5n-7$ Toffoli gates, $n_1+5$ CNOT gates, and $8n_1$ inverters
  ($2n_1$ inverters to set and reset ancillae before and after the computation).
  Figure \ref{fig:Mod21}a shows such a circuit and exhibits additional gate optimizations
  at the interface between the comparator and the subtractor. The inverter on $x_3$ is the
  result of simplifying a CNOT gate in a Cuccaro adder. Figure \ref{fig:Mod21}b shows
  further optimizations using Toffoli gates with negative controls.\footnote{In practice, CNOTs
  and Toffoli gates with negative controls may be as easy to implement as the gates with positive controls.
  Otherwise, additional inverters around the controls suffice. Negative controls not only result in more
  compact circuit diagrams, but can also help reading such circuits. Recall that positively-controlled \T
  gates with targets on 0-initialized ancillae compute the AND function $ab\oplus 0=ab$.
  Using negative controls and a 1-initialized ancilla computes the OR function:
  $a'b'\oplus 1 = (a+b)\oplus 0=a+b$.} Figure \ref{fig:controlled_Mod21}
  illustrates {\bf controlled modular reduction}, where the comparator and the subtractor remain intact,
  but the result of comparison is conditioned on the new control using a new ancilla. This ancilla is
  cleared at the end, but the garbage output $\gamma$ inherited from uncontrolled modular reduction remains. Modular reduction for $x\ll M$ is discussed in Section \ref{sec:mult} under division with remainder.

\ \\
{\bf Conditional modular addition of a constant.} The straightforward implementation $y= (x+a) \% M$
 by adding a constant and then performing mod-$M$ reduction clears the added carry bit,
 but leaves a garbage bit. This bit can be cleared by $(y<a)$. To avoid
 the carry, precompute $\alpha_M=M-a$, use $y=(x< \alpha_M ~?~ x+a : x-\alpha_M )$\label{fn:tern}\footnote{The ternary \texttt{conditional} $a ~?~ b : c$ is similar to \texttt{if(a) then b else c}, but is more flexible. It returns the value of $b$ or the value of $c$ (it can be an l-value).} and clear
 the ancilla via $(y<a)$. Comparators optimized for $x<\alpha_M$ may be smaller
 than those for $x<M$.

\section{Multiplicative circuit blocks}
\label{sec:mult}

\begin{figure}[tb]
\scriptsize
\scalebox{0.95}{
\input{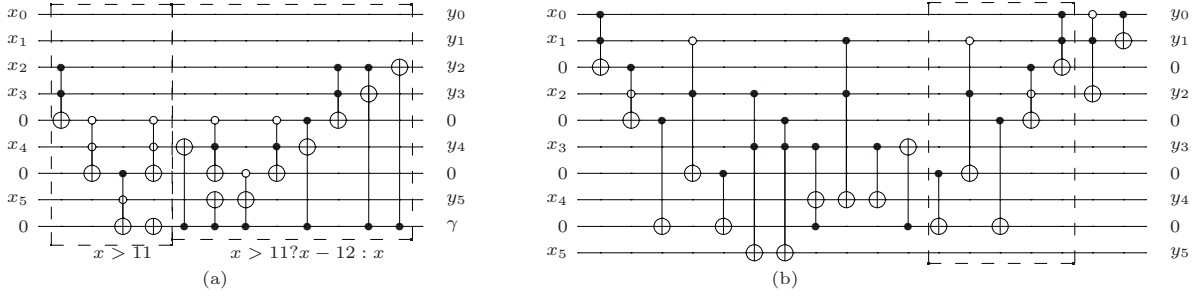}
}
\centering
\parbox{\length}{
\vspace{2mm}
\caption{
\label{fig:3x35-1}
Circuits for (a) modular reduction with $M=12$ (after optimization), and (b) computing $f(x)=3x$ for $x<16$.\ \
In (b), gates in the dashed box clear ancillae.}
}
\end{figure}

 We now develop several circuits for $Cx\%M$ and related operations,
 using additive building blocks from Section \ref{sec:blocks}.

  \ \\
  {\bf Circuits for $(2^k + 1)x$} {(not modular) can be constructed by shifts and adds, but the challenge is to avoid unnecessary garbage ancillae. Our circuits are structured as follows.
   For bit values $x_i$ $(i<n)$ of $x$, the bit values of $(2^k+1)x$, $S_i$, can be constructed by a $k$-bit shift of $x$ followed by an $n+k$ bit add (i.e., $2^k x + x$). The addition can be performed by a generic Cuccaro adder --- $2^k x$ on main qubits, $x$ on ancillae, --- but clearing these ancillae is difficult. Another approach is to construct logical sub-expressions for output bit $i$ based on the bit values of $x_1 \cdots x_i$. Formula \ref{eq:s_i} gives sub-expressions for each $S_i$ bit.
   To calculate each $S_i$, we precompute the incoming carry $c_i$ in  Formula \ref{eq:s_i} and
   store it on an ancilla. For $n$ such ancillae, we need at most $3n$ Toffoli gates. To construct $S_i$ values, at most $3n$ CNOT gates suffice.
   To clear the $c_i$ ancillae after use, the Toffoli gates that computed them are performed in reverse (their inputs did not change). With the additional $3n$ Toffoli gates to clear ancillae, a circuit for $(2^k+1)x$ needs up to $6n$ Toffoli gates and $3n$ CNOTs. Figure \ref{fig:3x35-1}b illustrates a 4-bit $3x$ circuit after two optimizations: $(i)$ literal reduction in $S_i$ and $c_i$ sub-expressions, and
   $(ii)$ absorbing inverters in Toffoli gates with negative controls.}

\begin{figure}[t]
\scriptsize
\input{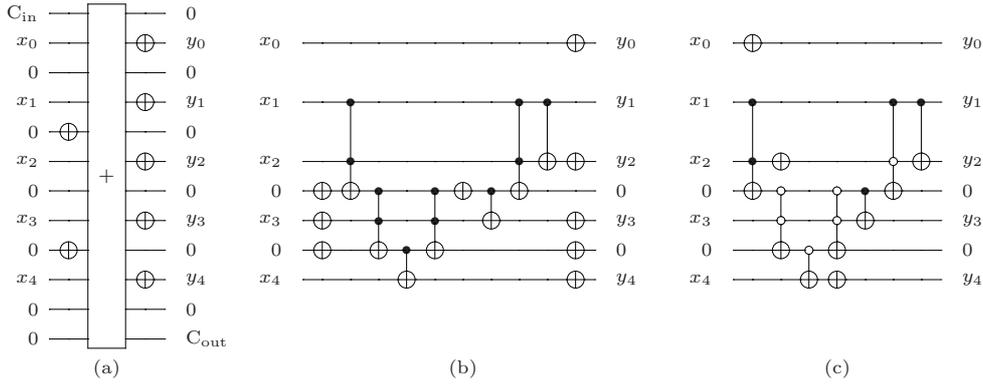}
\centering
\parbox{\length}{
\caption{
\label{fig:20x21}
Circuits for $-x \% 21$ based on a 5-bit Cuccaro adder: (a) Inverters on the zero-initialized ancillae prepare the value 10=31-21=21', (b) an optimized circuit, (c) further optimization using negative controls.}
}
\end{figure}

\begin{small}
\begin{equation}
\label{eq:s_i}
S_i  = \left\{ \begin{array}{l}
 x_i \,\,\,\,\,\,\,\,\,\,\,\,\,\,\,\,\,\,\,\,\,\,\,\,\,\,\,\,\,\,\,\,\,\,\,\,\,0 \le i < k \\
 x_i  \oplus x_{i - k}  \oplus c_i \,\,\,\,\,\,k \le i < n \\
 x_{i - k}  \oplus c_i \,\,\,\,\,\,\,\,\,\,\,\,\,\,\,\,\,\,n \le i < n + k \\
 x_{i - k - 1} c_{i-1} \,\,\,\,\,\,\,\,\,\,\,\,\,\,\,i = n + k \\
 \end{array} \right.\,c_i  = \left\{ \begin{array}{l}
 0\,\,\,\,\,\,\,\,\,\,\,\,\,\,\,\,\,\,\,\,\,\,\,\,\,\,\,\,\,\,\,\,\,\,\,\,\,\,\,\,\,\,\,\,\,\,\,\,\,\,\,\,\,\,\,\,\,\,\,0 \le i \le k \\
 x_{i - 1} x_{i - k - 1}  \oplus x_{i - 1} c_{i - 1}  \\
\,\,\,\,\,\,\,\,\oplus x_{i - k - 1} c_{i - 1} \,\,\,\,\,\,\,\,\,\,\,\,\,\,\,\,\,\,\,\,k + 1 \le i \le n \\
 x_{i - k - 1} c_{i - 1} \,\,\,\,\,\,\,\,\,\,\,\,\,\,\,\,\,\,\,\,\,\,\,\,\,\,\,\,\,\,\,\,\,\,n < i < n + k \\
 \end{array} \right.
\end{equation}
\end{small}

\ \\
 {\bf Circuits for $-x \% M$.} For bitwise negation $x'$, recall $x'=2^n-1-x$. Therefore,
 $(x+M')' = 2^n-1-(x+M') = 2^n-1 -x - (2^n-1-M) = M-x=-x\%M$. Therefore, for $x > M$, $-x \%M$ can be computed as $(x+M')'$ using one Cuccaro adder, as illustrated in Figure \ref{fig:20x21} for $M=21$ and $M'=31-21=10$. The proposed circuit maps $x=0$, $x=M$, and $x> M$ into $M$, $0$, and $2^n+(M-x)$, respectively. Note that
 inverting any circuit for $-x\%M$ will produce a circuit computing $-x\%M$ because $(M-(M-x))=x$. A circuit for  conditional $-x \% M$ can be constructed by converting each inverter to a CNOT gate and applying the conditional modular reduction discussed in Section \ref{sec:add} and illustrated in Figure \ref{fig:controlled_Mod21}.

\begin{figure}[b]
\scriptsize
\scalebox{0.9}{
\input{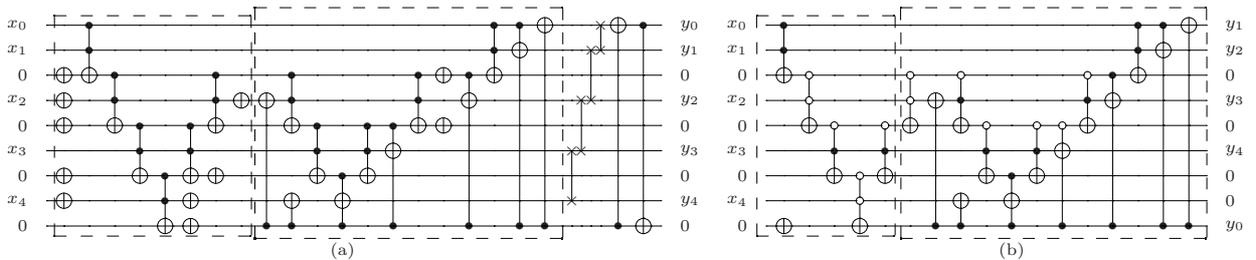}
}
\centering
\vspace{2mm}
\parbox{\length}{
\caption{\label{fig:2x21}
{Circuits for $f(x) = 2x\%21$, (a) module-based design, (b) after three optimizations. The first and second sub-circuits in these figures are for ($x>10$) and ($x>10$ ? $x-11$ : $x$), respectively. In (a), SWAP gates are used to compute $2x$, the second-to-last CNOT adds 1 (i.e., $2x+1$), and the last CNOT clears the bottom ancilla. These gates can be removed by reordering output labels as shown in (b).}
}
}
\end{figure}

 \ \\
 {\bf Circuits for  $2^k x \% M$} for odd $M>2$.
   We start with a linear-sized circuit for $2x\% M$ that clears its ancillae.\footnote{Circuits in the literature may exhibit quadratic size because,
   to clear ancillae, they implement $x/2~\%M$ by finding the modular
   inverse of 2 (see Section \ref{sec:blocks}) and decomposing it in binary.}
   The bulk of our $2x\%M$ circuit computes $x\% \lceil M/2 \rceil$ using a modular-reduction
   circuit we described earlier, which evaluates $x \geq \lceil M/2 \rceil$ on a 0-initialized ancilla,
   but also zeros out the most significant bit. To multiply $x\% \lceil M/2 \rceil$ by two, it suffices
   to rotate the bits, which moves the most significant zero into the least significant position.
   One also needs to (a) change the LSB to 1 conditional on the ancilla --- this can be done with a CNOT,
   (b) clear the ancilla conditional on the LSB --- this can be done with another CNOT.
   This circuit is illustrated in Figure \ref{fig:2x21}a. Further circuit optimization uses
   {\em three tricks}. {\bf One} is the merger of inverters into negative controls (shown with
   hollow circles) of Toffoli gates, this may benefit from creating pairs of canceling
   inverters and/or moving inverters through targets of CNOT/Toffoli gates.
   {\bf The second} optimization deals with the two CNOTs at the end of the circuit.
   It creates a pair of canceling CNOTs prior to them, so that three CNOTs can be combined
   into a SWAP. The remaining CNOT gate is controlled by a 0 value created by doubling, thus can be removed.
   We are left with a chain of SWAP gates that rotate the significant bits and onto
   an ancilla, in particular, the 0 value is rotated onto the ancilla
   (at which point all ancillae are cleared). {\bf The third} optimization interprets
   the bit rotation at the end of the circuit as a relabeling of outputs.
   The resulting circuit in Figure \ref{fig:2x21}b works correctly
   only for $x\leq M$, but $x\leq M$ can be guaranteed in Shor's algorithm.

   Since output relabeling cannot be used in a controlled $2x\%M$ circuit,
   controlled rotation can be implemented with controlled-SWAP gates.
   However, when multiple $2x\%M$ circuits are concatenated to implement $2^k\%M$,
   all controlled rotations can be merged into one such rotation at the end of the circuit.

\ \\
  {\bf Modular reduction $2^i x\%M$ for $M=2^k\pm1$} can be performed using a well-known algorithm. To compute $x\% (2^k-1)$, add $2^k$-ary digits of $n$-bit $x$ modulo $2^k-1$. To compute $x\% (2^k+1)$, alternate addition and subtraction of $2^k$-ary digits of $n$-bit $x$ modulo $2^k+1$. When $k>n$, no gates are needed. Otherwise, one can use $\lceil n/k\rceil$ Cuccaro adders on $\log_2 M$ bits. In the case $M=2^k - 1$, the output carry of each adder can be ignored. Hence, Toffoli and CNOT gate counts are $\lceil n/k\rceil \cdot 2k$ and $\lceil n/k\rceil (4k+1)$, respectively. For $M=2^k + 1$, the output carry of each Cuccaro adder should be considered. In this case, at most $\lceil n/2k\rceil$ mod-$M$ reduction modules on $\log_2 M$ bits are sufficient. Therefore, the numbers of Toffoli and CNOT gates are $\lceil n/k\rceil (2k+2) + \lceil n/2k\rceil (5k-2)$ and $\lceil n/k\rceil (4k+5) + \lceil n/2k\rceil (n_1+5)$ ($n_1 \leq k+1 $ represents the number of non-zero bits in $M$), respectively. Another approach to implement the required additions and subtractions is to implement the counters $(0,1,...,2^k\pm1-1)$ and $(2^k\pm1-1,2^k\pm1-2,...,0)$ conditional on bit values of $n$-bit $x$ as illustrated in Figure \ref{fig:3x35} for $x\%3$. Clearly, no $\%M$ modules will be required in this approach. A factor of $2^i$ only changes the indices of the bits read by the baseline algorithm. All circuits constructed here exhibit linear number of CNOT and Toffoli gates in terms of $n$.

\ \\
{\bf Special case} $2^kx\%M$ where $M=2^n-1\pm d$ and both $k$ and $d$ are very small.
 Breaking down $x$ into $k$ more significant bits and $n-k$ less significant bits, we write
$ 
 x=2^{n-k} x_k^{hi}+x_{n-k}^{lo}=2^{n-k} (x / 2^{n-k})+x \% 2^{n-k}
$ 
and then
\begin{small}
$$
 Cx \% M = \big( 2^n x_k^{hi} +2^k x_{n-k}^{lo} +0_{k} \big) \% M
$$
\begin{equation}
\label{eq:d}
 = \big(x_k^{hi} (1\pm d) + 2^k x_{n-k}^{lo} \big) \%M
   = \big((2^k x_{n-k}^{lo}+x_k^{hi} )\pm d x_k^{hi} \big) \%M
   = ( \rot_k(x) \pm  d x_k^{hi} ) \% M
\end{equation}
\end{small}

\noindent
where rot$_k(x)$ is a cyclic shift (rotation) of $x$ by $k$ bits
(rot$_k(x)$ may exceed $M$). Note that when $d=0$ or $d=2$, we get
a well-known special case described above.
When $|d|<2^{n-k}$ modular reduction can be computed by
subtracting $M$ if the number exceeds $M$, which allows one
to compute the product $dx_k^{hi}$ through a series of shifts,
additions and subtractions. For larger values of $d$, we can
write $2^k=2^{k_1+k_2}$ such that $|d|<2^{n-k_1}$ and $|d|<2^{n-k_2}$,
then compute $2^k x$ by multiplying by $2^{k_1}$ and by $2^{k_2}$ in separate steps.
Another approach would let the $i$-th bit of $x_k^{hi}$ control the modular addition
of a precomputed constant $d2^i~\%~M$, as shown in Section \ref{sec:add}.

\ \\
  {\bf Division with remainder} circuits convert $x$ into $x/\rho,x\%\rho$ without loss or gain of information. A simple example is given by $\rho=2^k$, where the quotient and the remainder are simply
  the $n-k$ high and the $k$ low bits of $x$. Previously, we have also shown linear-sized remainder circuits for $2^n\pm 1$. In general, division can be performed by a series of subtractive modular reductions, whose ancillae accumulate the bits of the quotient. When $x<2^k\rho$, the most significant bit is computed by a mod-$2^{k-1}\rho$ reduction, the next bit by a mod-$2^{k-2}\rho$ reduction,  etc for a total of $k+1$ reductions. The last reduction produces the remainder.

\section{$Cx\%M$ using division with remainder}
\label{sec:divrem}
 We propose the following computation of $Cx\%M$.

\begin{theorem} \label{thm:DR}
  Consider integers $0\leq x <M$, $1<C<M$ with $\mathrm{gcd}$$(C,M)=1$.
  Define $\rho=\lceil \frac{M}{C} \rceil$ and $\delta=C-M\%C$. Then
\begin{small}
\begin{equation}
\label{eq:mod-decomp}
    Cx\%M = \big( \delta \lfloor x/\rho \rfloor + C(x\%\rho) \big) \%M
\end{equation}
\end{small}
   Furthermore, when $C^2<M$,
\begin{small}
\begin{equation}
\label{eq:bound}
    \delta \lfloor x/\rho \rfloor + C(x\%\rho) < 2M
\end{equation}
\end{small}
   so that a single subtractive mod-$M$ reduction suffices.
\end{theorem}

\noindent
{\bf Proof.}  Clearly, $x=\rho \lfloor x/\rho \rfloor + x\%\rho$.
Then $Cx=C \rho \lfloor x/\rho \rfloor + C(x\%\rho)$. We leave the latter term as is
because $C(x\%\rho) \leq C (\rho-1) = C \lfloor M/C \rfloor < M$. To reduce the former term, note that
 $M < C \rho <2M$, thus $ C\rho \% M = C\rho - M$. Substitute $M=C\lfloor M/C \rfloor + M\%C = C\lceil M/C \rceil - C + M\%C $ to obtain $ C\rho \% M = C-M\%C = \delta$, proving Formula \ref{eq:mod-decomp}.

 Since $\delta < C$ and $\lfloor x/\rho\rfloor < Cx/M < C$, we have $\delta \lfloor x/\rho \rfloor$ $\leq C^2$ which proves Formula \ref{eq:bound}. \hfil \qed

\begin{figure}[tb]
\scriptsize
\input{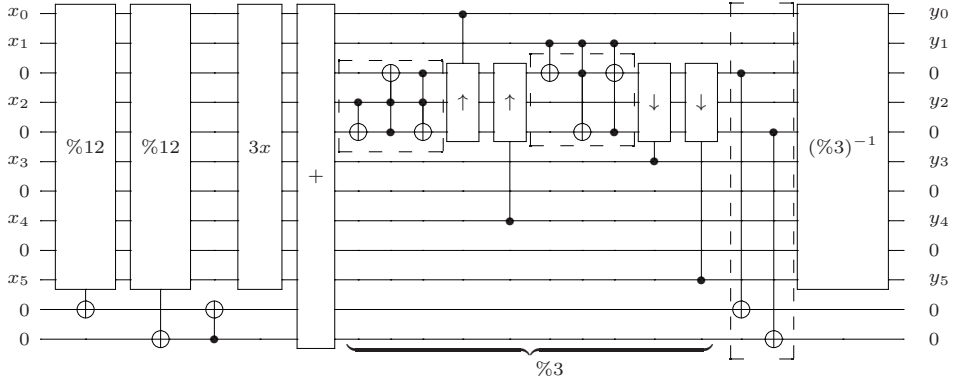}
\centering
\parbox{\length}{
\caption{
\label{fig:3x35}
A circuit for 3x\%35 in Example \ref{ex:3x35}. Circuits for $3x$ and modular reduction with $M = 12= \lceil 35/3 \rceil$ are illustrated in Figure \ref{fig:3x35-1}. The two mod-12 reductions compute $(x/12,x\%12)$, the latter is unconditionally multiplied by 3 (the $3x$ box) and the result is added to the former (the adder box). The
 \%3 module modifies two highest-placed ancilla bits, counting up (0,1,2) and down (0,2,1). It consists of three
 count-up ($\uparrow$) and three count-down ($\downarrow$) blocks --- the first block of each kind is shown in detail. The two CNOTs that target the last two lines clear ancillae set by \%12 modules. The $(\%3)^{-1}$ module clears ancillae set by the \%3 module.}
}
\end{figure}

  To construct reversible circuits using this result, use the circuits for division with remainder
  from Section \ref{sec:mult} to represent $x$ by the pair  ($\lfloor x/\rho \rfloor$, $x\%\rho$)
  without a loss or gain of information. This will require $\lceil \log_2 C \rceil$ subtractive
  mod-$2^i\rho$ reductions, with the $\lceil \log_2 C \rceil$-bit remainder stored in ancilla (for $C=3$, two mod-$\rho$ reductions
  are performed).
  A challenging part is to implement multiplications by constants $\delta \lfloor x/\rho \rfloor$ and $Cx\%\rho$ with reversible circuits, so that ancillae are cleared. This is illustrated in Section \ref{sec:mult} for $C=2^n\pm1$. After modular addition, ancillae can be cleared by computing $(Cx\%M)\%C$, which takes linear time for $C=2^n\pm 1$ as explained in Section \ref{sec:blocks}.
\begin{example} \label{ex:3x35}
    This approach is illustrated in Figure \ref{fig:3x35} for $3x\%35$
     where $\rho=12$, $\delta=1$. We implement $\lceil \log_2 3 \rceil=2$ subtractive mod-$12$ and mod-$24$ reductions by two successive \%12 modules. Accordingly, the second-to-last and the last ancillae evaluate to 1 when $12 \leq x$ and $24 \leq x$, respectively. After the first CNOT, the ancillae will be 1 for $12 \leq x < 24$ and $24 \leq x$. Therefore, the values of the lowest-placed ancillae are 0, 1 or 2 based on the value of $x$. Computing $3x$ (not modular) and adding 1 for $12 \leq x < 24$ or 2 for $24 \leq x$ (Formula \ref{eq:mod-decomp}) implement $3x\%35$. To clear the ancillae used, one needs to implement \%3 on two new ancillae (two highest-placed ancilla bits in Figure \ref{fig:3x35}) and uses the bits to control two CNOT gates. Since 2\%3 = −1, we can rewrite $x=2^5x_5+2^4x_4+\cdots+2x_1 + x_0$ \%3 as $-x_5 + x_4 - x_3 +  x_2 - x_1 + x_0$ \%3 which can be implemented by three up-counters conditioned on even bits and three down-counters conditioned on odd bits. The \%3 computation can be undone by applying \%3 in reverse (indicated by the $(\%3)^{-1}$ block in Figure \ref{fig:3x35}).
\end{example}

  Values $C=O(1)$ facilitate linear-time mod-$\rho$ decomposition by subtractive reductions
  and also imply $\delta=O(1)$, $x/\rho=O(1)$. Therefore the first multiplication
  can be performed through controlled additions of constants. Given our circuits for
  (not modular) multiplication by $(2^k+1)$ in Section \ref{sec:mult}, Formula \ref{eq:mod-decomp}
  can be used with $C=5,9,17,33,\ldots$. Circuits for multiplication by $\delta=C-M\%C$
  are available for $\delta$ of the form $2^k$ and $2^k+1$.

  Working with $C>\sqrt{M}$ directly can be difficult because many modular reductions may be required in Formula \ref{eq:mod-decomp}, and their ancillae must be cleared. It helps to postpone, until the end, clearing the ancilla that contain $x\%\rho$, and use them to clear the ancillae for modular reductions. Another trick is to avoid unnecessary modular reductions by interpreting each multiplication $\%M$. In particular, large $C$ values can be replaced by $M-C$ if the addition is replaced by subtraction. In this context, some large $C$ values may also be convenient when  $\delta=O(1)$ and $\rho=2^{O(1)}$, and thus the second multiplication can be performed through controlled additions of constants.

To count the number of Toffoli and CNOT gates for $x\mapsto (x/\rho,x\%\rho)$, we use $\lceil \log_2 C \rceil$ subtractive mod-$2^i\rho$ reductions and $\lceil \log_2 C \rceil$ ancillae. The reductions go from larger numbers to smaller numbers, ending with $\rho$. A $2^i\rho$-reduction module operates on $\lceil \log_2 \rho \rceil+i$ bits.
Hence, the number of Toffoli gates will be $\Sigma^{0}_{i=\lfloor \log_2(M/\rho) \rfloor} (5\lceil\log_2 \rho \rceil -7+5i)$ $< \lceil \log_2 C \rceil (5\lceil \log_2 M \rceil-7)$. To compute $(2^k+1)x \%M$ by division with remainder, one additionally uses a multiplicative module $Cx$ (not modular) to compute $C(x\%\rho)$. Consequently, one Cuccaro adder and one $\%M$ module are employed to add $\lfloor x/\rho \rfloor$ to the result and apply the mod-$M$ reduction. To clear ancillae by computing $(Cx\%M)\%C$, two $\%C$ modules and $O(1)$ gates are necessary.
Additionally, $\lceil \log_2 C \rceil$ and $\lceil \log_2 M \rceil+1$ ancillae are required for the first modular reductions and other blocks, respectively.

\begin{table}[tb]
 \parbox{\length}{
\caption{ \label{table:blocks}
Number of gates in circuit blocks.
In this table, $n_1$ and $n_1'$ represent the number of non-zero bits in $M$ and $C$ and $n=\lceil \log_2 M \rceil$, and $n'=\lceil \log_2 C \rceil$.
\T, \CC, and \AAA ~are the number of Toffoli, CNOT and ancillae, respectively.
More accurate gate count for division with remainder is discussed in text.
Further constant-specific optimizations may be possible. }
 }
    \scriptsize
\scalebox{1}{
    \begin{tabular}{|lr|p{2cm} p{2cm} l|}
        \hline
            & & \multicolumn{3}{c|}{Gates and ancillae}  \\
        Circuit block       & Formula & \T & \CC & \AAA \\
        \hline
        \hline
        Cuccaro adder   &  \begin{scriptsize}$\hspace{-0.5cm} (x,y)$ $\mapsto$ $(x,x + y)$\end{scriptsize}              & $2n$ & $4n+1$& $1$           \\ \hline
        Controlled addition & \begin{scriptsize} $\hspace{-0.5cm}(x,y)$ $\mapsto$ ctrl $?$ $(x,x + y)$ : $(x, y)$ \end{scriptsize}             & $4n+1$ & $2n$& $1$            \\ \hline
        Controlled addition of a constant & \begin{scriptsize}$\hspace{-0.5cm}(x,y)$ $\mapsto$ ctrl ? $(x,x + M)$ : $(x, y)$\end{scriptsize} & $3n-5$ & $n_1+2$& $n+1$     \\ \hline
        Comparator   & \begin{scriptsize} $\hspace{-0.5cm}(x>M)$ \end{scriptsize}                   & $2n-2$&  $3$& $n+1$        \\ \hline
        Subtractive modular reduction  & \begin{scriptsize}$\hspace{-0.5cm}(x,y)$ $\mapsto$ $x>M$ ? $(x,x -M)$ : $(x, y)$ \end{scriptsize}               & $5n-7$& $n_1+5$& $n+1$    \\ \hline
        Special-case multiplication & \begin{scriptsize}$\hspace{-0.5cm}(2^k + 1)x$   \end{scriptsize}                 & $6n$& $3n$& $n+1$            \\ \hline
        Negation & \begin{scriptsize}$\hspace{-0.5cm}-x \% M$ \end{scriptsize}             & $2n$& $4n+1$& $n+1$        \\ \hline
        Modular multiplication by powers of two & \begin{scriptsize}$\hspace{-0.5cm}2^k x \% M$     \end{scriptsize}      & $k(5n-7)$& $k(3n+n_1+5)$& $n+1$   \\ \hline
		Division with remainder & \begin{scriptsize}$\hspace{-0.5cm} x\mapsto (x/\rho,x\%\rho), \rho=\lceil \frac{M}{C} \rceil$ \end{scriptsize}& $n'(5n-7)$ & $n'(n_1+5)$& $n'$+$n$ \\ \hline
		Special-case modular multiplication	& \begin{scriptsize}$\hspace{-0.5cm}(2^k+1)x \%M$   \end{scriptsize}     & $(k+2)(5n-7)+8n+ 2(5n'- 7) +O(1)$ & $(k+2)(n_1+5)+ 7n+ 2(n_1'+5) +O(1)$ &  $k$+$n$+$2$                                     \\ \hline
\end{tabular}
}
\end{table}

\section{Circuit operators and decompositions} \label{sec:trans}
Given small circuits proposed earlier, we find additional $C$ values for which small $Cx\%M$ circuits exist.

\subsection{Multiplicative decompositions}
We employ two {\em circuit operators} --- inversion and negation --- that convert a reversible circuit for $Cx\%M$ into a circuit for $\tau(C)x\% M$, where the function $\tau(\cdot)$ characterizes the transform.

\ \\
\noindent
{\bf Inversion} reverses the order of the gates and replaces each gate
  with its inverse (inverters, CNOT gates, swaps and Toffoli gates are self-inverse).
  Circuit size is preserved. The generated circuit computes
  $\tau(C)x \% M$, where $\tau(C)$ is the mod-$M$ inverse of $C$, i.e., $\tau(C)C \% M =1$.
  When gcd$(C,M)=1$, modular inverse exists, is unique and can be computed
  by the extended Euclidean algorithm \cite{BrentZ2010}. When applied to small-power-of-two circuits ($C=2^k$),
  inversion produces negative-power-of-two circuits ($C=2^{-k} \% M$)
  and generates new convenient $C$ values unless $M=2^n-1$.

\ \\
\noindent
{\bf Negation} entails $\tau(C)= M-C = -C \%M$.
  Note that $-x\%M = M-x = (2^n-1)\pm d -x = x' \pm d$,
  where $'$ performs bitwise negation.
  Therefore, the circuit operators adds an inverter
  on every wire and performs one modular addition/subtraction
  with $d$, either before or after modular multiplication by $C$.
  Circuit size increases. When $M$ is odd, so are $M-2^k$, producing new {\em convenient values}.
Combining negation with inversion may produce additional
convenient values. Given that the two transforms commute,
applying inversion and negation to small powers of two
produces at most $4\lceil \log_2 M \rceil$ convenient values
(including small powers of two), which can be a lot smaller than $\varphi(M)$.

\ \\
{\bf Modular products.}
Composing compact circuits for convenient constants {\em in series},
one can often obtain additional convenient constants $C= C_1 C_2 \% M$.
However, when multiplying small positive and negative powers of two, no new values can be obtained.
Multiplying positive powers of two (or negative powers of two) does not help when $M=2^n-1$, e.g., for $M=15$.
Products with negated powers of two  do not give new convenient values when  $M=2^n+1$, e.g., for $M=33$.
 In general, since $(\mathbb{Z}/M\mathbb{Z})^\times$ is a product
of two cyclic groups, it suffices to build compact reversible circuits for
its two generators and compose them in various ways to produce reversible
circuits for all other group elements. This strategy is impractical because
$(i)$ the composed circuits will often be larger than necessary, $(ii)$ it is
not clear how to identify a pair of generators without knowing $p$ and $q$.

\subsection{Additive decompositions and a shortest-path formalism} \label{sec:formalism}
For large $C$, the multiplicative operators described above may be insufficient.
To also consider additive operators, we introduce a zero-initialized ancilla register which is cleared
after $Cx\%M$ is computed in the primary register. A value is copied into this register from the primary
register using a parallel chain of CNOT gates. Multiplicative operators can be applied to individual registers,
and additive operators replace the contents of one of the register with the modular sum or difference of two values (note
that these operations are reversible). The operators we consider are listed in Table \ref{tab:ops}, along
with their costs, measured as the number of \T ~gates (which dominate quantum cost).  We use the following circuit descriptions.

\begin{itemize}
\item Every step/operator takes exactly two characters
\item Odd-numbered characters are operator types: \texttt{c},\texttt{\~{}},\texttt{+},\texttt{-},\texttt{d},\texttt{h},\texttt{r},\texttt{t},\texttt{v},\texttt{f}
\item Even-numbered characters are register indices: 1 or 2.
\end{itemize}

For example, the literal \texttt{c2} represents a bit-wise CNOT operation with Register 2 as its target. It is meant to copy the contents of Register 1 to a zero-initialized Register 2 (or clear Register 2, when it duplicates Register 1). The same can be accomplished using the modular addition operator \texttt{+2} (the modular subtraction operator \texttt{-2}, respectively), but at a higher cost. The circuits \texttt{c2c2}, \texttt{+2-2} and \texttt{r1t1} do nothing, and the circuit \texttt{c2c1c2} swaps the contents of the two registers. As a more complex example, to compute
$(x,0) \mapsto (3x\%65,0)$ without the multiplicative $3x\%M$ operator we introduced earlier,
one might use the circuit  \texttt{c2+1+1+2+2d2+2d2d2c2}. It uses 154 \T ~gates, and is smaller than our generic $3x\%M$ circuit. However, such compact circuits need to be discovered for each $M$.
We reduce this task to finding a shortest path in a graph where the vertices represent possible
two-register states relative to the initial state $(x,0)$. For $0\leq a,b<M$, vertex $(a,b)$ represents $(ax,bx)$. The source vertex is $(1,0)$. The weighted edges represent operators from Table \ref{tab:ops} with respective costs. When traversing this graph, vertices and edges can be generated on the fly.
\begin{theorem} \label{thm:pathlen}
  For an $n$-bit value $M$ and any $0<C<M$ coprime with $M$, the worst-case gate count of optimal two-register circuits
  mapping $(x,0)\mapsto(Cx\%M,x)$ and $(x,0)\mapsto(Cx\%M,0)$ is $O(n^2)$.
\end{theorem}

\noindent
{\bf Proof.} Once the statement for $(x,0)\mapsto(Cx\%M,x)$ is proven, the statement for
$(x,0)\mapsto(Cx\%M,0)$ follows by Bennett's construction for clearing ancillae (Section \ref{sec:MM_pre_work}) which produces a circuit of the second kind by composing two circuits of the first kind.
Consider the binary decomposition of $x=\Sigma_i 2^i b_i$ and traverse it from the most significant bit.
Before considering a new bit, apply the \texttt{d2} operator, except when the second register holds value 0.
Upon seeing bit 1, apply the \texttt{+2} operator.  For example, $x=13=0b1101$ leads to operators \texttt{+2d2+2d2d2+2}, which produce

\begin{small}
$$(x,0)\mapsto(x,x)\mapsto(x,2x)\mapsto(x,3x)\mapsto(x,6x)\mapsto(x,12x)\mapsto(x,13x)$$
\end{small}
To swap the register values, one can apply \texttt{c2c1c2} or \texttt{c1c2c1}, but this may be unnecessary
within Bennett's construction.
Each operator uses $O(n)$ gates. The circuits use $n-1$ \texttt{d2} operators and up to $n$ \texttt{+2} operators, thus $O(n^2)$ gates total. \hfill  \qed

The upper bound on the \T-cost of $(x,0)\mapsto(Cx\%M,x)$ circuits implied by our proof is
$n(5n-7)+2n^2= 8n^2-7n$, with the average-case estimate $n(5n-7)+n^2= 7n^2-7n$ because half of the bits are 0
on average. These bounds can be improved by considering the canonical signed digit (CSD) decomposition,
which uses not only additions but also subtractions, and ensures that at least one of each two neighboring bits is a 0.
Thus $7n^2+O(n)$ becomes a worst-case bound, and the average case improves to $6\frac{2}{3}n^2+O(n)$.
For $(x,0)\mapsto(Cx\%M,0)$ circuits, doubling the above estimates due to the use of Bennett's construction
produces $14n^2+O(n)$ in the worst case and $13\frac{1}{3}n^2+O(n)$ on average. The smallest-cost circuits
we report in Section \ref{sec:largerM} improve upon these bounds by factors 2-4, but not asymptotically.
We also note that our shortest-path construction produces $O(n)$-sized circuits in some basic cases,
such as $C=2,3,4,..$. In contrast, resorting to Bennett's construction with binary or CSD expansion involves
the modular inverse of $C$ and typically leads to $n^2$-sized circuits.

\begin{table}
\caption{\label{tab:ops}Circuit operators for two $n$-bit registers and their costs
                        in terms of the number of \T ~gates. Note that for each operator,
                        its inverse is also listed in the table and has the same cost.}
\begin{center}
\scriptsize
\begin{tabular}{|r|l|r|l|}
\hline
  Op code & Binary/Unary & Cost & Transformation \\
\hline
  c1  &  Binary & 0 & $(0,y)\leftrightarrow (y,y)$ \\
  c2  &  Binary & 0 & $(x,0)\leftrightarrow (x,x)$ \\
  \~{}1 &  Unary  & $2n$     & $(x,y)\mapsto (-x\%M,y)$ \\
  \~{}2 &  Unary  & $2n$     & $(x,y)\mapsto (x,-y\%M)$ \\
  +1  &  Binary & $2n$     & $(x,y)\mapsto ((x+y)\%M,y)$ \\
  +2  &  Binary & $2n$     & $(x,y)\mapsto (x,(x+y)\%M)$ \\
  -1  &  Binary & $2n$     & $(x,y)\mapsto ((x-y)\%M,y)$ \\
  -2  &  Binary & $2n$     & $(x,y)\mapsto (x,(y-x)\%M)$ \\
  d1  &  Unary & $5n-7$   & $(x,y)\mapsto (2x\%M,y)$ \\
  d2  &  Unary & $5n-7$   & $(x,y)\mapsto (x,2y\%M)$ \\
  h1  &  Unary & $5n-7$   & $(x,y)\mapsto (x/2\%M,y)$ \\
  h2  &  Unary & $5n-7$   & $(x,y)\mapsto (x,y/2\%M)$ \\
  r1  &  Unary & $33n-35$ & $(x,y)\mapsto (3x\%M,y)$ \\
  r2  &  Unary & $33n-35$ & $(x,y)\mapsto (x,3y\%M)$ \\
  t1  &  Unary & $33n-35$ & $(x,y)\mapsto (x/3\%M,y)$ \\
  t2  &  Unary & $33n-35$ & $(x,y)\mapsto (x,y/3\%M)$ \\
  v1  &  Unary & $38n-42$ & $(x,y)\mapsto (5x\%M,y)$ \\
  v2  &  Unary & $38n-42$ & $(x,y)\mapsto (x,5y\%M)$ \\
  f1  &  Unary & $38n-42$ & $(x,y)\mapsto (x/5\%M,y)$ \\
  f2  &  Unary & $38n-42$ & $(x,y)\mapsto (x,y/5\%M)$ \\
\hline
\end{tabular}
\end{center}
\end{table}

\begin{figure}[tb]
\includegraphics[height=55mm]{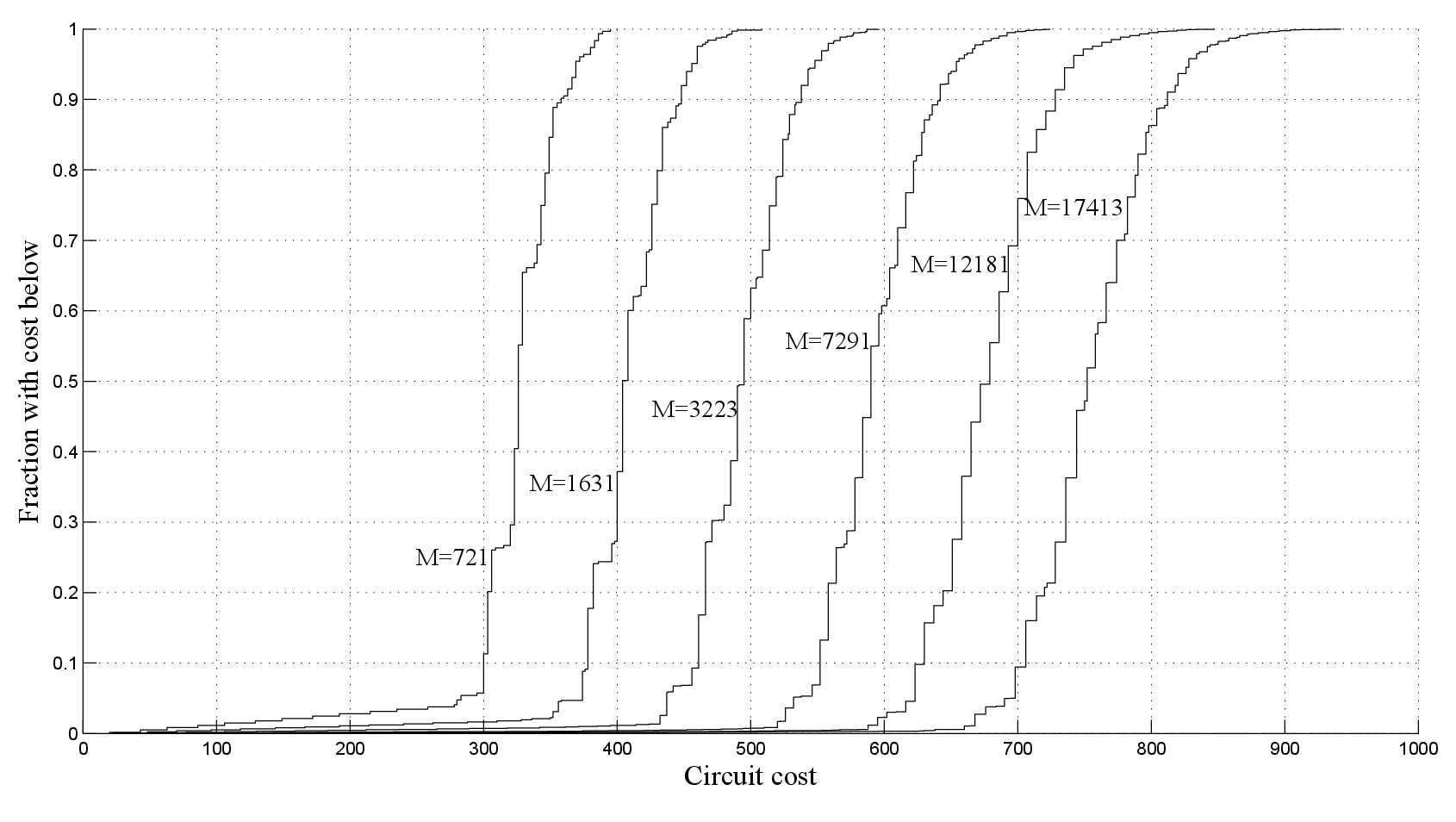}
\centering
\parbox{\length}{
\caption{
\label{fig:CDF-mult}
 $Cx\%M$ circuit costs for select $M$ values,  shown as cumulative distribution functions.
}
}
\end{figure}

\section{Examples of modular multiplication}
\label{sec:mod-mult-ex}

   Here we study $M=pq$ for small prime $p$ and $q$. One can argue that large classes
   of such $M$ values should be excluded from consideration in the context of Shor's algorithm
   because they offer no value for number-factoring. For example, numbers of the form
   $M=b^{2\pi}-1$ can be factorized quickly by computing gcd$(M,b^\pi\pm1)$, and this class
   includes the number 15, commonly used  in experimental demonstrations of Shor's algorithm.
   The same argument applies to some numbers that satisfy $mM=b^{2\pi}-1$, where $m$ has very few factors.
   This class includes the number 21, considered as the next example after 15 for quantum number-factoring.
   Indeed, $3\cdot 21=8^2-1=7\cdot9$ leads to gcd$(7,21)=7$. Nevertheless, we consider these cases for
   completeness and use them to illustrate general circuit constructions.\footnote{This does {\em not} justify the use of 15 and 21 in physical experiments where scalability must be demonstrated.}

\subsection{Very small moduli}
\label{sec:smallM}

Table \ref{table:mod-mult} describes small circuits for $Cx\%M$ functions with coprime $C$ and $M$ with 6 bits or less. Each circuit is described by a parenthesized triplet consisting of the Toffoli gate count, the CNOT gate count and the number of ancillae. An expression indicating circuit structure follows after a colon.
$C$ values where gcd$(C,M)>1$ and $C\geq M$ are marked by $\times$ and $-$, respectively. For each $M$, the last row reports circuits constructed by binary expansion of $x$ (Section \ref{sec:MM_pre_work}) with the smallest gate counts among different $C$ values.

In each case, we report the best circuit structure we could find. For example,
$3x\%35$ can be implemented as $-2^5$. At most one inverter may be used on each circuit line.
Of the techniques we presented, the most economical one is the use of
$2x\%M$ circuits, their repetitions, inverses and negations. In some cases
($M=21,39,55$), it suffices for all $C$ values. When additional circuit
  constructions are needed, we start with circuits for $3x\%M$ or $5x\%M$,
  except when the modulus is divisible by 3 or 5. By means of circuit operators,
  these additional {\em primitive circuits} generate a large number of composite
  circuits, especially that they can be composed with powers of two, etc.
  In Table \ref{table:mod-mult}, the first grayed cell of a column represents
  a primitive circuit that is not a power of two. The smallest circuits constructed by binary expansion of $x$ for each $M$ (shown in the bottom row) are typically larger than the largest circuits proposed.
The data suggest that divisibility of $M$ by 3 can lead to relatively large $Cx\%M$ circuits compared to other moduli $M$ with the same number of bits. This is because $C=3$ is the smallest $C$ value unrelated to powers of two, for which we can build compact multiplication circuits. Among $M$ values divisible by $3$, circuits for $M=39$ tend to be smaller because all $C$ values coprime with $M$ can be obtained through positive and negative powers of two, and their inverses.

\begin{table}
 \parbox{\length}{
\caption{\label{table:mod-mult}
The structure and gate count of the proposed circuits for $Cx\%M$ with coprime $C$ and $M$ with 6 bits or less.
Each circuit is described by a parenthesized triplet consisting of the Toffoli gate count, the CNOT gate count and the number of ancillae.
DR indicates direct use of division with remainder (Theorem \ref{thm:DR}).
Circuits for $M=15$ are illustrated in Figure \ref{fig:15}.
Gray cells indicate circuits that do not only use powers of two, negative powers of two, or their inverses.
All ancillae are cleared after the computation.
Additional optimizations are possible.
For each $M$, the last row reports circuits, constructed by binary expansion of $x$, with the smallest gate counts among different $C$ values. 
}
}
    \tiny
\scalebox{1}{
    \begin{tabular}{|c|c|c|c|c|c|c|c|}
    		\hline
            & \multicolumn{7}{c|}{$M=p \cdot q$ ($|(\mathbb{Z}/M\mathbb{Z})^\times|)$}  \\
    		  &\multicolumn{7}{c|}{} \\
    		 $C$ & $21=3\cdot7$ $(12) $ & $33=3\cdot11$ $(20)$ & $35=5\cdot7$ $(24)$ & $39=3\cdot13$ $(24)$ & $51=3\cdot17$ $(32)$ & $55=5\cdot11$ $(40)$ & $57=3\cdot 19$ $(36)$ \\
        \hline
        2 	& $\tca{15}{16}{4}$: $2^1$   	& $\tca{22}{24}{5}$: $2^1$    				& $\tca{17}{20}{4}$: $2^1$						 & $\tca{12}{19}{3}$: $2^1$		& $\tca{17}{19}{4}$: $2^1$   						&	$\tca{12}{18}{3}$: $2^1$ 			 &$\tca{22}{19}{5}$: $2^1$						\\ \hline
        3 	& $\times$  				& $\times$  								& \cellcolor{gray} $\tca{65}{39}{6}$: DR	 		& $\times$    					& $\times$   									 &	 $\tca{85}{127}{3}$: -$2^{-7}$ 		& $\times$									\\ \hline
        4 	& $\tca{30}{32}{4}$: $2^2$   	& $\tca{44}{48}{5}$: $2^2$ 					& $\tca{34}{40}{4}$: $2^2$ 						 & $\tca{24}{38}{3}$: $2^2$  		& $\tca{34}{38}{4}$: $2^2$    					&	$\tca{24}{36}{3}$: $2^2$ 			 &$\tca{44}{38}{5}$: $2^2$ 						\\ \hline
        5 	& $\tca{33}{34}{4}$: -$2^{-2}$& \cellcolor{gray} $\tca{123}{51}{7}$: DR 		& $\times$   									 & $\tca{36}{57}{3}$: $2^{-3}$ 	& \cellcolor{gray} $\tca{146}{62}{7}$: DR	   		&	$\times$							 & \cellcolor{gray}$\tca{138}{76}{8}$: DR			\\ \hline
        6 	& $\times$  				& $\times$  								& \cellcolor{gray} $\tca{82}{59}{6}$: $2 \cdot 3$		 & $\times$					& $\times$   									 &	 $\tca{73}{109}{3}$: -$2^{-6}$ 		&$\times$ 									 \\ \hline
        7 	& $\times$  				& \cellcolor{gray} $\tca{145}{75}{7}$: $2/5$& $\times$									 & $\tca{61}{98}{3}$: -$2^5$     	& \cellcolor{gray} $\tca{197}{119}{7}$: $5/8$   &	 $\tca{36}{54}{3}$: $2^{-3}$  			 &$\tca{71}{60}{5}$: -$2^{-3}$ 					\\ \hline
        8 	& $\tca{45}{48}{4}$: $2^3$	& $\tca{49}{51}{5}$: -$2^{-2}$    				& $\tca{51}{60}{4}$: $2^3$						 & $\tca{36}{57}{3}$: $2^3$     	& $\tca{51}{57}{4}$: $2^3$    	   				&	$\tca{36}{54}{3}$: $2^3$  			 &$\tca{66}{57}{5}$: $2^3$ 						\\ \hline
        9 	& $\times$  				& $\times$  								& $\tca{34}{40}{4}$: $2^{-2}$						 & $\times$					& $\times$   									&	$\tca{72}{108}{3}$: $2^6$  			 &$\times$ 									 \\ \hline
        10 	& $\tca{18}{18}{4}$: -$2^{-1}$& \cellcolor{gray} $\tca{145}{75}{7}$: $2\cdot5$ 	& $\times$									 & $\tca{24}{38}{3}$: $2^{-2}$    	& \cellcolor{gray} $\tca{149}{64}{7}$: -$5^{-1}$		&	$\times$						 	 &\cellcolor{gray}$\tca{160}{95}{8}$: $2\cdot5$ 		\\ \hline
        11 	& $\tca{15}{16}{4}$: $2^{-1}$ & $\times$  								& $\tca{68}{80}{4}$: $2^{-4}$ 					 & $\tca{60}{95}{3}$: $2^{-5}$    	& \cellcolor{gray} $\tca{180}{100}{7}$: $4/5$    	&	$\times$						 	 &\cellcolor{gray}$\tca{165}{98}{8}$: -$2/5$  		\\ \hline
        12 	& $\times$  				& $\times$  								& \cellcolor{gray}$\tca{65}{39}{6}$: $3^{-1}$ 		& $\times$					& $\times$   									 &	 $\tca{61}{91}{3}$: -$2^{-5}$  	 	&$\times$ 									 \\ \hline
        13 	& $\bf \tca{48}{50}{4}$: -$2^3$& \cellcolor{gray} $\tca{128}{54}{7}$: -$5^{-1}$	& $\tca{54}{62}{4}$: -$2^{-3}$ 					 & $\times$    					& $\tca{34}{38}{4}$: $2^{-2}$   					 &	$\tca{108}{162}{3}$: $2^{-9}$  	 	 &\cellcolor{gray}$\bf \tca{187}{117}{8}$: -$5/4$ 	\\ \hline
        14 	& $\times$  				& \cellcolor{gray} $\bf \tca{150}{78}{7}$: -$5/2$& $\times$									 & $\bf \tca{73}{117}{3}$: -$2^6$  	& \cellcolor{gray} $\tca{180}{100}{7}$: $5/4$	&	 $\tca{24}{36}{3}$: $2^{-2}$		 	 &$\tca{49}{41}{5}$: -$2^{-2}$ 					\\ \hline
        15 	& $\times$  				& $\times$  								& $\times$									 & $\times$					& $\times$   									&	$\times$						 	 &$\times$ 									 \\ \hline
        16 	& $\tca{30}{32}{4}$: $2^{-2}$	& $\tca{27}{27}{5}$: -$2^{-1}$    				& $\tca{68}{80}{4}$: $2^4$						 & $\tca{48}{76}{3}$: $2^4$     	& $\tca{68}{76}{4}$: $2^4$    	   				 &	$\tca{48}{72}{3}$: $2^4$  		 	 &$\tca{88}{76}{5}$: $2^4$ 						\\ \hline
        17 	& $\tca{33}{34}{4}$: -$2^2$	& $\tca{22}{24}{5}$: $2^{-1}$ 				& $\tca{20}{22}{4}$: -$2^{-1}$					 & $\tca{49}{79}{3}$: -$2^{-4}$   	& $\times$   									&	$\tca{108}{162}{3}$: $2^9$ 		 	 &\cellcolor{gray}$\tca{165}{98}{8}$: $\frac{-1}{2\cdot 5}$  		\\ \hline
        18 	& $\times$  				& $\times$  								& $\tca{17}{20}{4}$: $2^{-1}$						 & $\times$					& $\times$   									&	$\tca{84}{126}{3}$: $2^7$  		 	 &$\times$ 									 \\ \hline
        19 	& $\tca{18}{18}{4}$: -$2^1$   & \cellcolor{gray} $\tca{145}{75}{7}$: $5/2$	& $\tca{71}{82}{4}$: -$2^4$   					 & $\tca{13}{22}{3}$: -$2^{-1}$   	& $\tca{54}{59}{4}$: -$2^{-3}$    	   				 &	$\tca{97}{145}{3}$: -$2^8$ 		 	 &$\times$  									\\ \hline
        20 	& $\tca{3 }{2 }{1}$: -$1$	&\cellcolor{gray} $\tca{123}{51}{7}$: $5^{-1}$				& $\times$									 & $\tca{12}{19}{3}$: $2^{-1}$    	& \cellcolor{gray} $\tca{166}{83}{7}$: -$2/5$    &	$\times$						 	 &\cellcolor{gray}$\tca{182}{114}{8}$: $4\cdot5$  		\\ \hline
        21 	& -  					& $\times$  								& $\times$									 & $\times$					 & $\times$   									&	$\bf \tca{121}{181}{3}$: -$2^{-10}$  	&$\times$ 									 \\ \hline
        22 	& -  					& $\times$  								& $\tca{51}{60}{4}$: $2^{-3}$ 					 & $\tca{48}{76}{3}$: $2^{-4}$    	& \cellcolor{gray} $\tca{197}{119}{7}$: $8/5$    	&	$\times$						 	 &\cellcolor{gray}$\bf \tca{187}{117}{8}$: -$4/5$ 	\\ \hline
        23 	& -  					& \cellcolor{gray} $\bf \tca{150}{78}{7}$: $\frac{-1}{2\cdot 5}$ 	& \cellcolor{gray} $\tca{68}{41}{6}$: -$3^{-1}$		& $\tca{49}{79}{3}$: -$2^4$     	& \cellcolor{gray} $\tca{166}{83}{7}$: $-5/2$		&	 $\tca{61}{91}{3}$: -$2^5$  		 	 &\cellcolor{gray}$\tca{138}{76}{8}$: $5^{-1}$ 		\\ \hline
        24 	& -  					& $\times$  								& $\tca{71}{82}{4}$: -$2^{-4}$   					 & $\times$					& $\times$   									&	$\tca{49}{73}{3}$: -$2^{-4}$  	 	 &$\times$ 									 \\ \hline
        25 	& -  					& $\tca{44}{48}{5}$: $2^{-2}$					& $\times$									 & $\tca{72}{114}{3}$: $2^6$ 		& $\tca{20}{21}{4}$: -$2^{-1}$    	   				&	$\times$						 	 &$\tca{88}{76}{5}$: $2^{-4}$ 						\\ \hline
        26 	& -  					& \cellcolor{gray} $\bf \tca{150}{78}{7}$: -$2/5$		& $\tca{37}{42}{4}$: -$2^{-2}$   					 & $\times$    					& $\tca{17}{19}{4}$: $2^{-1}$						 &	$\tca{96}{144}{3}$: $2^{-8}$   	 	 &\cellcolor{gray}$\tca{165}{98}{8}$: -$5/2$ 		\\ \hline
        27 	& -  					& $\times$  								& $\tca{54}{62}{4}$: -$2^3$   					 & $\times$					& $\times$   									&	$\tca{13}{19}{3}$: -$2^{-1}$  	 	 &$\times$ 									 \\ \hline
        28 	& -  					& \cellcolor{gray} $\tca{128}{54}{7}$: -$5$   	& $\times$									 & $\tca{61}{98}{3}$: -$2^{-5}$   	& \cellcolor{gray} $\tca{163}{81}{7}$: $5/2$   	&	 $\tca{12}{18}{3}$: $2^{-1}$		 	 &$\tca{27}{22}{5}$: -$2^{-1}$ 					\\ \hline
        29 	& -  					& $\tca{49}{51}{5}$: -$2^2$  					& \cellcolor{gray} $\bf\tca{85}{61}{6}$: -$6$		& $\tca{25}{41}{3}$: -$2^{-2}$   	& \cellcolor{gray} $\bf \tca{200}{121}{7}$: -$8/5$    	&	$\tca{97}{145}{3}$: -$2^{-8}$ 	 	 &$\tca{22}{19}{5}$: $2^{-1}$ 						 \\ \hline
        30 	& -  					& $\times$  								& $\times$									 & $\times$					 & $\times$   									&	$\times$						 	 &$\times$ 									 \\ \hline
        31 	& -  					& $\tca{27}{27}{5}$: -$2^1$					& $\tca{37}{42}{4}$: -$2^2$						 & $\tca{37}{60}{3}$: -$2^3$     	& \cellcolor{gray} $\tca{163}{81}{7}$: $2/5$  	 	&	 $\tca{48}{72}{3}$: $2^{-4}$  		 	 &\cellcolor{gray}$\tca{160}{95}{8}$: $5/2$ 		\\ \hline
        32 	& -  					& $\tca{5}{4}{2}$: -$1$						& \cellcolor{gray} $\tca{68}{41}{4}$: -$3$ 			& $\tca{60}{95}{3}$: $2^5$     	& $\tca{51}{57}{4}$: $2^{-3}$ 	   				 &	 $\tca{60}{90}{3}$: $2^5$  		 	 &$\tca{93}{79}{5}$: -$2^{-4}$ 					\\ \hline
        33 	& -  					& -  									& $\tca{20}{22}{4}$: -$2^1$ 						 & $\times$					& $\times$   									&	$\times$						 	 &$\times$ 									 \\ \hline
        34 	& -  					& -  									& $\tca{3}{2}{1}$: -$1$ 							 & $\tca{37}{60}{3}$: -$2^{-3}$   	& $\times$									&	$\tca{120}{180}{3}$: $2^{-10}$  	 	 &\cellcolor{gray}$\tca{143}{79}{8}$: -$5^{-1}$ 		\\ \hline
        35 	& -  					& -  									& -											 & $\tca{25}{41}{3}$: -$2^2$     	& $\tca{71}{78}{4}$: -$2^4$    	   				&	$\times$						 	 &\cellcolor{gray}$\tca{182}{114}{8}$: $4/5$ 		\\ \hline
        36 	& -  					& -  									& -   										 & $\times$					 & $\times$   									&	$\tca{96}{144}{3}$: $2^8$  		 	 &$\times$ 									 \\ \hline
        37 	& -  					& -  									& -   										 & $\tca{13}{22}{3}$: -$2^1$     	& \cellcolor{gray} $ \tca{183}{102}{7}$: -$5/4$ 	&	 $\tca{85}{127}{3}$: -$2^7$  		 	 &\cellcolor{gray}$\bf \tca{187}{117}{8}$: -$4\cdot5$  		\\ \hline
        38 	& -  					& -  									& -   										 & $\tca{1}{3}{1}$: -$1$     		& $\tca{37}{40}{4}$: -$2^{-2}$   					&	 $\tca{109}{163}{3}$: -$2^9$  		 	 &$\times$ 									\\ \hline
        39 	& -  					& -  									& -   										 & -    						 & $\times$									&	$\tca{49}{73}{3}$: -$2^4$  		 	 &$\times$ 									 \\ \hline
        40 	& -  					& -  									& -   										 & -    						 & \cellcolor{gray} $\tca{183}{102}{7}$: -$4/5$   	&	$\times$						 	 &\cellcolor{gray}$\tca{160}{95}{8}$: $\frac{1}{2\cdot 5}$ 		\\ \hline
        41 	& -  					& -  									& -   										 & -    						 & \cellcolor{gray} $\tca{146}{62}{7}$: $5^{-1}$   	&	$\tca{25}{37}{3}$: -$2^{-2}$  	 	 &$\tca{93}{79}{5}$: -$2^4$ 						 \\ \hline
        42 	& -  					& -  									& -   										 & -    						 & $\times$									&	$\tca{109}{163}{3}$: -$2^{-9}$  	 	 &$\times$ 									 \\ \hline
        43 	& -  					& -  									& -   										 & -    						 & $\tca{54}{59}{4}$: -$2^3$   					&	$\tca{60}{90}{3}$: $2^{-5}$  		 	 &$\tca{44}{38}{5}$: $2^{-2}$ 						 \\ \hline
        44 	& -  					& -  									& -   										 & -    						 & \cellcolor{gray} $\bf \tca{200}{121}{7}$: -$5/8$ 		&	$\times$						 	 &\cellcolor{gray}$\tca{182}{114}{8}$: $5/4$ 			\\ \hline
        45 	& -  					& -  									& -   										 & -    						 & $\times$									&	$\times$						 	 &$\times$ 									 \\ \hline
        46 	& -  					& -  									& -   										 & -    						 & \cellcolor{gray} $\tca{149}{64}{7}$: -$5$ 			&	$\tca{73}{109}{3}$: -$2^6$  		 	 &\cellcolor{gray}$\tca{160}{95}{8}$: $2/5$  			\\ \hline
        47 	& -  					& -  									& -   										 & -    						 & $\tca{37}{40}{4}$: -$2^2$						&	$\tca{37}{55}{3}$: -$2^3$ 		 	 &\cellcolor{gray}$\tca{165}{98}{8}$: -$2\cdot5$ 			\\ \hline
        48 	& -  					& -  									& -   										 & -    						 & $\times$									&	$\tca{37}{55}{3}$: -$2^{-3}$  	 	 &$\times$ 									 \\ \hline
        49 	& -  					& -  									& -   										 & -    						 & $\tca{20}{21}{4}$: -$2^1$   					&	$\tca{72}{108}{3}$: -$2^{-6}$ 	 	 &$\tca{71}{60}{5}$: -$2^3$						 \\ \hline
        50 	& -  					& -  									& -   										 & -    						 & $\tca{3}{2}{1}$: -$1$ 							&	$\times$						 	 &$\tca{66}{57}{5}$: $2^{-3}$ 						 \\ \hline
        51 	& -  					& -  									& -   										 & -    						 & -   										&	$\tca{25}{37}{3}$: -$2^2$  		 	 &$\times$ 									 \\ \hline
        52 	& -  					& -  									& -   										 & -    						 & -   										&	$\tca{84}{126}{3}$: $2^{-7}$  	 	 &\cellcolor{gray}$\tca{143}{79}{8}$: -$5$ 			 \\ \hline
        53 	& -  					& -  									& -   										 & -    						 & -   										&	$\tca{13}{19}{3}$: -$2^1$  		 	 &$\tca{49}{41}{5}$: -$2^2$						 \\ \hline
        54 	& -  					& -  									& -   										 & -    						 & -   										&	$\tca{1}{1}{0}$: -$1$  			 	 &$\times$ 									 \\ \hline
        55 	& -  					& -  									& -   										 & -    						 & -   										&	-  							 	 &$\tca{27}{22}{5}$: -$2^1$ 						 \\ \hline
        56 	& -  					& -  									& -   										 & -    						 & -   										&	-  							 	 &$\tca{5}{3}{2}$: -$1$ 							 \\ \hline
	   \multicolumn{8}{c}{Smallest $Cx\%M$ circuits based on binary expansion of $x$ (with cleared ancillae)} \\ \hline
	   2		& $\bf \tca{136}{34}{11}$	&$\bf \tca{225}{51}{14}$						&$\bf \tca{241}{47}{14}$							 &$\bf \tca{225}{41}{14}$			&$\bf \tca{216}{36}{13}$							&$\bf \tca{202}{33}{13}$					 &$\bf \tca{202}{37}{13}$							\\ \hline
    \end{tabular}
}
\end{table}

\subsection{Larger moduli} \label{sec:largerM}

 We now illustrate the use of our shortest-path reduction to find two-register mod-mult circuits.
 Our C++ implementation of Dijkstra's algorithm operates on an $M\times M$ vertex array, but generates
 edges on the fly. In one pass, it finds all single-source shortest paths starting at $(1,0)$
 and produces $Cx\%M$ circuits for all $C$ coprime with $M$ (this is convenient, but not necessary
 when working with Shor's algorithm). The modular multiplication circuits with 7-14 bits produced by our techniques are available online at \texttt{http://www.eecs.umich.edu/\~{}imarkov/MME/}. In Table \ref{tab:65}, we show circuits for $Cx\%65$ with
 all coprime $C$. Figure \ref{fig:CDF-mult} shows circuit-cost distributions (for \T gate counts) for several $M$ values
 in terms of cumulative distribution functions (CDF).
 Maximum and average costs for all 6-14 bit semiprime $M$ values not divisible by 2 and 3
 are reported in Table \ref{tab:max_avg}. On a fast Linux workstation (3.0GHz Intel CPU with 8GB RAM),
 processing one 14-bit $M$ value takes one to six minutes, and less than three days for all 14-bit
 $M$ values.\footnote{We report timing for an implementation of Dijkstra's algorithm that uses a comparison-based priority queue from C++ STL. Given that all path lengths are integers below $10000$, we have also implemented an $O(1)$-time bin-based priority-queue. Runtime improvements were significant for smaller $n$, but memory usage increased somewhat. Since memory is the main bottleneck for larger $n$, we decided to use the more compact comparison-based priority queue.} Many 15-bit values require over 8GB memory, and runtime increases four- to eight-fold.
 Our implementation of Dijkstra's algorithm based on an explicit $2^n\times 2^n$ vertex array does not scale
 beyond 15-bit $M$. However, the shortest-path formalism can be applied in different ways to find optimal circuits
 for larger $M$ values, and also to perform heuristic optimization for {\em much larger} $M$ values.

  The sizes of $n$-bit modular multiplication circuits in Table \ref{tab:max_avg}
 fit very well ($R^2>0.999$) to quadratic functions, producing the worst-case bound
 $6n^2+O(n)$ and the average-case estimate $3.3n^2+O(n)$.
 Thus, our circuits are 4 times smaller on average than CSD-based circuits produced using Bennett's construction
 discussed after Theorem \ref{thm:pathlen}.
 \begin{table}
 \caption{\label{tab:65} Two-register circuits for $Cx\%65$. Cost is reported as the number of \T ~gates.}
 \centering
  \scriptsize
 \begin{tabular}{|r|c|l||r|c|l|}
 \hline
$C$ & \T-cost & Circuit & $C$ & \T-cost & Circuit \\
 \hline
2&28&d1				&	33&28&h1			\\
3&154&c2+2+2+1+1d1+1d1d1c2	&	34&140&c2+2+2+1+1d1+1d1-2c2\\
4&56&d1d1			&	36&126&c2+2+2+1+2+1+1d1-2c2\\
6&140&c2h1h1-1-2-1-1-2-1c2	&	37&140&c2+1d1+2+1+2d2+1+2\\
7&140&c2+2h1+1+2+1h2+2+2	&	38&126&c2+1+2+1+1+2+1d1-2c2\\
8&84&d1d1d1			&	41&126&c2+1+1+2+2+1+1+2+1+2\\
9&140&c2+2+1h1+1+2+1+2+2+2	&	42&140&c2+1+1+2h1+1+1+2+2+2\\
11&140&c2+1+2+1+1+2+1d1d1c2	&	43&168&c2+1+2+1+2d2-1-2d2c2d1\\
12&126&c2+1h2-2-1-2-2-1-2c2	&	44&140&c2+2h1-1h1-1-1-2c2\\
14&140&c2+2d2+1+2+1h2+2+2	&	46&126&c2+1+2+1+1+2+2+1+1+2\\
16&70&\~{}1h1h1			&	47&126&c2+2+2+1+2+1+2+1+1+2\\
17&140&c2+1+1+2+2+1+1+2d2+2	&	48&140&c2+2+2+1+1+2+2+1d1+2\\
18&126&c2+1+1+2+1+2+1+2+2+2	&	49&56&h1h1\\
19&126&c2+2+1+2+2+1+1+2+2+2	&	51&140&c2+1h2+2+1+2d2+1+2\\
21&154&c2+2+1h1+1+1+1+1+2+1+2	&	53&126&c2+2+1+2h1-1-2-1-2c2\\
22&154&c2h1h1-1-1-1-1-1-2-1c2	&	54&140&c2+1+2+1+2d2-1-2d2c2\\
23&140&c2+2+2+1h2+2+2+1+1+2	&	56&126&c2+1h2-2-2-1-2-1-1c2\\
24&126&c2+2+2+1+1+2+2+1+2+2	&	57&84&h1h1h1\\
27&126&c2+1+2+1+2d2-1-2-1c2	&	58&140&c2+1h2+2+1+2h1+1+2\\
28&140&c2+2d2+1+2+1d1+2+2	&	59&140&c2h2+2+1h2-2-1-2-1c2\\
29&140&c2+2+2+1+2+1+2+1d2+2	&	61&70&\~{}1d1d1\\
31&154&c2+2+2+1+1d1+2+1d2+2	&	62&168&c2h2+2+1h2-2-1-2-1c2h1\\
32&42&-1h1			&	63&42&\~{}1d1\\
&	&			&	64&14&\~{}1\\
\hline
\end{tabular}
\end{table}

 \begin{table}
\caption{\label{tab:max_avg} Costs of two-register circuits over $n$-bit semiprime $M$ values not divisible by 2 and 3.}
  \tiny
\scalebox{1}{
 \begin{tabular}{|c|c|c|c|lll|}
\hline
Bits	& \# of semiprimes & \multicolumn{2}{c|}{\T-costs} & \multicolumn{3}{l|}{Circuits with max costs}\\
$n$ & [smallest, largest] & max & avg & $C$& $M$ & circuit \\
 \hline
\multirow{16}{*}{7}	& \multirow{16}{*}{7 in [65, 119]} & \multirow{16}{*}{182}	& \multirow{16}{*}{134.3}   & $3$& $115$& c2+1+1+2+2d2d2d2d2+2\\
& 	& 	& 	&  $19$& $115$& c2+1+2+2+1+2+1d2d2d2+2 \\
& 	& 	& 	&  $38$& $115$& c2h1h1h1h1+1+1+2+2+2 \\
& 	& 	& 	&  $77$& $115$& c2h2h2h2h2+2+2+1+1+2 \\
& 	& 	& 	&  $109$& $115$& c2+1+2+2+1+2h1+1+1+1+1+2+2 \\
& 	& 	& 	&  $112$& $115$& c2+2+2+1+1d1d1d1d1+2 \\
& 	& 	& 	&  $39$& $119$& c2+1+2h2h2h2-2-2-1-2c2\~{}1 \\
& 	& 	& 	&  $58$& $119$& c2+1+2+1+1d1d1-2-1d2c2\~{}1 \\
& 	& 	& 	&  $79$& $119$& c2+1d1+1d1d1d1-2-2c2\~{}1 \\
& 	& 	& 	&  $95$& $119$& c2+2+2d2h1d2-1d2-1c2\~{}1 \\
& 	& 	& 	&  $99$& $119$& c2+1+2+1+1+2+1+2d2+1+2+1+2 \\
& 	& 	& 	&  $107$& $119$& c2+1+2+1d1+2+1+2+1+1+2+1+2 \\
& 	& 	& 	&  $109$& $119$& c2+1+2+1+1+2+1+2+1+1+1+2+2+2 \\
& 	& 	& 	&  $113$& $119$& c2+1+2+1h2+2+1+2+1+1+2+1+2 \\
& 	& 	& 	&  $114$& $119$& c2+2+1+2+1+1d1d1+1d1+2 \\
& 	& 	& 	&  $116$& $119$& c2+2+2h1h1h1-1h1-1c2\~{}1 \\
\hline
\multirow{2}{*}{8}	& \multirow{2}{*}{16 in [133, 253]}& \multirow{2}{*}{257}	& \multirow{2}{*}{194.3}   & $42$& $253$& c2+2+1+2+1+1+1+2+1+2+1+2+1d1+2+2 \\
&	& 	& 	& $247$& $253$& c2+2h1+1+2+1+2+1+2+1+1+1+2+1+2+2 \\
\hline
\multirow{12}{*}{9}	& \multirow{12}{*}{34 in [259, 511]} & \multirow{12}{*}{326}	& \multirow{12}{*}{258.0} 	& $431$& $485$& c2+2h1-1-1-2-1-2-2-1-1-2-1-2-1-2-2-1c2\\
&	& 	& 	& $476$& $485$& c2+1+2+2+1+2+1+2+1+1+2+2+1+2+1+1d1-2c2 \\
&	& 	& 	& $14$& $505$& c2+1+2+2+1+1+2+2+1+1h2+2+2+2+1+1+2+2 \\
&	& 	& 	& $18$& $505$& c2+1+2+2+1+1+1h2+2+2+1+1+2+2+1+1+2+2 \\
&	& 	& 	& $28$& $505$& c2+1+2+2+1+1+2+2+1+1d1+2+2+2+1+1+2+2 \\
&	& 	& 	& $36$& $505$& c2+1+2+2+1+1+1d1+2+2+1+1+2+2+1+1+2+2 \\
&	& 	& 	& $459$& $505$& c2+1+2+2+1+1+1+1d1+2+2+1+2+1+2+1+2+2 \\
&	& 	& 	& $469$& $505$& c2+2+1+1+2+2+2d2+1+1+2+2+1+1+2+2+1+2 \\
&	& 	& 	& $477$& $505$& c2+2+1+1+2+2+1+1+2+2d2+1+1+1+2+2+1+2 \\
&	& 	& 	& $487$& $505$& c2+2+1+1+2+2+2h1+1+1+2+2+1+1+2+2+1+2 \\
&	& 	& 	& $491$& $505$& c2+2+1+1+2+2+1+1+2+2h1+1+1+1+2+2+1+2\\
&	& 	& 	& $494$& $505$& c2+2+1+2+1+2+1+2+2h1+1+1+1+1+2+2+1+2\\
\hline
\multirow{6}{*}{10} & \multirow{6}{*}{72 in [515, 1007]}	& \multirow{6}{*}{418}	& \multirow{6}{*}{327.3}   & $935$& $1007$& c2+1+1+1d1+1d1h2h2d1-2d1-2-2c2\~{}1\\
& 	& 	&    	& $951$& $1007$& c2+1+1d1+1h2d1h2d1-2d1-2-2-2c2\~{}1\\
& 	& 	&    	& $971$& $1007$& c2+1+1+1d1+1d1h2h2d1-2h2-2-2c2\~{}1\\
& 	& 	&    	& $979$& $1007$& c2+2+2d2+2d2d2h1h1-1h1-1-1-1c2\~{}1\\
& 	& 	&    	& $989$& $1007$& c2+2+2+2h1+2h1h1h1d2-1d2-1-1c2\~{}1\\
& 	& 	&    	& $993$& $1007$& c2+2+2h1+2h1h1h1d2-1d2-1-1-1c2\~{}1\\
\hline
\multirow{2}{*}{11}	& \multirow{2}{*}{152 in [1027, 2047]} & \multirow{2}{*}{518}	& \multirow{2}{*}{405.0}   & $292$& $2045$& c2+1+2+1+1+2+2+2+2d2+1+1+2+1+2+1+1+1d1d1-2c2\\
&	&	&	& $2038$& $2045$& c2+2h1h1-1-1-1-2-1-2-1-1h2-2-2-2-2-1-1-2-1c2\\
\hline
\multirow{2}{*}{12} & \multirow{2}{*}{299 in [2051, 4087]}	& \multirow{2}{*}{635}	& \multirow{2}{*}{488.8}   & $2229$& $3901$& c2+1+1+1d1d1+1d1+2+1d2+1+2d2+1d2d2+1+2\\
&	& 	&    	& $3894$& $3901$& c2+2h1+2h1-1-2-1h2h2+2h2+2h2h2+2+2+2+2\\
\hline
\multirow{18}{*}{13} & 	& \multirow{18}{*}{750}	& \multirow{18}{*}{580.3} & $6347$& $7405$& c2h1h1+1h1h1+1h1h1h1-1h1-1-1-1-2-1-1-2c2\~{}1 \\
&	& 	&    	& $7398$& $7405$& c2+2+1+1+2+1+1+1d1+1d1d1d1-1d1d1-1d1d1c2\~{}1 \\
&	& 	&    	& $1060$& $7421$& c2h1h1+1h1h1h1+1h1h1h1-1-1-1-1-2-2-2-1c2\~{}1 \\
&	& 	&    	& $7414$& $7421$& c2+2+1+1+1+2+2+2+2d2d2d2-2d2d2d2-2d2d2c2\~{}1 \\
&	& 	&    	& $5352$& $7493$& c2+1+1+1d1d1+1d1d1-1d1d1-2-2-1-1d1+2d1+2 \\
&	& 	&    	& $7486$& $7493$& c2+1+1+2+2h1-1h1h1h1+1h1+1h1h1h1+1+1+1+2 \\
&	& 	&    	& $2163$& $7571$& c2+1+1+1d1d1+1d1d1d1-1d1+2+2+1+2+1d2d2+2 \\
&	 621 in&	&    	& $7564$& $7571$& c2h1+2h1-1-1-2-1-2h2h2h2h2+2h2h2+2+2+2+2 \\
&	[4097, 8189]& 	&    	& $2211$& $7739$& c2+1+1+1d1d1+1d1d1+1d1d1+2+2+1+2+1d2d2+2 \\
&	& 	&    	& $7732$& $7739$& c2+2+2+1+2+1h1h1h1h1+1h1h1+1h1h1+1+1+1+2 \\
&	& 	&    	& $2223$& $7781$& c2+2+2+2d2d2+2d2+2d2d2-2d2d2d2-1-1-2-2c2\~{}1 \\
&	& 	&    	& $7774$& $7781$& c2+2+2+1+1h2h2h2-2h2-2h2h2-2h2h2-2-2-2c2\~{}1 \\
&	& 	&    	& $6839$& $7979$& c2h1h1-1h1h1h1h1+1h1+1h1+1+1+1+2+1+1+2+2 \\
&	& 	&    	& $7972$& $7979$& c2+2+1+1+2+1+1+1d1+1d1+1d1d1d1d1-1d1d1+2 \\
&	& 	&    	& $6959$& $8119$& c2h1h1h1h1+1h1h1+1h1+1h1+1+1+1+2+1+1+2+2 \\
&	& 	&    	& $8112$& $8119$& c2+2+1+1+2+1+1+1+1d1d1d1-1d1-1d1d1d1d1+2 \\
&	& 	&    	& $4076$& $8159$& c2+2+2+1+2+1d1+1d1d1+1d1+1d1d1+1d1d1+2+2 \\
&	& 	&    	& $4662$& $8159$& c2+2h1h1+1h1h1+1h1+1h1h1+1h1+1+2+1+2+2+2 \\
\hline
\multirow{44}{*}{14} & 	& \multirow{44}{*}{882}  &  \multirow{44}{*}{678.6} & $2020$& $14141$& c2+1+2+1+2+1+1+1d1+2+1+2d2+1+2d2+1+1+2d2d2+2d2+1+2\\
&&&&	$ 14134$& $14141$&  c2+2+2h1+2h1h1+1h1h1h1-1h1h1d2h1+1+1+1+2 \\
&&&&	$ 12143$& $14167$&  c2h1h1h1h1+1h1+1h1h1h1h1+1+1+1+2h1+1+1+2 \\
&&&&	$ 14160$& $14167$&  c2+1+1+2+2+1d1+1d1d1d1d1+1d1+1d1d1d1d1+2 \\
&&&&	$ 12467$& $14545$&  c2+2+2+2d2d2d2d2+2d2d2d2d2-2h1-1d2-1-1c2\~{}1 \\
&&&&	$ 14538$& $14545$&  c2+1+1h2+1h2+2h2h2h2h2-2h2h2d1h2-2-2-2c2\~{}1 \\
&&&&	$ 12503$& $14587$&  c2+2+2+1+2+1+1+1+1d1d1d1+2+1+2d2+1+2d2+1+2+1+1d1+2 \\
&&&&	$ 14580$& $14587$&  c2h1+1+1+2+1h2+2+1h2+2+1+2h1h1h1+1+1+1+1+2+1+2+2+2 \\
&&&&	$ 12755$& $14881$&  c2+1+2+1+2+2+2+2+2+2d2d2d2d2+1+2+1d1-2-2-1-1-2d1-2c2 \\
&&&&	$ 14874$& $14881$&  c2+2+2h1-1-2-1-2-1-1-2h2h2h2h2h2-2-2-2-2-2-2-1-2-1c2 \\
&&&&	$ 12863$& $15007$&  c2+1+2+1+2+1+2+2d2d2d2d2-1-2-1-2-1-2-1-2-1d1-2d1c2\~{}1 \\
&&&&	$ 15000$& $15007$&  c2+2+2h1+2h1h1h1h1h1-1h1-1h1d2h1-1-1-1c2\~{}1\\
&&&&	$ 13151$& $15343$&  c2+2+1+2+1+2+1+1d1d1d1d1d1+2+1+2+1+2+2+2+1+1d1+2+2 \\
&&&&	$ 15336$& $15343$&  c2+2h1+1+1+2+2+2+1+2+1+2h1h1h1h1h1+1+1+2+1+2+1+2+2 \\
&&&&	$ 13259$& $15469$&  c2+1+2+1+2+1+1+1+2+1+2d2-1-1-2-1d1-1d1d1-2d1-2-1d1c2 \\
&&&&	$ 15462$& $15469$&  c2h1+1+2h1+2h1h1+1h1+1+2+1+1h2-2-1-2-1-1-1-2-1-2-1c2 \\
&&&&	$  2236$& $15653$&  c2+1+2+1+2+1+2+2d2d2+1+1+1+2+1d1d1-2-1-2-1-2-2d2d2c2 \\
&&&&	$ 15646$& $15653$&  c2+1+2+2+1+1+2+2+1+1+2+2+2d2d2d2+1+2+1+2+2+1d1d1d1c2 \\
&&&&	$ 13463$& $15707$&  c2+1+2+1+2+1+2+2d2d2d2+1+2+1d1+2+2+2+1+1d1-2-1d1-2c2 \\
&&&&	$ 15700$& $15707$&  c2+1h2+2+1h2-2-2-1-1-1h2-2-1-2h1h1h1-1-1-2-1-2-1-2c2 \\
&1212 in&&&	$ 13511$& $15763$&  c2+2+1+2+2+1+1+1+1d1d1d1d1+2+2+1+2+2+1+1+1d1d1+2+2 \\
& [8197, 16379]&&&	$ 15756$& $15763$&  c2+2h1h1+1+1+1+2+2+1+2+2h1h1h1h1+1+1+1+1+2+2+1+2+2 \\
&&&&	$  2254$& $15779$&  c2h1h1+1h1h1h1h1h1+1h1h1+1h1+1+2+1+2+2+2 \\
&&&&	$ 15772$& $15779$&  c2+2+2+1+2+1d1+1d1d1+1d1d1d1d1d1+1d1d1+2 \\
&&&&	$  2260$& $15821$&  c2+2+1+2+2+1+1+1+1d1d1d1d1+2+1+2+1+1+2+2+2d2d2+1+2 \\
&&&&	$ 15814$& $15821$&  c2+1h2h2+2+2+2+1+1+2+1+2h1h1h1h1+1+1+1+1+2+2+1+2+2 \\
&&&&	$ 14079$& $15839$&  \~{}1t1t1 \\
&&&&	$ 15830$& $15839$&  \~{}1r1r1 \\
&&&&	$  2272$& $15905$&  c2+1+1+1d1+1d1d1d1-1d1d1d1d1h2+2d1+2+2+2 \\
&&&&	$ 15898$& $15905$&  c2+2+2h1+2h1h1h1h1h1-1h1d2h1+1h1+1+1+1+2 \\
&&&&	$ 13655$& $15931$&  c2+1+2+1+2+2+2+2+2+2d2+2d2d2d2d2+1+1+2+1+2+1d1-2-2c2 \\
&&&&	$ 15924$& $15931$&  c2+2+1+2h1-1-1-2-1-1-1h2h2h2h2-2h2-2-2-2-2-1-2-1-1c2 \\
&&&&	$  2280$& $15961$&  c2h2h2h2h2h2-2h2h2-2h2h2-2h2-2-1-2-1-1c2\~{}1 \\
&&&&	$ 15954$& $15961$&  2+2+2+1+2+1d1+1d1d1+1d1d1+1d1d1d1d1d1c2\~{}1 \\
&&&&	$     7$& $15989$&  c2+2+2+1+2+1d1+1d1d1+1d1d1+1d1d1d1+1d1d1c2 \\
&&&&	$  2284$& $15989$&  c2+1+2+1+1+2+2+2+2d2d2+1+1+2d2+1+1+1+1+2+1d1d1d1-2c2 \\
&&&&	$ 13705$& $15989$&  c2+1+1+2+2+2+1+2+2d2d2d2d2-1-1-2-1-2-1-2-1-2d2-1d2c2 \\
&&&&	$ 15982$& $15989$&  c2+1h2h2-2-2-2-1-2-1-2-2-1h1h1h1h1-1-1-1-1-2-2-1-2c2 \\
&&&&	$  4584$& $16045$&  c2+2+2+1+2+1+2h1h1h1h1+1h1h1+1+2+1+1+2+1+1+2+2+1+2 \\
&&&&	$  8019$& $16045$&  c2+1+2+2+1+1+2+1+1+2+1d1d1+1d1d1d1d1+2+1+2+1+2+2+2 \\
&&&&	$  4652$& $16283$&  c2+2+2h1h1+2h1-1-1-2-1h2h2-2-2-2-2-1-1-2-1h2-2-2-1c2 \\
&&&&	$  8138$& $16283$&  c2+2+2+2+2+1+2+1+1+2+2+2+2d2d2+1+2+1+1d1-2d1d1-2d1c2 \\
&&&&	$ 14015$& $16351$&  c2+2+2+2d2+2d2d2d2+2d2d2d2d2h1-1d2-1-1c2\~{}1 \\
&&&&	$ 16344$& $16351$&  c2+1+1h2+1h2h2h2h2h2-2h2h2h2-2d1-2-2-2c2\~{}1 \\
\hline
\multicolumn{7}{c}{Trend lines --- ~~~~~ Max($n$)=$5.309n^2 - 11.59n + 4.5$ ~~~~~ Avg($n$)=$3.351n^2 + 7.127n - 78.57$} \\
\hline
\multicolumn{7}{|c|}{Extrapolated values --- 20: (1896,1404); 50: (12697,8655); 100: (51935,34144); 200: (210046,135386); 300: (474337,303649)}\\
\hline
\end{tabular}
}
\end{table}

\section{Circuits for modular exponentiation}
\label{sec:mod-exp}

When implementing $f(y)=b^y\% M$, one deals with modular multiplications $C_kx\%M,~C_k=b^{2^k}\%M$ conditional on bits $y_k$, as outlined in Section \ref{sec:intro}.

\subsection{Reordering and factoring of modular multiplications}
\label{sec:reorder}

\begin{figure}[t]
\scriptsize
\scalebox{0.8}{
\input{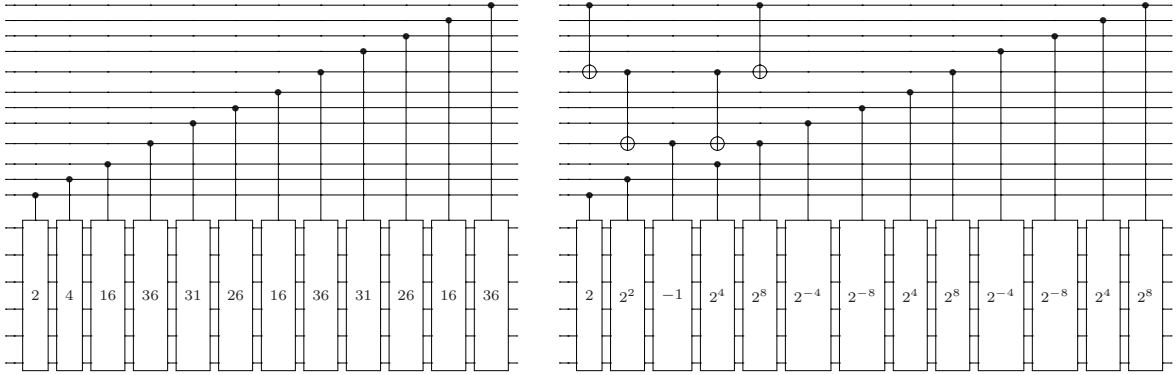}
}
\centering
\vspace{2mm}
    \parbox{\length}{
        \caption{\label{fig:factor} Reordering and factoring in modular exponentiation for $b=2, M=55$. The initial circuit (left). Merging conditional $-x\%55$ operations into one block conditional on the XOR of related controls based on $Cx\%M$ blocks from Table \ref{table:mod-mult} (right). The correspondence between blocks in the two circuits can be established by matching control lines.
}
    }
\end{figure}

  The order of conditional modular multiplications does not affect the result,
  and this becomes useful after some of them are factored. As we have shown earlier,
  sometimes $C_kx\%M$ is easiest to implement as $-(M-C_k)x\%M$. In this case, we can
  factor out $-x\%M$ conditional on $y_k$. Any number of conditional
  $-x\%M$ operations can be consolidated into one such operation, conditional
  on the XOR of relevant control bits. This XOR value can be computed using a chain of CNOTs
  without ancillae, and uncomputed by the same chain after use. Figure \ref{fig:factor} illustrates these optimizations for $b = 2$ and $M = 55$. Modular exponentiation with base 2 and $M=55$, requires conditional multiplications by 2, 4, 16, $36=16 \cdot 16\%55$, $31=36 \cdot 36 \% 55$, and $26=31 \cdot 31\% 55$.

  Reordering also allows one to move a small set of the most difficult multiplications to the front
  of the circuit, where the initial value is 1 and generic multiplication circuits can be avoided,
  as shown below.

\subsection{$k$-bit look-up tables} \label{sec:lut}

 A $(k,m)$ look-up table (LUT) takes $k$ read-only input bits
 and $m>\log_2 k$  zero-initialized ancillae.
 For each $2^k$ input combination, a LUT produces
 a pre-determined $m$-bit value. For example,
 a (2,4)-LUT may be defined by values (1,2,4,8)
 or (1,4,1,4).

 Look-up tables arise in implementations of Shor's algorithm
 (with initialized bits) where the first conditional modular
 multiplication is applied to the constant 1, and can therefore
 produce only two possible values --- 1 and the multiplier.
 Such a circuit can be implemented with at most $m$
 CNOT gates ($m/2$ on average). When two conditional
 multiplications are considered, four output combinations are possible.
 For every bit of the result, this defines a two-input Boolean function,
 which can be implemented with at most two reversible gates
 (possibly with negative controls) and no ancillae.
All these gates operate in
 parallel, although most existing technologies are not able
 to use such amount of parallelism. When two output bits
 implement the same function using Toffoli gates, one of
 them can be replaced by a CNOT that copies the computed
 value.

\begin{figure}[tb]
\scriptsize
\scalebox{0.9}{
\input{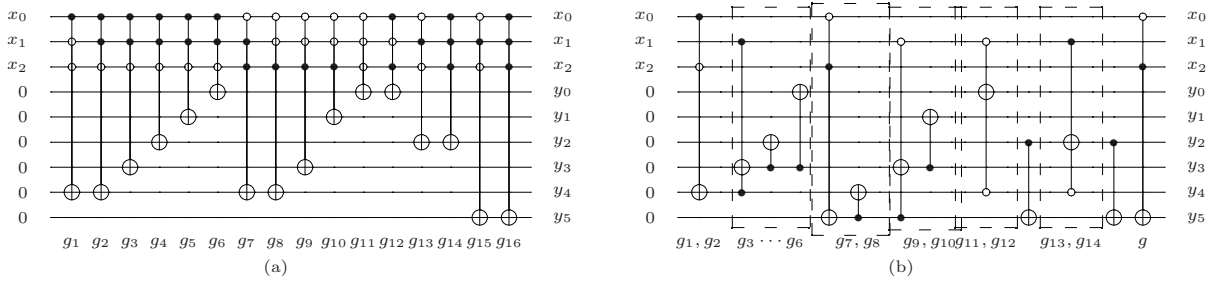}
}
\centering
\parbox{\length}{
\vspace{2mm}
\caption{
\label{fig:LUT55}
Implementation of conditional modular multiplications by 16, 36 and 31 in modular exponentiation for $b=2$, $M=55$ as a (3,6)-LUT. (a) A straightforward implementation, (b) an optimized circuit. The gates $g_{11},g_{12}$ and $g_{13},g_{14}$ use $y_4'$ (i.e., $x_1 x_3 \oplus x_1' x_3'$) as a control. Applying the last two CNOT gates in (b) equals to applying $g_{15}$ and $g_{16}$ in (a). In (b) $g$ is added to clear $y_5$ used to simplify $g_7,g_8$.}
}
\end{figure}

Reconsider conditional mod-mults by 16, 36 and 31 required in modular exponentiation for $b=2$, $M=55$ as shown in Figure \ref{fig:factor}. Depending on the three input bits, the output may be 1, 16, 36, 31, $26=16 \cdot 36\%55$, $1=16 \cdot 31\%55$, $16=31 \cdot 36\%55$, $36=16 \cdot 31 \cdot 36\%55$. Figure \ref{fig:LUT55}a illustrates a simple realization based on the following Boolean expressions for output variables where $\bigoplus\{i,j\}$ is used to denote $m_i \oplus m_j$ for minterms\footnote{For a Boolean function of $n$ variables, a minterm is a product term of all $n$ variables (either complemented or un-complemented). Each minterm can be labeled by an integer by interpreting
  negated literals as 0 bits in the label. For example, expanding minterms for $y_0$ leads to $x_1' x_2' x_3' \oplus x_1' x_2 x_3 \oplus x_1' x_2 x_3'$.} $m_i$ and $m_j$.

\begin{figure}[t]
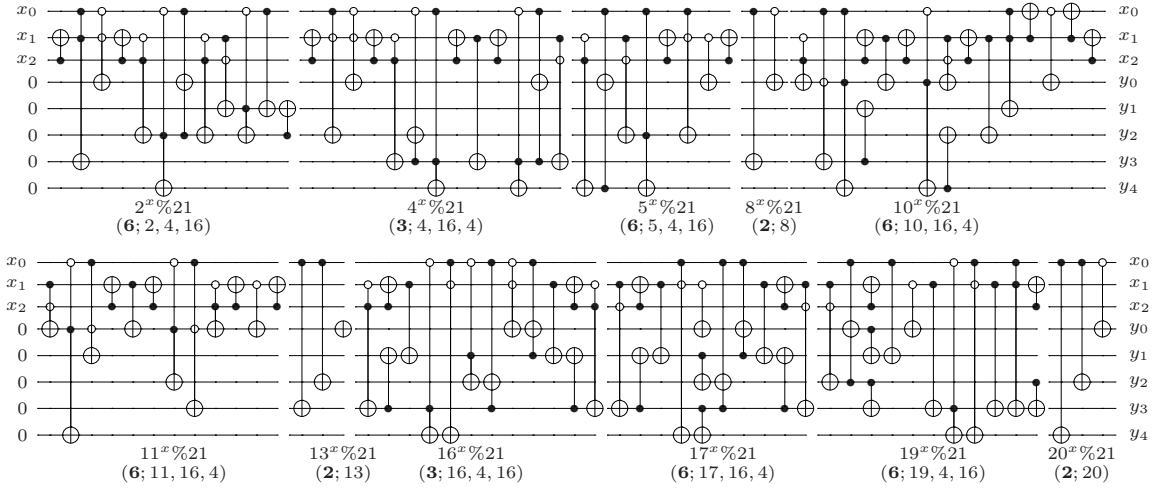

    \begin{center}$
        \scriptsize
        \begin{array}{cc}
\scalebox{0.95}{
            \input{qcircuits/ModExp21-1}
}\\\\
\scalebox{0.95}{
            \input{qcircuits/ModExp21-2}
}
        \end{array}$
    \end{center}
    \centering
    \vspace{2mm}
     \parbox{\length}{
    \caption{\label{fig:ModExp21}
Modular exponentiation with $M=21$ and all coprime base values implemented as $(3,5)$-LUTs. For each circuit, the parenthesized label includes the period of modular exponentiation (boldfaced) and the multipliers of conditional multiplications. The four $b$ values where $b^2\%M=1$ ($b=1,8,13,20$) lead to particularly compact circuits, but finding such values ($\neq\pm1$) for large $M$ is as hard as factoring $M$. Coprime values 4, 5, 20, 16 and 17 trigger restarts in Shor's algorithm and are given only to illustrate the circuits.
}
     }
\end{figure}

\begin{small}
$$
y_0 =\bigoplus\{0,3,5\},
y_1 =\bigoplus\{3,4\},
y_2 =\bigoplus\{2,3,7\},
y_3 =\bigoplus\{3,4\} ,
y_4 = \bigoplus\{1,3,4,6\},
y_5 = \bigoplus\{2,7\}
$$
\end{small}
Since some Boolean functions with $>2$ gates repeat, they can be computed once and then copied. Some Boolean functions can be used to compute other Boolean functions too. Following these optimizations, an improved circuit for circuit in Figure \ref{fig:LUT55}(a) is shown in Figure \ref{fig:LUT55}(b) which is smaller than three conditional modular multiplications by 16, 36 and 31 as reported in Table \ref{table:mod-mult}. Figure \ref{fig:ModExp21} illustrates the LUT-implementation of modular exponentiation for $M=21$ with different coprime base values. Predictably, the cases with $b^2\%M=1$ result in the most compact circuits.\footnote{In general, for a semiprime $M$, there are four values $b$ such that $b^2\%M=1$, two of them being $b=\pm1$. The other two are as difficult to find as factoring $M$.}

{\bf Systematic synthesis}. We now construct circuits to implement each output of a reversible LUT. Viewing each output as a Boolean function of read-only inputs, one can write the Shannon decomposition $F = xF_x \oplus x'F_{x'}$ where $F_x$ and $F_{x'}$ are positive and negative cofactors of $F$. This equation can be written as Formula \ref{eq:pDavio}, the positive Davio decomposition, or as Formula \ref{eq:nDavio}, the negative Davio decomposition.

\begin{small}
\begin{equation}\label{eq:pDavio}
F=F_{x'} \oplus x(F_x \oplus F_{x'})
\end{equation}
\begin{equation}\label{eq:nDavio}
F=F_{x} \oplus x'(F_x \oplus F_{x'})
\end{equation}
\end{small}

Table \ref{table:LUT2} shows that each 2-input Boolean function can be implemented by a reversible circuit with read-only inputs using at most three gates, of which at most one is a Toffoli gate. To implement a three-input function, cofactor it with respect to one of its inputs. Implement the first cofactor without controls and then implement a controlled version of the XOR of the two cofactors. This approach leads to at most one 4-input Toffoli gate and at most 6 smaller gates. Circuit costs can be minimized by choosing the cofactoring variable (pivot) so as to minimize the total costs of cofactors based on Table \ref{table:LUT2}. Working with four-input functions, one can implement four modular-multiplication modules by one (4,$n$)-LUT by implementing cofactors as three-input functions. However, using two separate cofactoring steps may require five-input Toffoli gates. An alternative approach is to consider the four double-cofactors (each a two-input function) with respect to two variables as shown in Formula \ref{eq:Davio4}, and introduce an ancilla to enable the fourth cofactor. This ancilla will be set and unset by a Toffoli gate and will enable the cofactor using a single control. One of the following formulas can be selected based on the costs of double-cofactors obtained from Table \ref{table:LUT2}.

\begin{table}[tb]
\parbox{16cm}{
\caption{ \label{table:LUT2}
Circuits for all 16 two-input functions. \N, \CC, and \T ~are used for NOT, CNOT and Toffoli gates. Variables $a$ and $b$ are inputs and $z$ is the output. No ancillae are used.
}
}
    \centering
    \scriptsize
    \begin{tabular}{|c|c|l|}
        \hline
        Function & Minterms      & Circuit  \\
        \hline
        \hline
	0000 &	- 		& 	-					\\
        \hline		
	0001 &	0 		& 	\T$(a',b',z)$			\\
        \hline		
	0100 &	1 		& 	\T$(a',b,z)$				\\
        \hline		
	0010 &	2 		& 	\T$(a,b',z)$				\\
        \hline		
	1000 &	3 		& 	\T$(a,b,z)$				\\
        \hline		
	0101 &	0,1  	&	\CC$(a',z)$				\\
        \hline		
	0011 &	0,2  	&	\CC$(b',z)$			\\
        \hline		
	0110 &	1,2  	&	\CC$(a,z),$ \CC$(b,z)$			\\
        \hline		
	1001 &	0,3  	&	\CC$(a,z),$ \CC$(b,z),$ \N$(z)$	\\
        \hline		
	1100 &	1,3  	&	\CC$(b,z)$				\\
        \hline		
	1010 &	2,3  	&	\CC$(a,z)$				\\
        \hline		
	0111 &	0,1,2 	&	\T$(a,b,z),$ \N$(z)$			\\
        \hline		
	1101 &	0,1,3 	&	\T$(a,b',z),$ \N$(z)$		\\
        \hline		
	1011 &	0,2,3 	&	\T$(a',b,z),$ \N$(z)$		\\
        \hline		
	1110 &	1,2,3 	&	\T$(a',b',z),$ \N$(z)$		\\
        \hline		
	1111 &	0,1,2,3	& 	\N$(z)$					\\
        \hline		
\end{tabular}
\end{table} 

\begin{small}
\begin{equation}\label{eq:Davio4}
\begin{array}{l}
F = F_{x'y'} \oplus x(F_{xy'}\oplus F_{x'y'}) \oplus y(F_{x'y}\oplus F_{x'y'}) \oplus xy (F_{x'y'} \oplus F_{x'y} \oplus F_{xy'} \oplus F_{xy})\\
F = F_{x'y} \oplus x(F_{xy}\oplus F_{x'y}) \oplus y'(F_{x'y}\oplus F_{x'y'}) \oplus xy' (F_{x'y'} \oplus F_{x'y} \oplus F_{xy'} \oplus F_{xy})\\
F = F_{xy'} \oplus x'(F_{xy'}\oplus F_{x'y'}) \oplus y(F_{xy}\oplus F_{xy'}) \oplus x'y (F_{x'y'} \oplus F_{x'y} \oplus F_{xy'} \oplus F_{xy})\\
F = F_{xy} \oplus x'(F_{xy}\oplus F_{x'y}) \oplus y'(F_{xy}\oplus F_{xy'}) \oplus x'y' (F_{x'y'} \oplus F_{x'y} \oplus F_{xy'} \oplus F_{xy})\\
 \end{array}
\end{equation}
\end{small}

\begin{figure}
\scriptsize
\scalebox{0.9}{
\input{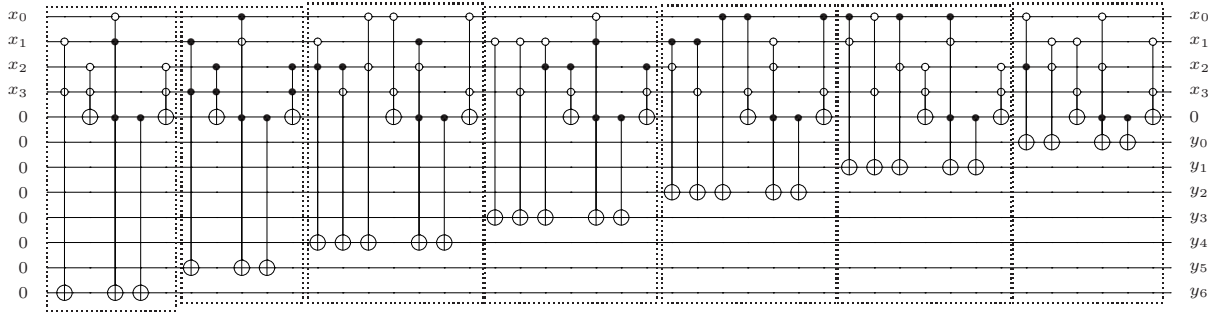}
}
\centering
\vspace{2mm}
        \caption{\label{fig:LUT87} Implementation of conditional modular multiplications by 4, 16, 82, and 25 in modular exponentiation for $b = 2$, $M = 87$ as a $(4,7)$-LUT. Each sub-circuit is the result of applying the systematic synthesis approach for one output. Each computation uses one ancilla and clears it. Further optimization is possible.
    }
\end{figure}

Of the four cofactors in each formula, one can be implemented without control, two with a single control without ancillae, and one with a single control with an ancilla. This approach leads to at most 12 gates, of which at most three are four-input Toffoli gates. The depth and gate count of a $(4,n)$-LUT are $O(n)$. Figure \ref{fig:LUT87} illustrates the result of applying the systematic synthesis to $2^x\%87$.
Selecting the cofactoring variable carefully, implementing the appropriate cofactor without control, and sharing cofactors among different functions can reduce the number of gates. Davio decompositions were used in \cite{Wille2009} to synthesize a given reversible function. However, the technique in \cite{Wille2009} implements the Davio decompositions by assuming that the factors have already been computed on dedicated ancillae. Therefore, the resulting circuits require numerous ancillae. The work in \cite{Wille2009} does not clear these ancillae. Our (4,$n$)-LUT circuits use at most one ancilla, and we clear it.

\subsection{Control optimization using 2-to-2 multiplexors}
\label{sec:control}

\begin{figure}
\scriptsize
\scalebox{0.85}{
\input{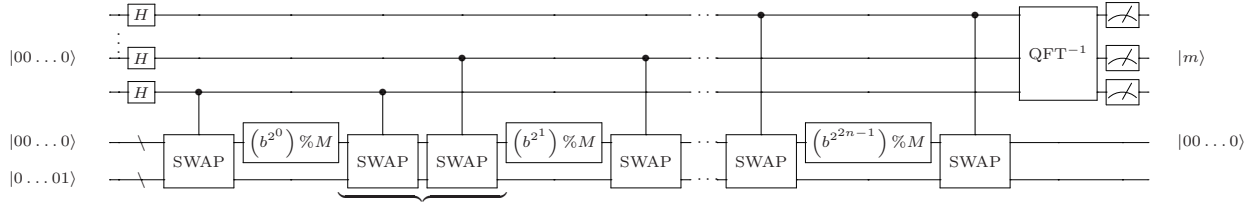}
}
\centering
\vspace{2mm}
\parbox{\length}{
\caption{\label{fig:shoropt}
Control optimization for modular exponentiation.
{\em Conditional multiplications} by $C_i=b^{2^i} \% M$ are replaced by
{\em multiplexing} that conditionally swaps constant zeros into the input of
multiplication and swaps the resulting bits out. Pairs of adjacent 2-to-2 multiplexors
are optimized further in Figure \ref{fig:mux}.}
}
\end{figure}

\begin{figure}
\scriptsize
\scalebox{0.9}{
\input{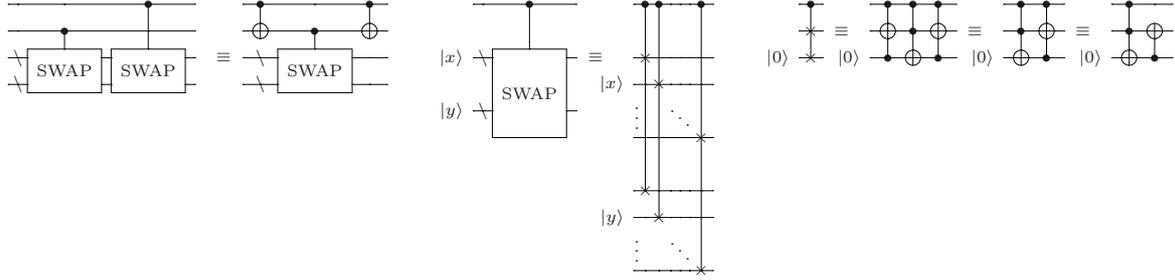}
}
\centering
\vspace{2mm}
    \parbox{\length}{
        \caption{\label{fig:mux}
        Merging neighboring 2-to-2 multiplexors (left);
        implementing a 2-to-2 multiplexor with three-input controlled-SWAP/Fredkin gates
        (middle); optimizing Fredkin gates (right) \cite[Section 2.6]{Beauregard2003}.}
    }
\end{figure}

  A large fraction of quantum-gate costs can be attributed to controls (read-only bits) \cite{BarencoEtAl1995}, and this is particularly true for mod-exp circuits, where each $C_i x \% M$-block is enabled (controlled) by one bit of Register 1.\footnote{Relevant optimizations in \cite[Section III.C]{VedralEtAl1995} and \cite[Section IV.D]{Beckman1996} are costlier than ours.}
    To avoid propagating these controls to each gate of the $C_i x \% M$-block, we observe that the binary 000...0 is a fixed point of every such block. Control can be implemented {\em indirectly} by conditionally swapping a constant zero into the register before the block and swapping the result out after the block
(Figure \ref{fig:shoropt}).  For $n$ qubits, this technique requires an
additional $n$-qubit zero-initialized {\em swap register} and $2n$ Fredkin
(controlled-SWAP) gates. We merge pairs of adjacent Fredkin gates with controls
from Register 1 and common target bits in Register 2.  Indeed,  Register 2 must be
swapped with the swap register only when the two control bits carry mutually
exclusive values. Therefore, we first apply a CNOT gate to the two controls
from Register 1, then (optimized) Fredkin gates (for each qubit of Register 2)
controlled by the target bit of the CNOT, and then we repeat the same CNOT gate
to restore the modified control bit. This is illustrated in Figure
\ref{fig:mux} and is similar to optimizations in \cite[Section 2.6]{Beauregard2003}. Each Fredkin gate can be broken down into a single-controlled
Toffoli surrounded by two CNOT gates.  However, when one of the swapped inputs
always carries a zero, the first CNOT gate can be removed. Given that $C_i x \%
M$-blocks in the literature contain $\Theta(n^2)$ gates, our two optimizations
bring substantial savings and simplify the structure of mod-exp circuits.

\ \\
{\bf Ancillae sharing.}  \label{sec:sharing}
 Our proposed optimizations trade off the overhead of control logic for a number of additional ancillae. In addition to the control register (where the Hadamard gates are applied) and the results register (Figure \ref{fig:shoropt}), multiplexing requires a swap register of size $n$. This is separate from the ancillae required by our mod-mult circuits shown in Table \ref{table:blocks}. Fortunately, many ancillae already used by the $Cx\%M$ circuits can be reused for multiplexing under some conditions. To this end, our multiplexing construction guarantees that {\em either the results register or the swap register is holding all zeros}. In the latter case, the swap register bits can clearly be used as zero-initialized ancillae in mod-mult circuits, as long as we restore them before the next multiplexing which we do (at least for $x<C$, as discussed for additive and multiplicative circuit blocks in Sections \ref{sec:add} and \ref{sec:mult}). In the former case, we need to make sure that when the $Cx\%M$ computation is performed with $x=0$, the ancillae are restored to their (possibly non-zero) initial values. Consider the modular reduction in Section \ref{sec:blocks} with one comparator and one conditional subtraction where the comparison and conditional subtraction are performed on the value stored in ancillae. Consider the zero-initialized ancilla $\varsigma$ that carries the condition bit. For any value $A \neq 0$ in the ancillae, $x=0 < A$ and we have $\varsigma=0$. Hence, the conditional subtraction is not applied. Since the Cuccaro adder recovers the values in the second register (and changes the first register to the result), the possibly non-zero initial values in the ancillae will be recovered. Therefore, we need to add one ancilla to save $n-1$ ancillae. The added ancilla will be cleared as before. Now consider the following individual blocks used in our circuits.

 \begin{itemize}
 \item $-x \% M$. Our construction maps $x = 0$ into $M$.
 \item $2^k x \% M$ for odd $M>2$. This block contains one modular reduction but has a fixed point at $x = 0$.
 \item $2^i x\%M$ for $M=2^k\pm1$. Our construction contains several modular reductions and additions based on Cuccaro adder. $x=0$ is a fixed point for Cuccaro adder. However, a non-zero value in the ancillae changes $x=0$ after addition.
 \item Division with remainder circuits. Our construction includes a set of modular reductions followed by a circuit for $(2^k +1)x$ (not modular). Assigning $x=0$ in Formula \ref{eq:s_i} reveals that the $(2^k +1)x$ circuit does not change $x=0$ as far as the ancillae used for $c_i$ carry zero. Next, we use a set of Cuccaro-based modular reductions and additions. Overall, $x=0$ is a fixed point for division with remainder circuits with zero-initialized ancillae.
 \end{itemize}

This analysis indicates that for $2^k x \% M$ for odd $M>2$, ancillae can be shared. However, the $-x \% M$ block complicates the proposed sharing of ancillae with 2-to-2 multiplexors. Therefore, we factor out such blocks, aggregate them into one as described in Section \ref{sec:reorder}, and implement one conditional $-x\%M$ as described in Section \ref{sec:mult} directly. Without multiplexing, the swap register must hold all zeros and can thus hold the ancillae of the $-x\%M$ block. For other cases with $2^i x\%M$ for $M=2^k\pm1$ and division-with-remainder circuits, ancilla sharing cannot be applied and separate ancillae are needed for mod-mult and multiplexer modules.

\subsection{Circuit structure for modular exponentiation}

{\bf Overall structure.} Summarizing the content of the above subsections, we propose mod-exp circuits consisting of three modules: $(i)$ an initial LUT, $(ii)$ an XOR-controlled negation, and $(iii)$ remaining conditional modular multiplications. The first two modules will use a linear number of gates, while the bulk of the circuit will be in modular  multiplications. To simplify the implementation of control, we use uncontrolled modular multiplications with multiplexors. The size of mod-mult circuits is moderated by factoring out negations and by implementing the most difficult multiplications in the LUT module. For an $n$-bit modulus $M$, our circuits use an $n$-qubit results register, a $2n$-qubit control register, an $n$-qubit swap register and an $n$-qubit ancilla register for each modular multiplication. In addition to these $5n$ qubits, less than $n$ ancillae may be needed for arithmetic operations (such as doubling and trippling), but these ancillae can also be shared with the swap register.
{\em In toto}, $5n$ to $6n$ qubits are used.

\ \\
{\bf Base selection.} Recall that in Shor's algorithm not all values $b>1$ for the base of exponentiation succeed --- for some values the period $2\pi$ of $f(x)=b^x\%M$ is even and $b^{\pi}\%M \neq -1$ and for others it is either odd or $b^{\pi}\%M = -1$. It is proven \cite{NielsenC2000} that the successful case occurs with probability at least 50\% (also check Table \ref{table:srate}). Therefore, common descriptions of Shor's algorithm make a random choice of $1<b<M$, invoke period-finding, and repeat the entire process for another $b$ if the period is either odd or $b^{\pi}\%M = -1$. Obviously, when gcd$(b,M)>1$, there is no need for quantum circuits, but this occurs increasingly rarely for large semiprime $M$. Therefore, when illustrating mod-exp circuits in our work, we observe gcd$(b,M)=1$. The set of reasonable $b$ values can be further restricted as follows.

\begin{theorem} \label{thm:period}
Define admissible $b$ values as those satisfying $1<b<M$ and $\mathrm{gcd}(b,M)=1$. Consider the function $f_b(x)=b^x\%M$ with an admissible $b$ value.
\begin{itemize}
    \item  For an integer $k$, $\mathrm{Period}[f_b]=\mathrm{gcd}(\mathrm{Period}[f_b],k) \cdot \mathrm{Period}[f_{b^k}]$. In particular, if $b$ results in an even period, so does $b^{2k+1}$ for $k>0$. \item For two admissible $b$ values $b_0$ and $b_1$, if $b_0$ and $b_1$ produce odd periods, so does
        $b_0b_1\%M$.
        \item If $b^{\mathrm{Period}[f_b]/2}\%M =-1$, then the same holds for $b^{2k+1}$ for $k>0$.
  \end{itemize}
\end{theorem}

\noindent
{\bf Proof.} Assume that $p$ is the smallest positive number to satisfy $b^p\%M=1$ and $p_k$ is the smallest positive number to satisfy $(b^k)^{p_k} \%M=b^{kp_k}\%M = 1$. Then $kp_k$ must be a multiple of $p$, or else  $kp_k \% p < p$ would be the period of $b$ (since we can factor out multiples of $p$ at will). The smallest positive multiple of $p$ of this form is $pk/\mathrm{gcd}(p,k)$. Therefore, $p/\mathrm{gcd}(p,k)$ is the period of $b^k$.

As for the second case, consider the smallest positive values $p_0$ and $p_1$ to satisfy $b_0^{p_0}\% M=1$ and $b_1^{p_1}\% M=1$, and also the smallest positive value $p$ to satisfy $ (b_0 b_1)^p \%M =1$. Since $p_*=p_0 p_1 / \mathrm{gcd}(p_0,p_1)$ is a multiple of both $p_0$ and $p_1$, it must satisfy the latter equation. Therefore, $p_*=pm$ for some integer $m>0$ (or else $p_*\% p <p$ would satisfy the equation, since we can factor $p$ out). If $p_0$ and $p_1$ are odd, then so is $p_*$, and thus $p$ cannot be even.
\noindent
Substituting $\mathrm{Period}[f_{b^{2k+1}}]=\mathrm{Period}[f_b]/\mathrm{gcd}(\mathrm{Period}[f_b],2k+1)$ into $b^{(2k+1)\cdot{\mathrm{Period}[f_{b^{2k+1}}]/2}}\%M$ leads to the equation $(-1)^{(2k+1)/\gcd(\mathrm{Period}[f_b], 2k + 1)}=-1$ which proves the third case.\hfill \qed

Theorem \ref{thm:period} suggests using odd powers of primes for $b$.\footnote{Fortunately, primality testing is in \P\ and can be performed very efficiently in practice.} Straightforward computational experiments show that small primes have much greater probability of success than 50\%. Assuming that success for different primes is not strongly correlated, trying only $b=2$, $b=3$ and $b=5$ can be expected to work in a majority of the cases. To illustrate this, Table \ref{table:srate} reports the percentage of semiprime $M=p \cdot q$ values with $p,q <2^{13}$  where the resulting function $f_b(x)=b^x\%M$ for $b=2,3,5$, their products $b=6,10,15$, and their squares $b=4,9,25$ has an even period $r$ where $b^{r/2} \%M \neq -1$. In this table we exclude easy values of $M$ with small $p$ and $q$ factors. Percentage statistics for unrestricted $p, q$ factors are very similar. The rows $2|3|5$, $6|10|15$, and $4|9|25$ show the percentage of semiprime $M$ values for which at least one of $b=2,3,$ or 5, $b=6,10,$ or 15, and $b=4,9,$ and 25 produces a useful period. The rows \#Failed show the total number of $M$ values considered for each $n$ that do not yield a useful period with $b=2, 3, 5$, $b=6, 10, 15$, and $b=4, 9, 25$. Adding primes $<43$ as a base for the first set and $<61$ for the second and third sets is sufficient to ensure that a useful period can be observed in all cases. The last row (\#Total) shows the total number of $M$ for each $n$. In addition to the results reported in Table \ref{table:srate}, we discovered that choosing larger primes than $b=2,3,5$ as bases
leads to more failed $M$ values. Therefore, the smallest bases are the most promising and can be tried first.

\begin{table}
\caption{\label{table:srate}
The percentage of $n$-bit semiprimes $M=p \cdot q$ with $p,q <2^{13}$ and $|\lceil \log_2 p \rceil - \lceil \log_2 q \rceil|<2$ for which $b=2,3,5$, their products $b=6,10,15$, and their squares $b=4,9,25$ result in $f_b(x)=b^x\%M$ having an even period $r$ where $b^{r/2} \%M \neq -1$.
The rows $2|3|5$, $6|10|15$, and $4|9|25$ show the percentage of semiprimes $M$ where at least one of $b=2,3,$ or 5, $b=6,10,$ or 15, and $b=4,9,$ or 25 produces a useful period.
}
    \scriptsize
\scalebox{0.9}{
    \begin{tabular}{|c|ccccccccccccccccccccc|}
		\hline
		& \multicolumn{21}{c|}{Number of bits ($n$) in $M=p \cdot q$ and success rates in \%} \\
    		 &  \multicolumn{21}{c|}{} \\
		$b$		&6		&$\!\!\!\!\!$7		&$\!\!\!\!\!$8		&$\!\!\!\!\!$9		&$\!\!\!\!\!$10		&$\!\!\!\!\!$11		 &$\!\!\!\!\!$12		 &$\!\!\!\!\!$13		&$\!\!\!\!\!$14		&$\!\!\!\!\!$15		&$\!\!\!\!\!$16		&$\!\!\!\!\!$17		 &$\!\!\!\!\!$18		&$\!\!\!\!\!$19		 &$\!\!\!\!\!$20		&$\!\!\!\!\!$21		&$\!\!\!\!\!$22		&$\!\!\!\!\!$23		 &$\!\!\!\!\!$24		&$\!\!\!\!\!$25		&$\!\!\!\!\!$26		 \\
		\hline	
		2		&100		&$\!\!\!\!\!$100		&$\!\!\!\!\!$100		&$\!\!\!\!\!$100		&$\!\!\!\!\!$95		&$\!\!\!\!\!$90		 &$\!\!\!\!\!$92		&$\!\!\!\!\!$85		&$\!\!\!\!\!$85		&$\!\!\!\!\!$87		&$\!\!\!\!\!$90		&$\!\!\!\!\!$89		&$\!\!\!\!\!$91		 &$\!\!\!\!\!$91		&$\!\!\!\!\!$91		&$\!\!\!\!\!$89		&$\!\!\!\!\!$89		&$\!\!\!\!\!$90		&$\!\!\!\!\!$90		&$\!\!\!\!\!$90		 &$\!\!\!\!\!$90	 	\\
		\hline	
		3		&100		&$\!\!\!\!\!$100		&$\!\!\!\!\!$67		&$\!\!\!\!\!$89		&$\!\!\!\!\!$100		&$\!\!\!\!\!$91		 &$\!\!\!\!\!$92		&$\!\!\!\!\!$89		&$\!\!\!\!\!$89		&$\!\!\!\!\!$86		&$\!\!\!\!\!$86		&$\!\!\!\!\!$84		&$\!\!\!\!\!$86		 &$\!\!\!\!\!$87		&$\!\!\!\!\!$88		&$\!\!\!\!\!$87		&$\!\!\!\!\!$87		&$\!\!\!\!\!$87		&$\!\!\!\!\!$88		&$\!\!\!\!\!$87		 &$\!\!\!\!\!$86		\\
		\hline	
		5		&-		&$\!\!\!\!\!$100		&$\!\!\!\!\!$83		&$\!\!\!\!\!$89		&$\!\!\!\!\!$94		&$\!\!\!\!\!$90		 &$\!\!\!\!\!$100		&$\!\!\!\!\!$90		&$\!\!\!\!\!$83		&$\!\!\!\!\!$81		&$\!\!\!\!\!$80		&$\!\!\!\!\!$84		 &$\!\!\!\!\!$87		 &$\!\!\!\!\!$87		&$\!\!\!\!\!$90		&$\!\!\!\!\!$89		&$\!\!\!\!\!$88		&$\!\!\!\!\!$87		 &$\!\!\!\!\!$87		&$\!\!\!\!\!$86		 &$\!\!\!\!\!$88		\\
		\hline	
		$2|3|5$	&\bf 100	&\bf$\!\!\!\!\!$100	&\bf$\!\!\!\!\!$100	&\bf$\!\!\!\!\!$100	&\bf$\!\!\!\!\!$100	&\bf$\!\!\!\!\!$96.4	 &\bf$\!\!\!\!\!$97.9	&\bf$\!\!\!\!\!$100	&\bf$\!\!\!\!\!$95.9	&\bf$\!\!\!\!\!$96.7	&\bf$\!\!\!\!\!$95.6	&\bf$\!\!\!\!\!$96.3	 &\bf$\!\!\!\!\!$95.9	&\bf$\!\!\!\!\!$96.1	&\bf$\!\!\!\!\!$96.1	&\bf$\!\!\!\!\!$95.8	&\bf$\!\!\!\!\!$95.9	 &\bf$\!\!\!\!\!$95.8	 &\bf$\!\!\!\!\!$95.8	&\bf$\!\!\!\!\!$95.8	&\bf$\!\!\!\!\!$96.6	\\
		\hline	
		\#Failed	&0		&$\!\!\!\!\!$0		&$\!\!\!\!\!$0		&$\!\!\!\!\!$0		&$\!\!\!\!\!$0		&$\!\!\!\!\!$1		 &$\!\!\!\!\!$1		&$\!\!\!\!\!$0		&$\!\!\!\!\!$7		&$\!\!\!\!\!$9		&$\!\!\!\!\!$24		&$\!\!\!\!\!$35		&$\!\!\!\!\!$74		 &$\!\!\!\!\!$119		&$\!\!\!\!\!$223		&$\!\!\!\!\!$421		&$\!\!\!\!\!$796		&$\!\!\!\!\!$1404	&$\!\!\!\!\!$2765	 &$\!\!\!\!\!$4817	 &$\!\!\!\!\!$2218	\\
		\hline	
		\hline	
		6		&100		&$\!\!\!\!\!$100		&$\!\!\!\!\!$100		&$\!\!\!\!\!$86		&$\!\!\!\!\!$84		&$\!\!\!\!\!$92		 &$\!\!\!\!\!$90		&$\!\!\!\!\!$99		&$\!\!\!\!\!$89		&$\!\!\!\!\!$92		&$\!\!\!\!\!$86		&$\!\!\!\!\!$87		&$\!\!\!\!\!$85		 &$\!\!\!\!\!$84		&$\!\!\!\!\!$85		&$\!\!\!\!\!$87		&$\!\!\!\!\!$87		&$\!\!\!\!\!$88		&$\!\!\!\!\!$87		&$\!\!\!\!\!$87		 &$\!\!\!\!\!$87		\\
		\hline	
		10		&-		&$\!\!\!\!\!$-		&$\!\!\!\!\!$100		&$\!\!\!\!\!$100		&$\!\!\!\!\!$100		&$\!\!\!\!\!$67		 &$\!\!\!\!\!$64		 &$\!\!\!\!\!$85		&$\!\!\!\!\!$84		&$\!\!\!\!\!$92		&$\!\!\!\!\!$88		&$\!\!\!\!\!$88		 &$\!\!\!\!\!$86		&$\!\!\!\!\!$88		 &$\!\!\!\!\!$87		&$\!\!\!\!\!$87		&$\!\!\!\!\!$89		&$\!\!\!\!\!$89		 &$\!\!\!\!\!$89		&$\!\!\!\!\!$87		&$\!\!\!\!\!$89		 \\
		\hline	
		15		&-		&$\!\!\!\!\!$50		&$\!\!\!\!\!$100		&$\!\!\!\!\!$100		&$\!\!\!\!\!$100		&$\!\!\!\!\!$100		 &$\!\!\!\!\!$72		 &$\!\!\!\!\!$79		&$\!\!\!\!\!$86		&$\!\!\!\!\!$83		&$\!\!\!\!\!$86		&$\!\!\!\!\!$87		 &$\!\!\!\!\!$88		&$\!\!\!\!\!$87		 &$\!\!\!\!\!$87		&$\!\!\!\!\!$88		&$\!\!\!\!\!$88		&$\!\!\!\!\!$87		 &$\!\!\!\!\!$87		&$\!\!\!\!\!$87		&$\!\!\!\!\!$90		 \\
		\hline	
		$6|10|15$	&\bf 100	&\bf$\!\!\!\!\!$66.7	&\bf$\!\!\!\!\!$100	&\bf$\!\!\!\!\!$88.9	&\bf$\!\!\!\!\!$95	 &\bf$\!\!\!\!\!$96.4	 &\bf$\!\!\!\!\!$95.8	&\bf$\!\!\!\!\!$94.6	&\bf$\!\!\!\!\!$92.4	&\bf$\!\!\!\!\!$92.7	 &\bf$\!\!\!\!\!$92.9	&\bf$\!\!\!\!\!$93.1	 &\bf$\!\!\!\!\!$92.7	&\bf$\!\!\!\!\!$92.9	&\bf$\!\!\!\!\!$93.1	 &\bf$\!\!\!\!\!$92.6	&\bf$\!\!\!\!\!$92.6	&\bf$\!\!\!\!\!$92.7	 &\bf$\!\!\!\!\!$92.7	&\bf$\!\!\!\!\!$93.6	 &\bf$\!\!\!\!\!$97.8	\\
		\hline	
		\#Failed	&0		&$\!\!\!\!\!$1		&$\!\!\!\!\!$0		&$\!\!\!\!\!$1		&$\!\!\!\!\!$1		&$\!\!\!\!\!$1		 &$\!\!\!\!\!$2		&$\!\!\!\!\!$5		&$\!\!\!\!\!$13		&$\!\!\!\!\!$20		&$\!\!\!\!\!$39		&$\!\!\!\!\!$65		&$\!\!\!\!\!$132		 &$\!\!\!\!\!$215		&$\!\!\!\!\!$395		&$\!\!\!\!\!$738		&$\!\!\!\!\!$1449	&$\!\!\!\!\!$2468	&$\!\!\!\!\!$4818	 &$\!\!\!\!\!$7309	 &$\!\!\!\!\!$1409	\\
		\hline
		\hline
		4		&100		&$\!\!\!\!\!$50		&$\!\!\!\!\!$67		&$\!\!\!\!\!$75		&$\!\!\!\!\!$75		&$\!\!\!\!\!$68		 &$\!\!\!\!\!$67		&$\!\!\!\!\!$55		&$\!\!\!\!\!$54		&$\!\!\!\!\!$60		&$\!\!\!\!\!$63		&$\!\!\!\!\!$64		&$\!\!\!\!\!$66		 &$\!\!\!\!\!$65		&$\!\!\!\!\!$63		&$\!\!\!\!\!$60		&$\!\!\!\!\!$60		&$\!\!\!\!\!$61		&$\!\!\!\!\!$62		&$\!\!\!\!\!$61		 &$\!\!\!\!\!$63		\\
		\hline	
		9		&100		&$\!\!\!\!\!$33		&$\!\!\!\!\!$33		&$\!\!\!\!\!$56		&$\!\!\!\!\!$65		&$\!\!\!\!\!$52		 &$\!\!\!\!\!$48		&$\!\!\!\!\!$60		&$\!\!\!\!\!$57		&$\!\!\!\!\!$55		&$\!\!\!\!\!$54		&$\!\!\!\!\!$53		&$\!\!\!\!\!$55		 &$\!\!\!\!\!$53		&$\!\!\!\!\!$55		&$\!\!\!\!\!$52		&$\!\!\!\!\!$54		&$\!\!\!\!\!$54		&$\!\!\!\!\!$55		&$\!\!\!\!\!$54		 &$\!\!\!\!\!$55		\\
		\hline	
		25		&-		&$\!\!\!\!\!$50		&$\!\!\!\!\!$67		&$\!\!\!\!\!$67		&$\!\!\!\!\!$60		&$\!\!\!\!\!$46		 &$\!\!\!\!\!$67		 &$\!\!\!\!\!$54		&$\!\!\!\!\!$51		&$\!\!\!\!\!$51		&$\!\!\!\!\!$54		&$\!\!\!\!\!$51		 &$\!\!\!\!\!$51		&$\!\!\!\!\!$51		 &$\!\!\!\!\!$56		&$\!\!\!\!\!$55		&$\!\!\!\!\!$54		&$\!\!\!\!\!$54		 &$\!\!\!\!\!$54		&$\!\!\!\!\!$54		&$\!\!\!\!\!$54		 \\
		\hline	
		$4|9|25$	&\bf 100	&\bf$\!\!\!\!\!$66.7	&\bf$\!\!\!\!\!$66.7	&\bf$\!\!\!\!\!$77.8	&\bf$\!\!\!\!\!$75	 &\bf$\!\!\!\!\!$75	 &\bf$\!\!\!\!\!$77.1	&\bf$\!\!\!\!\!$74.2	&\bf$\!\!\!\!\!$68	&\bf$\!\!\!\!\!$70.4	&\bf$\!\!\!\!\!$70.5	 &\bf$\!\!\!\!\!$72.1	 &\bf$\!\!\!\!\!$73.7	&\bf$\!\!\!\!\!$72.4	&\bf$\!\!\!\!\!$73.6	&\bf$\!\!\!\!\!$72.3	 &\bf$\!\!\!\!\!$72.6	&\bf$\!\!\!\!\!$72.6	 &\bf$\!\!\!\!\!$73.8	&\bf$\!\!\!\!\!$73.3	&\bf$\!\!\!\!\!$74.9	\\
		\hline	
		\#Failed	&0		&$\!\!\!\!\!$1		&$\!\!\!\!\!$2		&$\!\!\!\!\!$2		&$\!\!\!\!\!$5		&$\!\!\!\!\!$7		 &$\!\!\!\!\!$11		&$\!\!\!\!\!$24		&$\!\!\!\!\!$55		&$\!\!\!\!\!$81		&$\!\!\!\!\!$163		&$\!\!\!\!\!$263		 &$\!\!\!\!\!$473		 &$\!\!\!\!\!$835		&$\!\!\!\!\!$1522	&$\!\!\!\!\!$2771	&$\!\!\!\!\!$5359	&$\!\!\!\!\!$9259	 &$\!\!\!\!\!$17845	&$\!\!\!\!\!$30381	 &$\!\!\!\!\!$16215	\\
		\hline
		\hline
		\#Total	&2		&$\!\!\!\!\!$3		&$\!\!\!\!\!$6		&$\!\!\!\!\!$9		&$\!\!\!\!\!$20		&$\!\!\!\!\!$28		 &$\!\!\!\!\!$48		 &$\!\!\!\!\!$93		&$\!\!\!\!\!$172		&$\!\!\!\!\!$274		&$\!\!\!\!\!$553		&$\!\!\!\!\!$944		 &$\!\!\!\!\!$1802	&$\!\!\!\!\!$3023	 &$\!\!\!\!\!$5764	&$\!\!\!\!\!$10015	&$\!\!\!\!\!$19543	&$\!\!\!\!\!$33761	 &$\!\!\!\!\!$66548	&$\!\!\!\!\!$114015	&$\!\!\!\!\!$64719	 \\
		\hline
    \end{tabular}
}
\end{table}

\ \\
{\bf Selecting the number of controls.}  If we find that $b^k \%M=b^m \%M$ for some $k\neq m$, that allows us to upper-bound the period and then find it by binary search. When factoring large integers $M$ using Shor's algorithm,
we can pursue different strategies for selecting the number of control qubits. Most of the literature shows that selecting twice as many qubits as bits in $M$ is sufficient. However, fewer bits suffice in many cases. Given that physical implementations of Shor's algorithm are typically limited by the number of qubits, a more practical strategy is to start with a small number of qubits, perform number-factoring, and increase the number of qubits in case of failure. This adds at most a poly-time factor to runtime complexity, but also reduces circuit sizes. Assuming that modular exponentiation circuits generally have size on the order of $n^3$, the difference in sizes $n^3-(n-1)^3$ is on the order of $n^2$, which can be significant.

\section{Examples of modular exponentiation}
\label{sec:modexp-ex}

Our first series of experiments illustrates the proposed construction of mod-exp circuits but uses only {\em multiplicative} circuit decompositions for individual mod-mult blocks. Multiplicative decompositions do not require an ancilla register used by two-register circuits, but in some cases generate larger circuits, and for larger $M$ values may not be able to generate some $Cx\%M$ circuits at all. Therefore, this approach is more relevant for small $M$ values and an environment with a very limited number of qubits. Gate counts and the structure of the proposed circuits for $b^x\%M$ for $M$ with 9 functional qubits or less are reported in Table \ref{table:mod-exp2}.
 In this table, the notation $x(y)$ represents modular multiplication by $x$ controlled on the line $y$. Each circuit is described by a parenthesized triplet consisting of the Toffoli gate count, the CNOT gate count and the number of ancillae. For each $M$, we initially selected $b=2$ in modular exponentiation. If this triggered a restart in Shor's algorithm, we tried $b = 3$, and if that failed we selected $b=5$. For each $M$ value, we calculated all parameters $C_i = b^{2^i}\%M$ and found the least costly multiplicative decomposition for each $C_i$
   according to Table \ref{table:blocks}. The four costliest modular multiplications were selected
   for each $M$ value, and a $(4,n)$-LUT was synthesized for these multiplications using our {\em systematic synthesis} procedure. The remaining controlled modular multiplications were implemented directly and connected through 2-to-2 multiplexers. The number of Toffoli and CNOT gates required for each mod-mult sub-circuit is reported in Table \ref{table:blocks}. Each controlled SWAP in a 2-to-2 multiplexer can be implemented by one Toffoli and one CNOT gates (Figure \ref{fig:mux}). For the first and last controlled SWAP gates, $n=\lceil \log_2 M \rceil$ Toffoli and the same number of CNOT gates are applied. For intermediate controlled SWAPs, two additional CNOT gates are essential. Gate counts for $Cx\%M$ modules used in circuits of Table \ref{table:mod-exp2} are computed by adding up gate counts from Table \ref{table:blocks}.
   To simplify circuits for 4-LUTs, we applied the rule-based optimization method in \cite{Arabzadeh2010} which optimizes sub-circuits with common-target gates and uses both negative and positive control Toffoli gates during the optimization. For each $M$ in Table \ref{table:mod-exp2}, another $b$ value may admit a smaller circuit, but finding the best $b$ (for a given large $M$) that is useful in number-factoring $M$ is, in general, no easier than number-factoring.

\begin{figure}[tb]
\includegraphics[height=55mm]{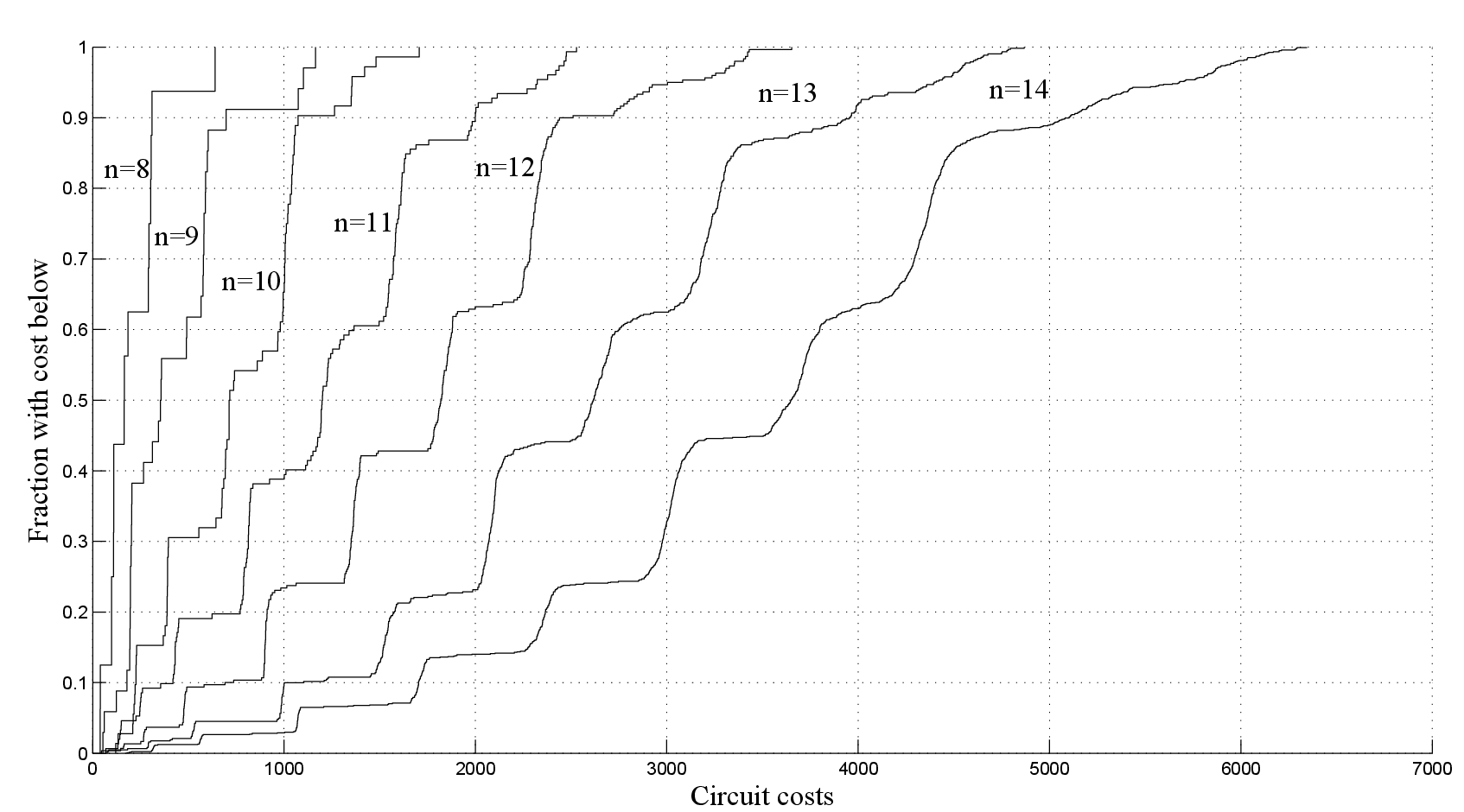}
\centering
\parbox{\length}{
\caption{
\label{fig:CDF-exp}
Costs of modular exponentiation circuits for $n$-bit $M$ values $8 \leq n \leq 14$ shown as cumulative distribution functions. The default base of exponentiation $b=2$ was replaced by 3 or 5 as needed.
}
}
\end{figure}

Our further experiments focused on scalable minimization of gate counts, but were allowed to use an additional $n$-qubit ancilla register to facilitate two-register mod-mult circuits. Figure \ref{fig:CDF-exp} shows the distributions of mod-exp circuit sizes for $n=7..14$. Each line represents the cumulative density function for \T-cost of mod-exp circuits constructed for all $n$-bit semiprime $M$ not divisible by 2 or 3. We note that for a given $n$ the median cost is about 2/3 of the maximal cost, but the smallest cost is only a fraction of the median cost. Table \ref{tab:modexp} reports min, max and average costs numerically, as well as the $M$ values for which extreme circuit costs were observed. The data for average and max costs are amendable to polynomial extrapolation ($R^2>0.999$), allowing us to estimate achievable circuit costs for much greater values of $n$ without necessarily having practical synthesis algorithms.
However, the costs of smallest-seen circuits are too erratic for reliable extrapolation. Notably, our experiments optimize the number of {\em control qubits}, typically assumed to be $2n$. For each $M$, we use the smallest number that does not lead to failures in Shor's algorithm and report it in Table \ref{tab:modexp}, along with the period found by Shor's algorithm.\footnote{For each semiprime $M$, there are two non-trivial $b$ values, such that $b^2\%M=1$. While these bases lead to the most compact mod-exp circuits, finding them is as hard as number-factoring. To this end, the data Table \ref{tab:modexp} suggest bases $b=2,3,5$ sometimes lead to unusually small circuits and short periods.}

Comparing to mod-exp circuits in \cite{VanMeter2005} that use $100n$ ancillae, our circuits use only $5n$ to $6n$ ancillae and are several orders of magnitude smaller in terms of gate counts. Circuit depth seems comparable for $n=14$. However, considering circuit depth as a measure of circuit speed assumes that any number of gates can be implemented in parallel, which does not hold for many existing physical implementations. In an environment with a limited supply of qubits and limited parallelism, our circuits appear far superior to those proposed earlier.
Whether or not many gates can be applied in parallel, larger circuits may require heavier quantum error correction, and this trend favors circuits with fewer gates.

\begin{table}
\caption{\label{table:mod-exp2}
The structure and gate count of circuits for $b^x\%M$ for $M$ with 9 bits or less.
The notation $x(y)$ represents modular multiplication by $x$ controlled on the line $y$.
Each circuit is described by a parenthesized triplet consisting of the \T ~gate count, the \CC ~gate count and the number of ancillae.
Circuits for $M=15$ and $M=21$ are illustrated in Figure \ref{fig:exp15} and Figure \ref{fig:ModExp21}, respectively.
All ancillae are cleared.
Gray cells indicate circuits without ancillae sharing.
$M$ values with $b \neq 2$ are boldfaced.
}
    \centering
    \tiny
\scalebox{0.95}{
    \begin{tabular}{|l|c|c|cccc|c|ccc|c|}
    		\hline
    		 \multirow{2}{*}{$M$} 	&  \multirow{2}{*}{b} 	&  \multirow{2}{*}{Per} 	& \multicolumn{5}{c|}{Structure of $b^x\%M$ circuits} 		& \multicolumn{4}{c|}{Gate/ancilla costs $($\T,\CC$,A)$}   \\
    		 					& 						& 						& \multicolumn{4}{c|}{Look-up tables (4-LUTs)}  & $Cx\%M$  					 & LUT 		& $Cx\%M$ 	& Mux	& Total \\
        \hline
\bf{33} & 5 &$\!\!\!\!\!$ 10 &$\!\!\!\!\!$ 5(1) &$\!\!\!\!\!$ 25(2) &$\!\!\!\!\!$ 31(3) &$\!\!\!\!\!$ 4(4) &$\!\!\!\!\!$ - &$\!\!\!\!\!$ (49,7,1) &$\!\!\!\!\!$- &$\!\!\!\!\!$- &$\!\!\!\!\!$(49,7,1)\\
\hline
35 &$\!\!\!\!\!$ 2 &$\!\!\!\!\!$ 12 &$\!\!\!\!\!$ 2(1) &$\!\!\!\!\!$ 4(2) &$\!\!\!\!\!$ 16(3) &$\!\!\!\!\!$ 11(4) &$\!\!\!\!\!$ - &$\!\!\!\!\!$ (51,7,1) &$\!\!\!\!\!$- &$\!\!\!\!\!$- &$\!\!\!\!\!$(51,7,1)\\
\hline
39 &$\!\!\!\!\!$ 2 &$\!\!\!\!\!$ 12 &$\!\!\!\!\!$ 2(1) &$\!\!\!\!\!$ 4(2) &$\!\!\!\!\!$ 16(3) &$\!\!\!\!\!$ 22(4) &$\!\!\!\!\!$ - &$\!\!\!\!\!$ (44,4,1) &$\!\!\!\!\!$- &$\!\!\!\!\!$- &$\!\!\!\!\!$(44,4,1)\\
\hline
51 &$\!\!\!\!\!$ 2 &$\!\!\!\!\!$ 8 &$\!\!\!\!\!$ 2(1) &$\!\!\!\!\!$ 4(2) &$\!\!\!\!\!$ 16(3) &$\!\!\!\!\!$ - &$\!\!\!\!\!$- &$\!\!\!\!\!$ (27,4,1) &$\!\!\!\!\!$- &$\!\!\!\!\!$- &$\!\!\!\!\!$(27,4,1)\\
\hline
55 &$\!\!\!\!\!$ 2 &$\!\!\!\!\!$ 20 &$\!\!\!\!\!$ 4(2) &$\!\!\!\!\!$ 16(3) &$\!\!\!\!\!$ 36(4) &$\!\!\!\!\!$ 31(5) &$\!\!\!\!\!$ 2(1)&$\!\!\!\!\!$ (47,9,1) &$\!\!\!\!\!$(23,29,7) &$\!\!\!\!\!$(12,12,6) &$\!\!\!\!\!$(82,50,7)\\
\hline
\bf{57} &$\!\!\!\!\!$ 5 &$\!\!\!\!\!$ 18 &$\!\!\!\!\!$ 5(1) &$\!\!\!\!\!$ 25(2) &$\!\!\!\!\!$ 4(4) &$\!\!\!\!\!$ 16(5) &$\!\!\!\!\!$ 55(3); $55=-2$ &$\!\!\!\!\!$ (51,6,1) &$\!\!\!\!\!$(35,54,7) &$\!\!\!\!\!$(12,12,6) &$\!\!\!\!\!$(98,72,7)\\
\hline
\hline
\bf{65} &$\!\!\!\!\!$ 3 &$\!\!\!\!\!$ 12 &$\!\!\!\!\!$ 3(1) &$\!\!\!\!\!$ 9(2) &$\!\!\!\!\!$ 16(3) &$\!\!\!\!\!$ 61(4) &$\!\!\!\!\!$ - &$\!\!\!\!\!$ (41,12,1) &$\!\!\!\!\!$- &$\!\!\!\!\!$- &$\!\!\!\!\!$(41,12,1)\\
\hline
69 &$\!\!\!\!\!$ 2 &$\!\!\!\!\!$ 22 &$\!\!\!\!\!$ 4(2) &$\!\!\!\!\!$ 16(3) &$\!\!\!\!\!$ 49(4) &$\!\!\!\!\!$ 55(5) &$\!\!\!\!\!$ 2(1)&$\!\!\!\!\!$ (50,7,1) &$\!\!\!\!\!$(28,33,8) &$\!\!\!\!\!$(14,14,7) &$\!\!\!\!\!$(92,54,8)\\
\hline
77 &$\!\!\!\!\!$ 2 &$\!\!\!\!\!$ 30 &$\!\!\!\!\!$ 4(2) &$\!\!\!\!\!$ 16(3) &$\!\!\!\!\!$ 25(4) &$\!\!\!\!\!$ 9(5) &$\!\!\!\!\!$ 2(1)&$\!\!\!\!\!$ (55,6,1) &$\!\!\!\!\!$(28,33,8) &$\!\!\!\!\!$(14,14,7) &$\!\!\!\!\!$(97,53,8)\\
\hline
85 &$\!\!\!\!\!$ 2 &$\!\!\!\!\!$ 8 &$\!\!\!\!\!$ 2(1) &$\!\!\!\!\!$ 4(2) &$\!\!\!\!\!$ 16(3) &$\!\!\!\!\!$ - &$\!\!\!\!\!$- &$\!\!\!\!\!$ (36,2,1) &$\!\!\!\!\!$- &$\!\!\!\!\!$- &$\!\!\!\!\!$(36,2,1)\\
\hline
87 &$\!\!\!\!\!$ 2 &$\!\!\!\!\!$ 28 &$\!\!\!\!\!$ 4(2) &$\!\!\!\!\!$ 16(3) &$\!\!\!\!\!$ 82(4) &$\!\!\!\!\!$ 25(5) &$\!\!\!\!\!$ 2(1)&$\!\!\!\!\!$ (56,9,1) &$\!\!\!\!\!$(28,33,8) &$\!\!\!\!\!$(14,14,7) &$\!\!\!\!\!$(98,56,8)\\
\hline
91 &$\!\!\!\!\!$ 2 &$\!\!\!\!\!$ 12 &$\!\!\!\!\!$ 2(1) &$\!\!\!\!\!$ 4(2) &$\!\!\!\!\!$ 16(3) &$\!\!\!\!\!$ 74(4) &$\!\!\!\!\!$ - &$\!\!\!\!\!$ (56,6,1) &$\!\!\!\!\!$- &$\!\!\!\!\!$- &$\!\!\!\!\!$(56,6,1)\\
\hline
93 &$\!\!\!\!\!$ 2 &$\!\!\!\!\!$ 10 &$\!\!\!\!\!$ 2(1) &$\!\!\!\!\!$ 4(2) &$\!\!\!\!\!$ 16(3) &$\!\!\!\!\!$ 70(4) &$\!\!\!\!\!$ - &$\!\!\!\!\!$ (50,3,1) &$\!\!\!\!\!$- &$\!\!\!\!\!$- &$\!\!\!\!\!$(50,3,1)\\
\hline
95 &$\!\!\!\!\!$ 2 &$\!\!\!\!\!$ 36 &$\!\!\!\!\!$ 16(3) &$\!\!\!\!\!$ 66(4) &$\!\!\!\!\!$ 81(5) &$\!\!\!\!\!$ 6(6) &$\!\!\!\!\!$ 2(1),4(2)&$\!\!\!\!\!$ (43,9,1) &$\!\!\!\!\!$(84,99,8) &$\!\!\!\!\!$(21,23,7) &$\!\!\!\!\!$(148,131,8)\\
\hline
111 &$\!\!\!\!\!$ 2 &$\!\!\!\!\!$ 36 &$\!\!\!\!\!$ 16(3) &$\!\!\!\!\!$ 34(4) &$\!\!\!\!\!$ 46(5) &$\!\!\!\!\!$ 7(6) &$\!\!\!\!\!$ 2(1),4(2)&$\!\!\!\!\!$ (51,7,1) &$\!\!\!\!\!$(84,99,8) &$\!\!\!\!\!$(21,23,7) &$\!\!\!\!\!$(156,129,8)\\
\hline
115 &$\!\!\!\!\!$ 2 &$\!\!\!\!\!$ 44 &$\!\!\!\!\!$ 16(3) &$\!\!\!\!\!$ 26(4) &$\!\!\!\!\!$ 101(5) &$\!\!\!\!\!$ 81(6) &$\!\!\!\!\!$ 2(1),4(2)&$\!\!\!\!\!$ (45,11,1) &$\!\!\!\!\!$(84,99,8) &$\!\!\!\!\!$(21,23,7) &$\!\!\!\!\!$(150,133,8)\\
\hline
119 &$\!\!\!\!\!$ 2 &$\!\!\!\!\!$ 24 &$\!\!\!\!\!$ 4(2) &$\!\!\!\!\!$ 16(3) &$\!\!\!\!\!$ 18(4) &$\!\!\!\!\!$ 86(5) &$\!\!\!\!\!$ 2(1)&$\!\!\!\!\!$ (57,6,1) &$\!\!\!\!\!$(28,33,8) &$\!\!\!\!\!$(14,14,7) &$\!\!\!\!\!$(99,53,8)\\
\hline
123 &$\!\!\!\!\!$ 2 &$\!\!\!\!\!$ 20 &$\!\!\!\!\!$ 4(2) &$\!\!\!\!\!$ 16(3) &$\!\!\!\!\!$ 10(4) &$\!\!\!\!\!$ 100(5) &$\!\!\!\!\!$ 2(1)&$\!\!\!\!\!$ (61,6,1) &$\!\!\!\!\!$(28,33,8) &$\!\!\!\!\!$(14,14,7) &$\!\!\!\!\!$(103,53,8)\\
\hline
\hline
133 &$\!\!\!\!\!$ 2 &$\!\!\!\!\!$ 18 &$\!\!\!\!\!$ 4(2) &$\!\!\!\!\!$ 16(3) &$\!\!\!\!\!$ 123(4) &$\!\!\!\!\!$ 100(5) &$\!\!\!\!\!$ 2(1)&$\!\!\!\!\!$ (50,14,1) &$\!\!\!\!\!$(33,37,9) &$\!\!\!\!\!$(16,16,8) &$\!\!\!\!\!$(99,67,9)\\
\hline
141 &$\!\!\!\!\!$ 2 &$\!\!\!\!\!$ 46 &$\!\!\!\!\!$ 16(3) &$\!\!\!\!\!$ 115(4) &$\!\!\!\!\!$ 112(5) &$\!\!\!\!\!$ 136(6) &$\!\!\!\!\!$ 2(1),4(2)&$\!\!\!\!\!$ (57,8,1) &$\!\!\!\!\!$(99,111,9) &$\!\!\!\!\!$(24,26,8) &$\!\!\!\!\!$(180,145,9)\\
\hline
143 &$\!\!\!\!\!$ 2 &$\!\!\!\!\!$ 60 &$\!\!\!\!\!$ 16(3) &$\!\!\!\!\!$ 113(4) &$\!\!\!\!\!$ 42(5) &$\!\!\!\!\!$ 48(6) &$\!\!\!\!\!$ 2(1),4(2)&$\!\!\!\!\!$ (49,10,1) &$\!\!\!\!\!$(99,111,9) &$\!\!\!\!\!$(24,26,8) &$\!\!\!\!\!$(172,147,9)\\
\hline
155 &$\!\!\!\!\!$ 2 &$\!\!\!\!\!$ 20 &$\!\!\!\!\!$ 4(2) &$\!\!\!\!\!$ 16(3) &$\!\!\!\!\!$ 101(4) &$\!\!\!\!\!$ 126(5) &$\!\!\!\!\!$ 2(1)&$\!\!\!\!\!$ (62,11,1) &$\!\!\!\!\!$(33,37,9) &$\!\!\!\!\!$(16,16,8) &$\!\!\!\!\!$(111,64,9)\\
\hline
159 &$\!\!\!\!\!$ 2 &$\!\!\!\!\!$ 52 &$\!\!\!\!\!$ 16(3) &$\!\!\!\!\!$ 97(4) &$\!\!\!\!\!$ 28(5) &$\!\!\!\!\!$ 148(6) &$\!\!\!\!\!$ 2(1),4(2)&$\!\!\!\!\!$ (52,13,1) &$\!\!\!\!\!$(99,111,9) &$\!\!\!\!\!$(24,26,8) &$\!\!\!\!\!$(175,150,9)\\
\hline
\bf{161} &$\!\!\!\!\!$ 3 &$\!\!\!\!\!$ 66 &$\!\!\!\!\!$ 3(1) &$\!\!\!\!\!$ 9(2) &$\!\!\!\!\!$ 100(6) &$\!\!\!\!\!$ 18(7) &$\!\!\!\!\!$ 81(3),121(4),151(5)&$\!\!\!\!\!$ (58,11,1) &$\!\!\!\!\!$(231,259,9) &$\!\!\!\!\!$(32,36,8) &$\!\!\!\!\!$(321,306,9)\\
&$\!\!\!\!\!$ &$\!\!\!\!\!$ &$\!\!\!\!\!$ &$\!\!\!\!\!$ &$\!\!\!\!\!$ &$\!\!\!\!\!$ &$\!\!\!\!\!$ $81 = 2^{-1}, 121 = 2^{-2}, 151=2^{-4}$  &$\!\!\!\!\!$ &$\!\!\!\!\!$ &$\!\!\!\!\!$ &$\!\!\!\!\!$  \\
\hline
\bf{177} &$\!\!\!\!\!$ 5 &$\!\!\!\!\!$ 58 &$\!\!\!\!\!$ 25(2) &$\!\!\!\!\!$ 163(4) &$\!\!\!\!\!$ 19(5) &$\!\!\!\!\!$ 7(6) &$\!\!\!\!\!$ 5(1),94(3); $94=-2^{-5}$ &$\!\!\!\!\!$ (48,8,1) &$\!\!\!\!\!$(443,352,12) &$\!\!\!\!\!$(24,26,8) &$\!\!\!\!\!$\cellcolor{gray} (515,386,20)\\
\hline
183 &$\!\!\!\!\!$ 2 &$\!\!\!\!\!$ 60 &$\!\!\!\!\!$ 16(3) &$\!\!\!\!\!$ 73(4) &$\!\!\!\!\!$ 22(5) &$\!\!\!\!\!$ 118(6) &$\!\!\!\!\!$ 2(1),4(2)&$\!\!\!\!\!$ (67,11,1) &$\!\!\!\!\!$(99,111,9) &$\!\!\!\!\!$(24,26,8) &$\!\!\!\!\!$(190,148,9)\\
\hline
\bf{185} &$\!\!\!\!\!$ 3 &$\!\!\!\!\!$ 36 &$\!\!\!\!\!$ 3(1) &$\!\!\!\!\!$ 9(2) &$\!\!\!\!\!$ 81(3) &$\!\!\!\!\!$ 86(4) &$\!\!\!\!\!$ 181(5),16(6); $181 = -2^2 $ &$\!\!\!\!\!$ (61,7,1) &$\!\!\!\!\!$(214,255,9) &$\!\!\!\!\!$(24,26,8) &$\!\!\!\!\!$(299,288,9)\\
\hline
187 &$\!\!\!\!\!$ 2 &$\!\!\!\!\!$ 40 &$\!\!\!\!\!$ 16(3) &$\!\!\!\!\!$ 69(4) &$\!\!\!\!\!$ 86(5) &$\!\!\!\!\!$ 103(6) &$\!\!\!\!\!$ 2(1),4(2)&$\!\!\!\!\!$ (70,9,1) &$\!\!\!\!\!$(99,111,9) &$\!\!\!\!\!$(24,26,8) &$\!\!\!\!\!$(193,146,9)\\
\hline
203 &$\!\!\!\!\!$ 2 &$\!\!\!\!\!$ 84 &$\!\!\!\!\!$ 53(4) &$\!\!\!\!\!$ 170(5) &$\!\!\!\!\!$ 74(6) &$\!\!\!\!\!$ 198(7) &$\!\!\!\!\!$ 2(1),4(2),16(3)&$\!\!\!\!\!$ (63,12,1) &$\!\!\!\!\!$(231,259,9) &$\!\!\!\!\!$(32,36,8) &$\!\!\!\!\!$(326,307,9)\\
\hline
\bf{205} &$\!\!\!\!\!$ 3 &$\!\!\!\!\!$ 8 &$\!\!\!\!\!$ 3(1) &$\!\!\!\!\!$ 9(2) &$\!\!\!\!\!$ 81(3) &$\!\!\!\!\!$ - &$\!\!\!\!\!$- &$\!\!\!\!\!$ (40,3,1) &$\!\!\!\!\!$- &$\!\!\!\!\!$- &$\!\!\!\!\!$(40,3,1)\\
\hline
\bf{209} &$\!\!\!\!\!$ 3 &$\!\!\!\!\!$ 90 &$\!\!\!\!\!$ 9(2) &$\!\!\!\!\!$ 82(4) &$\!\!\!\!\!$ 36(5) &$\!\!\!\!\!$ 92(7) &$\!\!\!\!\!$ 3(1),81(3),42(6)&$\!\!\!\!\!$ (60,12,1) &$\!\!\!\!\!$(738,547,12) &$\!\!\!\!\!$(32,36,8) &$\!\!\!\!\!$\cellcolor{gray} (830,595,20)\\
&$\!\!\!\!\!$ &$\!\!\!\!\!$ &$\!\!\!\!\!$ &$\!\!\!\!\!$ &$\!\!\!\!\!$ &$\!\!\!\!\!$ &$\!\!\!\!\!$ $81 = -2^7, 42 = 5^{-1}$  &$\!\!\!\!\!$ &$\!\!\!\!\!$ &$\!\!\!\!\!$  &$\!\!\!\!\!$ \cellcolor{gray}  \\
\hline
213 &$\!\!\!\!\!$ 2 &$\!\!\!\!\!$ 70 &$\!\!\!\!\!$ 43(4) &$\!\!\!\!\!$ 145(5) &$\!\!\!\!\!$ 151(6) &$\!\!\!\!\!$ 10(7) &$\!\!\!\!\!$ 2(1),4(2),16(3)&$\!\!\!\!\!$ (63,13,1) &$\!\!\!\!\!$(231,259,9) &$\!\!\!\!\!$(32,36,8) &$\!\!\!\!\!$(326,308,9)\\
\hline
215 &$\!\!\!\!\!$ 2 &$\!\!\!\!\!$ 28 &$\!\!\!\!\!$ 4(2) &$\!\!\!\!\!$ 16(3) &$\!\!\!\!\!$ 41(4) &$\!\!\!\!\!$ 176(5) &$\!\!\!\!\!$ 2(1)&$\!\!\!\!\!$ (62,13,1) &$\!\!\!\!\!$(33,37,9) &$\!\!\!\!\!$(16,16,8) &$\!\!\!\!\!$(111,66,9)\\
\hline
\bf{217} &$\!\!\!\!\!$ 5 &$\!\!\!\!\!$ 6 &$\!\!\!\!\!$ 5(1) &$\!\!\!\!\!$ 25(2) &$\!\!\!\!\!$ 191(3) &$\!\!\!\!\!$ - &$\!\!\!\!\!$- &$\!\!\!\!\!$ (39,5,1) &$\!\!\!\!\!$- &$\!\!\!\!\!$- &$\!\!\!\!\!$(39,5,1)\\
\hline
219 &$\!\!\!\!\!$ 2 &$\!\!\!\!\!$ 18 &$\!\!\!\!\!$ 4(2) &$\!\!\!\!\!$ 16(3) &$\!\!\!\!\!$ 37(4) &$\!\!\!\!\!$ 55(5) &$\!\!\!\!\!$ 2(1)&$\!\!\!\!\!$ (53,9,1) &$\!\!\!\!\!$(33,37,9) &$\!\!\!\!\!$(16,16,8) &$\!\!\!\!\!$(102,62,9)\\
\hline
221 &$\!\!\!\!\!$ 2 &$\!\!\!\!\!$ 24 &$\!\!\!\!\!$ 4(2) &$\!\!\!\!\!$ 16(3) &$\!\!\!\!\!$ 35(4) &$\!\!\!\!\!$ 120(5) &$\!\!\!\!\!$ 2(1)&$\!\!\!\!\!$ (60,9,1) &$\!\!\!\!\!$(33,37,9) &$\!\!\!\!\!$(16,16,8) &$\!\!\!\!\!$(109,62,9)\\
\hline
235 &$\!\!\!\!\!$ 2 &$\!\!\!\!\!$ 92 &$\!\!\!\!\!$ 21(4) &$\!\!\!\!\!$ 206(5) &$\!\!\!\!\!$ 136(6) &$\!\!\!\!\!$ 166(7) &$\!\!\!\!\!$ 2(1),4(2),16(3)&$\!\!\!\!\!$ (56,16,1) &$\!\!\!\!\!$(231,259,9) &$\!\!\!\!\!$(32,36,8) &$\!\!\!\!\!$(319,311,9)\\
\hline
237 &$\!\!\!\!\!$ 2 &$\!\!\!\!\!$ 78 &$\!\!\!\!\!$ 19(4) &$\!\!\!\!\!$ 124(5) &$\!\!\!\!\!$ 208(6) &$\!\!\!\!\!$ 130(7) &$\!\!\!\!\!$ 2(1),4(2),16(3)&$\!\!\!\!\!$ (62,10,1) &$\!\!\!\!\!$(231,259,9) &$\!\!\!\!\!$(32,36,8) &$\!\!\!\!\!$(325,305,9)\\
\hline
247 &$\!\!\!\!\!$ 2 &$\!\!\!\!\!$ 36 &$\!\!\!\!\!$ 16(3) &$\!\!\!\!\!$ 9(4) &$\!\!\!\!\!$ 81(5) &$\!\!\!\!\!$ 139(6) &$\!\!\!\!\!$ 2(1),4(2)&$\!\!\!\!\!$ (51,11,1) &$\!\!\!\!\!$(99,111,9) &$\!\!\!\!\!$(24,26,8) &$\!\!\!\!\!$(174,148,9)\\
\hline
253 &$\!\!\!\!\!$ 2 &$\!\!\!\!\!$ 110 &$\!\!\!\!\!$ 3(4) &$\!\!\!\!\!$ 9(5) &$\!\!\!\!\!$ 81(6) &$\!\!\!\!\!$ 236(7) &$\!\!\!\!\!$ 2(1),4(2),16(3)&$\!\!\!\!\!$ (47,12,1) &$\!\!\!\!\!$(231,259,9) &$\!\!\!\!\!$(32,36,8) &$\!\!\!\!\!$(310,307,9)\\
\hline
\hline
259 &$\!\!\!\!\!$ 2 &$\!\!\!\!\!$ 36 &$\!\!\!\!\!$ 16(3) &$\!\!\!\!\!$ 256(4) &$\!\!\!\!\!$ 9(5) &$\!\!\!\!\!$ 81(6) &$\!\!\!\!\!$ 2(1),4(2)&$\!\!\!\!\!$ (47,12,1) &$\!\!\!\!\!$(114,123,10) &$\!\!\!\!\!$(27,29,9) &$\!\!\!\!\!$(188,164,10)\\
\hline
267 &$\!\!\!\!\!$ 2 &$\!\!\!\!\!$ 22 &$\!\!\!\!\!$ 4(2) &$\!\!\!\!\!$ 16(3) &$\!\!\!\!\!$ 256(4) &$\!\!\!\!\!$ 121(5) &$\!\!\!\!\!$ 2(1)&$\!\!\!\!\!$ (62,7,1) &$\!\!\!\!\!$(38,41,10) &$\!\!\!\!\!$(18,18,9) &$\!\!\!\!\!$(118,66,10)\\
\hline
287 &$\!\!\!\!\!$ 2 &$\!\!\!\!\!$ 60 &$\!\!\!\!\!$ 16(3) &$\!\!\!\!\!$ 256(4) &$\!\!\!\!\!$ 100(5) &$\!\!\!\!\!$ 242(6) &$\!\!\!\!\!$ 2(1),4(2)&$\!\!\!\!\!$ (63,17,1) &$\!\!\!\!\!$(114,123,10) &$\!\!\!\!\!$(27,29,9) &$\!\!\!\!\!$(204,169,10)\\
\hline
291 &$\!\!\!\!\!$ 2 &$\!\!\!\!\!$ 48 &$\!\!\!\!\!$ 16(3) &$\!\!\!\!\!$ 256(4) &$\!\!\!\!\!$ 61(5) &$\!\!\!\!\!$ 229(6) &$\!\!\!\!\!$ 2(1),4(2)&$\!\!\!\!\!$ (58,16,1) &$\!\!\!\!\!$(114,123,10) &$\!\!\!\!\!$(27,29,9) &$\!\!\!\!\!$(199,168,10)\\
\hline
295 &$\!\!\!\!\!$ 2 &$\!\!\!\!\!$ 116 &$\!\!\!\!\!$ 256(4) &$\!\!\!\!\!$ 46(5) &$\!\!\!\!\!$ 51(6) &$\!\!\!\!\!$ 241(7) &$\!\!\!\!\!$ 2(1),4(2),16(3)&$\!\!\!\!\!$ (76,17,1) &$\!\!\!\!\!$(266,287,10) &$\!\!\!\!\!$(36,40,9) &$\!\!\!\!\!$(378,344,10)\\
\hline
299 &$\!\!\!\!\!$ 2 &$\!\!\!\!\!$ 132 &$\!\!\!\!\!$ 256(4) &$\!\!\!\!\!$ 55(5) &$\!\!\!\!\!$ 35(6) &$\!\!\!\!\!$ 29(7) &$\!\!\!\!\!$ 2(1),4(2),16(3),243(8); $243 = 2^{-4}$ &$\!\!\!\!\!$ (56,12,1) &$\!\!\!\!\!$(418,451,10) &$\!\!\!\!\!$(45,51,9) &$\!\!\!\!\!$(519,514,10)\\
\hline
301 &$\!\!\!\!\!$ 2 &$\!\!\!\!\!$ 42 &$\!\!\!\!\!$ 16(3) &$\!\!\!\!\!$ 256(4) &$\!\!\!\!\!$ 219(5) &$\!\!\!\!\!$ 102(6) &$\!\!\!\!\!$ 2(1),4(2)&$\!\!\!\!\!$ (65,8,1) &$\!\!\!\!\!$(114,123,10) &$\!\!\!\!\!$(27,29,9) &$\!\!\!\!\!$(206,160,10)\\
\hline
303 &$\!\!\!\!\!$ 2 &$\!\!\!\!\!$ 100 &$\!\!\!\!\!$ 256(4) &$\!\!\!\!\!$ 88(5) &$\!\!\!\!\!$ 169(6) &$\!\!\!\!\!$ 79(7) &$\!\!\!\!\!$ 2(1),4(2),16(3)&$\!\!\!\!\!$ (54,5,1) &$\!\!\!\!\!$(266,287,10) &$\!\!\!\!\!$(36,40,9) &$\!\!\!\!\!$(356,332,10)\\
\hline
\bf{305} &$\!\!\!\!\!$ 3 &$\!\!\!\!\!$ 20 &$\!\!\!\!\!$ 3(1) &$\!\!\!\!\!$ 9(2) &$\!\!\!\!\!$ 81(3) &$\!\!\!\!\!$ 156(4) &$\!\!\!\!\!$ 241(5); $241 = -2^6 $ &$\!\!\!\!\!$ (59,9,1) &$\!\!\!\!\!$(246,283,10) &$\!\!\!\!\!$(18,18,9) &$\!\!\!\!\!$(323,310,10)\\
\hline
309 &$\!\!\!\!\!$ 2 &$\!\!\!\!\!$ 102 &$\!\!\!\!\!$ 256(4) &$\!\!\!\!\!$ 28(5) &$\!\!\!\!\!$ 166(6) &$\!\!\!\!\!$ 55(7) &$\!\!\!\!\!$ 2(1),4(2),16(3)&$\!\!\!\!\!$ (59,17,1) &$\!\!\!\!\!$(266,287,10) &$\!\!\!\!\!$(36,40,9) &$\!\!\!\!\!$(361,344,10)\\
\hline
319 &$\!\!\!\!\!$ 2 &$\!\!\!\!\!$ 140 &$\!\!\!\!\!$ 141(5) &$\!\!\!\!\!$ 103(6) &$\!\!\!\!\!$ 82(7) &$\!\!\!\!\!$ 25(8) &$\!\!\!\!\!$ 2(1),4(2),16(3),256(4)&$\!\!\!\!\!$ (65,13,1) &$\!\!\!\!\!$(570,615,10) &$\!\!\!\!\!$(45,51,9) &$\!\!\!\!\!$(680,679,10)\\
\hline
323 &$\!\!\!\!\!$ 2 &$\!\!\!\!\!$ 72 &$\!\!\!\!\!$ 256(4) &$\!\!\!\!\!$ 290(5) &$\!\!\!\!\!$ 120(6) &$\!\!\!\!\!$ 188(7) &$\!\!\!\!\!$ 2(1),4(2),16(3)&$\!\!\!\!\!$ (74,12,1) &$\!\!\!\!\!$(266,287,10) &$\!\!\!\!\!$(36,40,9) &$\!\!\!\!\!$(376,339,10)\\
\hline
327 &$\!\!\!\!\!$ 2 &$\!\!\!\!\!$ 36 &$\!\!\!\!\!$ 16(3) &$\!\!\!\!\!$ 256(4) &$\!\!\!\!\!$ 136(5) &$\!\!\!\!\!$ 184(6) &$\!\!\!\!\!$ 2(1),4(2)&$\!\!\!\!\!$ (62,11,1) &$\!\!\!\!\!$(114,123,10) &$\!\!\!\!\!$(27,29,9) &$\!\!\!\!\!$(203,163,10)\\
\hline
\bf{329} &$\!\!\!\!\!$ 3 &$\!\!\!\!\!$ 138 &$\!\!\!\!\!$ 81(3) &$\!\!\!\!\!$ 310(4) &$\!\!\!\!\!$ 53(7) &$\!\!\!\!\!$ 177(8) &$\!\!\!\!\!$ 3(1),9(2),32(5),37(6)&$\!\!\!\!\!$ (59,13,1) &$\!\!\!\!\!$(1136,1090,13) &$\!\!\!\!\!$(45,51,9) &$\!\!\!\!\!$\cellcolor{gray} (1240,1154,22)\\
&$\!\!\!\!\!$ &$\!\!\!\!\!$ &$\!\!\!\!\!$ &$\!\!\!\!\!$ &$\!\!\!\!\!$ &$\!\!\!\!\!$ &$\!\!\!\!\!$ $9 = 2^{-8}, 37 = 2^{10}$ &$\!\!\!\!\!$ &$\!\!\!\!\!$ &$\!\!\!\!\!$ &$\!\!\!\!\!$ \cellcolor{gray}  \\
\hline
335 &$\!\!\!\!\!$ 2 &$\!\!\!\!\!$ 132 &$\!\!\!\!\!$ 256(4) &$\!\!\!\!\!$ 211(5) &$\!\!\!\!\!$ 301(6) &$\!\!\!\!\!$ 151(7) &$\!\!\!\!\!$ 2(1),4(2),16(3),21(8)&$\!\!\!\!\!$ (54,11,1) &$\!\!\!\!\!$(418,451,10) &$\!\!\!\!\!$(45,51,9) &$\!\!\!\!\!$(517,513,10)\\
\hline
339 &$\!\!\!\!\!$ 2 &$\!\!\!\!\!$ 28 &$\!\!\!\!\!$ 4(2) &$\!\!\!\!\!$ 16(3) &$\!\!\!\!\!$ 256(4) &$\!\!\!\!\!$ 109(5) &$\!\!\!\!\!$ 2(1)&$\!\!\!\!\!$ (67,8,1) &$\!\!\!\!\!$(38,41,10) &$\!\!\!\!\!$(18,18,9) &$\!\!\!\!\!$(123,67,10)\\
\hline
341 &$\!\!\!\!\!$ 2 &$\!\!\!\!\!$ 10 &$\!\!\!\!\!$ 2(1) &$\!\!\!\!\!$ 4(2) &$\!\!\!\!\!$ 16(3) &$\!\!\!\!\!$ 256(4) &$\!\!\!\!\!$ - &$\!\!\!\!\!$ (61,5,1) &$\!\!\!\!\!$- &$\!\!\!\!\!$- &$\!\!\!\!\!$(61,5,1)\\
\hline
355 &$\!\!\!\!\!$ 2 &$\!\!\!\!\!$ 140 &$\!\!\!\!\!$ 216(5) &$\!\!\!\!\!$ 151(6) &$\!\!\!\!\!$ 81(7) &$\!\!\!\!\!$ 171(8) &$\!\!\!\!\!$ 2(1),4(2),16(3),256(4)&$\!\!\!\!\!$ (75,13,1) &$\!\!\!\!\!$(570,615,10) &$\!\!\!\!\!$(45,51,9) &$\!\!\!\!\!$(690,679,10)\\
\hline
365 &$\!\!\!\!\!$ 2 &$\!\!\!\!\!$ 36 &$\!\!\!\!\!$ 16(3) &$\!\!\!\!\!$ 256(4) &$\!\!\!\!\!$ 201(5) &$\!\!\!\!\!$ 251(6) &$\!\!\!\!\!$ 2(1),4(2)&$\!\!\!\!\!$ (62,10,1) &$\!\!\!\!\!$(114,123,10) &$\!\!\!\!\!$(27,29,9) &$\!\!\!\!\!$(203,162,10)\\
\hline
371 &$\!\!\!\!\!$ 2 &$\!\!\!\!\!$ 156 &$\!\!\!\!\!$ 240(5) &$\!\!\!\!\!$ 95(6) &$\!\!\!\!\!$ 121(7) &$\!\!\!\!\!$ 172(8) &$\!\!\!\!\!$ 2(1),4(2),16(3),256(4)&$\!\!\!\!\!$ (61,13,1) &$\!\!\!\!\!$(570,615,10) &$\!\!\!\!\!$(45,51,9) &$\!\!\!\!\!$(676,679,10)\\
\hline
\bf{377} &$\!\!\!\!\!$ 3 &$\!\!\!\!\!$ 84 &$\!\!\!\!\!$ 9(2) &$\!\!\!\!\!$ 152(4) &$\!\!\!\!\!$ 107(5) &$\!\!\!\!\!$ 139(6) &$\!\!\!\!\!$ 3(1),81(3),94(7)&$\!\!\!\!\!$ (70,10,1) &$\!\!\!\!\!$(660,580,12) &$\!\!\!\!\!$(36,40,9) &$\!\!\!\!\!$\cellcolor{gray} (766,630,21)\\
&$\!\!\!\!\!$ &$\!\!\!\!\!$ &$\!\!\!\!\!$ &$\!\!\!\!\!$ &$\!\!\!\!\!$ &$\!\!\!\!\!$ &$\!\!\!\!\!$ $81 = 2^{-8}, 94 = -2^{-2}$ &$\!\!\!\!\!$ &$\!\!\!\!\!$ &$\!\!\!\!\!$ &$\!\!\!\!\!$  \cellcolor{gray} \\
\hline
381 &$\!\!\!\!\!$ 2 &$\!\!\!\!\!$ 14 &$\!\!\!\!\!$ 2(1) &$\!\!\!\!\!$ 4(2) &$\!\!\!\!\!$ 16(3) &$\!\!\!\!\!$ 256(4) &$\!\!\!\!\!$ - &$\!\!\!\!\!$ (56,7,1) &$\!\!\!\!\!$- &$\!\!\!\!\!$- &$\!\!\!\!\!$(56,7,1)\\
\hline
391 &$\!\!\!\!\!$ 2 &$\!\!\!\!\!$ 88 &$\!\!\!\!\!$ 256(4) &$\!\!\!\!\!$ 239(5) &$\!\!\!\!\!$ 35(6) &$\!\!\!\!\!$ 52(7) &$\!\!\!\!\!$ 2(1),4(2),16(3)&$\!\!\!\!\!$ (70,12,1) &$\!\!\!\!\!$(266,287,10) &$\!\!\!\!\!$(36,40,9) &$\!\!\!\!\!$(372,339,10)\\
\hline
\bf{393} &$\!\!\!\!\!$ 5 &$\!\!\!\!\!$ 130 &$\!\!\!\!\!$ 376(4) &$\!\!\!\!\!$ 289(5) &$\!\!\!\!\!$ 205(6) &$\!\!\!\!\!$ 367(7) &$\!\!\!\!\!$ 5(1),25(2),232(3),283(8)&$\!\!\!\!\!$ (61,20,1) &$\!\!\!\!\!$(1936,1223,13) &$\!\!\!\!\!$(45,51,9) &$\!\!\!\!\!$\cellcolor{gray} (2042,1294,22)\\
&$\!\!\!\!\!$ &$\!\!\!\!\!$ &$\!\!\!\!\!$ &$\!\!\!\!\!$ &$\!\!\!\!\!$ &$\!\!\!\!\!$ &$\!\!\!\!\!$ $232 = -2^{-11}, 283 = 5^{-2}$ &$\!\!\!\!\!$ &$\!\!\!\!\!$ &$\!\!\!\!\!$ &$\!\!\!\!\!$   \cellcolor{gray} \\
\hline
395 &$\!\!\!\!\!$ 2 &$\!\!\!\!\!$ 156 &$\!\!\!\!\!$ 361(5) &$\!\!\!\!\!$ 366(6) &$\!\!\!\!\!$ 51(7) &$\!\!\!\!\!$ 231(8) &$\!\!\!\!\!$ 2(1),4(2),16(3),256(4)&$\!\!\!\!\!$ (63,14,1) &$\!\!\!\!\!$(570,615,10) &$\!\!\!\!\!$(45,51,9) &$\!\!\!\!\!$(678,680,10)\\
\hline
403 &$\!\!\!\!\!$ 2 &$\!\!\!\!\!$ 60 &$\!\!\!\!\!$ 16(3) &$\!\!\!\!\!$ 256(4) &$\!\!\!\!\!$ 250(5) &$\!\!\!\!\!$ 35(6) &$\!\!\!\!\!$ 2(1),4(2)&$\!\!\!\!\!$ (72,9,1) &$\!\!\!\!\!$(114,123,10) &$\!\!\!\!\!$(27,29,9) &$\!\!\!\!\!$(213,161,10)\\
\hline
407 &$\!\!\!\!\!$ 2 &$\!\!\!\!\!$ 180 &$\!\!\!\!\!$ 9(5) &$\!\!\!\!\!$ 81(6) &$\!\!\!\!\!$ 49(7) &$\!\!\!\!\!$ 366(8) &$\!\!\!\!\!$ 2(1),4(2),16(3),256(4)&$\!\!\!\!\!$ (52,10,1) &$\!\!\!\!\!$(570,615,10) &$\!\!\!\!\!$(45,51,9) &$\!\!\!\!\!$(667,676,10)\\
\hline
411 &$\!\!\!\!\!$ 2 &$\!\!\!\!\!$ 68 &$\!\!\!\!\!$ 16(3) &$\!\!\!\!\!$ 256(4) &$\!\!\!\!\!$ 187(5) &$\!\!\!\!\!$ 34(6) &$\!\!\!\!\!$ 2(1),4(2),334(7); $334 = 2^{-4}$ &$\!\!\!\!\!$ (64,9,1) &$\!\!\!\!\!$(266,287,10) &$\!\!\!\!\!$(36,40,9) &$\!\!\!\!\!$(366,336,10)\\
\hline
413 &$\!\!\!\!\!$ 2 &$\!\!\!\!\!$ 174 &$\!\!\!\!\!$ 282(5) &$\!\!\!\!\!$ 228(6) &$\!\!\!\!\!$ 359(7) &$\!\!\!\!\!$ 25(8) &$\!\!\!\!\!$ 2(1),4(2),16(3),256(4)&$\!\!\!\!\!$ (71,11,1) &$\!\!\!\!\!$(570,615,10) &$\!\!\!\!\!$(45,51,9) &$\!\!\!\!\!$(686,677,10)\\
\hline
415 &$\!\!\!\!\!$ 2 &$\!\!\!\!\!$ 164 &$\!\!\!\!\!$ 381(5) &$\!\!\!\!\!$ 326(6) &$\!\!\!\!\!$ 36(7) &$\!\!\!\!\!$ 51(8) &$\!\!\!\!\!$ 2(1),4(2),16(3),256(4)&$\!\!\!\!\!$ (58,14,1) &$\!\!\!\!\!$(570,615,10) &$\!\!\!\!\!$(45,51,9) &$\!\!\!\!\!$(673,680,10)\\
\hline
\bf{417} &$\!\!\!\!\!$ 5 &$\!\!\!\!\!$ 138 &$\!\!\!\!\!$ 25(2) &$\!\!\!\!\!$ 259(6) &$\!\!\!\!\!$ 361(7) &$\!\!\!\!\!$ 217(8) &$\!\!\!\!\!$ 5(1),208(3),313(4),391(5)&$\!\!\!\!\!$ (66,16,1) &$\!\!\!\!\!$(584,471,13) &$\!\!\!\!\!$(45,51,9) &$\!\!\!\!\!$\cellcolor{gray} (695,538,22)\\
&$\!\!\!\!\!$ &$\!\!\!\!\!$ &$\!\!\!\!\!$ &$\!\!\!\!\!$ &$\!\!\!\!\!$ &$\!\!\!\!\!$ &$\!\!\!\!\!$  $208 = -2^{-1}, 313 = 2^{-2}, 391 = 2^{-4}$ &$\!\!\!\!\!$ &$\!\!\!\!\!$ &$\!\!\!\!\!$ &$\!\!\!\!\!$ \cellcolor{gray} \\
\hline
427 &$\!\!\!\!\!$ 2 &$\!\!\!\!\!$ 60 &$\!\!\!\!\!$ 16(3) &$\!\!\!\!\!$ 256(4) &$\!\!\!\!\!$ 205(5) &$\!\!\!\!\!$ 179(6) &$\!\!\!\!\!$ 2(1),4(2)&$\!\!\!\!\!$ (71,11,1) &$\!\!\!\!\!$(114,123,10) &$\!\!\!\!\!$(27,29,9) &$\!\!\!\!\!$(212,163,10)\\
\hline
437 &$\!\!\!\!\!$ 2 &$\!\!\!\!\!$ 198 &$\!\!\!\!\!$ 423(5) &$\!\!\!\!\!$ 196(6) &$\!\!\!\!\!$ 397(7) &$\!\!\!\!\!$ 289(8) &$\!\!\!\!\!$ 2(1),4(2),16(3),256(4)&$\!\!\!\!\!$ (61,15,1) &$\!\!\!\!\!$(570,615,10) &$\!\!\!\!\!$(45,51,9) &$\!\!\!\!\!$(676,681,10)\\
\hline
445 &$\!\!\!\!\!$ 2 &$\!\!\!\!\!$ 44 &$\!\!\!\!\!$ 16(3) &$\!\!\!\!\!$ 256(4) &$\!\!\!\!\!$ 121(5) &$\!\!\!\!\!$ 401(6) &$\!\!\!\!\!$ 2(1),4(2)&$\!\!\!\!\!$ (65,10,1) &$\!\!\!\!\!$(114,123,10) &$\!\!\!\!\!$(27,29,9) &$\!\!\!\!\!$(206,162,10)\\
\hline
447 &$\!\!\!\!\!$ 2 &$\!\!\!\!\!$ 148 &$\!\!\!\!\!$ 274(5) &$\!\!\!\!\!$ 427(6) &$\!\!\!\!\!$ 400(7) &$\!\!\!\!\!$ 421(8) &$\!\!\!\!\!$ 2(1),4(2),16(3),256(4)&$\!\!\!\!\!$ (60,14,1) &$\!\!\!\!\!$(570,615,10) &$\!\!\!\!\!$(45,51,9) &$\!\!\!\!\!$(675,680,10)\\
\hline
451 &$\!\!\!\!\!$ 2 &$\!\!\!\!\!$ 20 &$\!\!\!\!\!$ 4(2) &$\!\!\!\!\!$ 16(3) &$\!\!\!\!\!$ 256(4) &$\!\!\!\!\!$ 141(5) &$\!\!\!\!\!$ 2(1)&$\!\!\!\!\!$ (68,9,1) &$\!\!\!\!\!$(38,41,10) &$\!\!\!\!\!$(18,18,9) &$\!\!\!\!\!$(124,68,10)\\
\hline
453 &$\!\!\!\!\!$ 2 &$\!\!\!\!\!$ 30 &$\!\!\!\!\!$ 4(2) &$\!\!\!\!\!$ 16(3) &$\!\!\!\!\!$ 256(4) &$\!\!\!\!\!$ 304(5) &$\!\!\!\!\!$ 2(1)&$\!\!\!\!\!$ (63,12,1) &$\!\!\!\!\!$(38,41,10) &$\!\!\!\!\!$(18,18,9) &$\!\!\!\!\!$(119,71,10)\\
\hline
469 &$\!\!\!\!\!$ 2 &$\!\!\!\!\!$ 66 &$\!\!\!\!\!$ 16(3) &$\!\!\!\!\!$ 256(4) &$\!\!\!\!\!$ 345(5) &$\!\!\!\!\!$ 368(6) &$\!\!\!\!\!$ 2(1),4(2),352(7); $352 = 2^{-2}$ &$\!\!\!\!\!$ (58,16,1) &$\!\!\!\!\!$(190,205,10) &$\!\!\!\!\!$(36,40,9) &$\!\!\!\!\!$(284,261,10)\\
\hline
471 &$\!\!\!\!\!$ 2 &$\!\!\!\!\!$ 52 &$\!\!\!\!\!$ 16(3) &$\!\!\!\!\!$ 256(4) &$\!\!\!\!\!$ 67(5) &$\!\!\!\!\!$ 250(6) &$\!\!\!\!\!$ 2(1),4(2)&$\!\!\!\!\!$ (82,8,1) &$\!\!\!\!\!$(114,123,10) &$\!\!\!\!\!$(27,29,9) &$\!\!\!\!\!$(223,160,10)\\
\hline
\bf{473} &$\!\!\!\!\!$ 3 &$\!\!\!\!\!$ 210 &$\!\!\!\!\!$ 81(3) &$\!\!\!\!\!$ 185(6) &$\!\!\!\!\!$ 169(7)&$\!\!\!\!\!$ 181(8) &$\!\!\!\!\!$ 3(1),9(2),412(4),410(5) &$\!\!\!\!\!$ (69,18,1) &$\!\!\!\!\!$(2042,1193,12) &$\!\!\!\!\!$(45,51,9) &$\!\!\!\!\!$\cellcolor{gray} (2156,1262,21)\\
&$\!\!\!\!\!$ &$\!\!\!\!\!$ &$\!\!\!\!\!$ &$\!\!\!\!\!$ &$\!\!\!\!\!$ &$\!\!\!\!\!$ &$\!\!\!\!\!$  $412 =-2^{-3} \cdot 3 \cdot 5, 410=3^{-1}\cdot 5^{-1}$ &$\!\!\!\!\!$ &$\!\!\!\!\!$ &$\!\!\!\!\!$ &$\!\!\!\!\!$ \cellcolor{gray} \\
\hline
\bf{481} &$\!\!\!\!\!$ 3 &$\!\!\!\!\!$ 18 &$\!\!\!\!\!$ 9(2) &$\!\!\!\!\!$ 81(3) &$\!\!\!\!\!$ 308(4) &$\!\!\!\!\!$ 107(5) &$\!\!\!\!\!$ 3(1)&$\!\!\!\!\!$ (64,13,1) &$\!\!\!\!\!$(262,133,12) &$\!\!\!\!\!$(18,18,9) &$\!\!\!\!\!$\cellcolor{gray} (344,164,21)\\
\hline
485 &$\!\!\!\!\!$ 2 &$\!\!\!\!\!$ 48 &$\!\!\!\!\!$ 16(3) &$\!\!\!\!\!$ 256(4) &$\!\!\!\!\!$ 61(5) &$\!\!\!\!\!$ 326(6) &$\!\!\!\!\!$ 2(1),4(2)&$\!\!\!\!\!$ (74,9,1) &$\!\!\!\!\!$(114,123,10) &$\!\!\!\!\!$(27,29,9) &$\!\!\!\!\!$(215,161,10)\\
\hline
493 &$\!\!\!\!\!$ 2 &$\!\!\!\!\!$ 56 &$\!\!\!\!\!$ 16(3) &$\!\!\!\!\!$ 256(4) &$\!\!\!\!\!$ 460(5) &$\!\!\!\!\!$ 103(6) &$\!\!\!\!\!$ 2(1),4(2)&$\!\!\!\!\!$ (64,14,1) &$\!\!\!\!\!$(114,123,10) &$\!\!\!\!\!$(27,29,9) &$\!\!\!\!\!$(205,166,10)\\
\hline
\bf{497} &$\!\!\!\!\!$ 3 &$\!\!\!\!\!$ 210 &$\!\!\!\!\!$ 81(3) &$\!\!\!\!\!$ 121(6) &$\!\!\!\!\!$ 228(7) &$\!\!\!\!\!$ 296(8) &$\!\!\!\!\!$ 3(1),9(2),100(4),60(5)&$\!\!\!\!\!$ (61,15,1) &$\!\!\!\!\!$(1766,1130,13) &$\!\!\!\!\!$(45,51,9) &$\!\!\!\!\!$\cellcolor{gray} (1872,1196,22)\\
&$\!\!\!\!\!$ &$\!\!\!\!\!$ &$\!\!\!\!\!$ &$\!\!\!\!\!$ &$\!\!\!\!\!$ &$\!\!\!\!\!$ &$\!\!\!\!\!$  $100 = 5^{-1} \cdot 3, 60 = 2^{11}$ &$\!\!\!\!\!$ &$\!\!\!\!\!$ &$\!\!\!\!\!$ &$\!\!\!\!\!$ \cellcolor{gray} \\
\hline
501 &$\!\!\!\!\!$ 2 &$\!\!\!\!\!$ 166 &$\!\!\!\!\!$ 406(5) &$\!\!\!\!\!$ 7(6) &$\!\!\!\!\!$ 49(7) &$\!\!\!\!\!$ 397(8) &$\!\!\!\!\!$ 2(1),4(2),16(3),256(4)&$\!\!\!\!\!$ (62,16,1) &$\!\!\!\!\!$(570,615,10) &$\!\!\!\!\!$(45,51,9) &$\!\!\!\!\!$(677,682,10)\\
\hline
\bf{511} &$\!\!\!\!\!$ 3 &$\!\!\!\!\!$ 12 &$\!\!\!\!\!$ 3(1) &$\!\!\!\!\!$ 9(2) &$\!\!\!\!\!$ 81(3) &$\!\!\!\!\!$ 429(4) &$\!\!\!\!\!$ - &$\!\!\!\!\!$ (54,6,1) &$\!\!\!\!\!$- &$\!\!\!\!\!$- &$\!\!\!\!\!$(54,6,1)\\
\hline
\end{tabular}
}
    \end{table}

\begin{table}
\caption{\label{tab:modexp} \T-Costs for modular exponentiation circuits for $n$-bit $M$ values not divisible by 2 and 3. All $M$ values divisible by 5 were factored with $b=2$ or $b=3$. Trend lines were extrapolated using $b=2$ and $n=9..14$. Parameters $l$ and $\pi$ represent the number of controls and the period, respectively. Max and min $l$ may not correspond to max-cost and min-cost circuits for a given $n$. Numbers in [~] are $C$ in LUT.}
\scriptsize
\scalebox{1}{
\begin{tabular}{|c|l|l|l|l|l|l|l|l|l|}
\hline
$\!\!\!\!\!$Bits$\!\!\!\!\!$ & \multicolumn{3}{c|}{\T-costs for $b=2$} & \multicolumn{3}{c|}{\# of lines for $b=2$} & \multicolumn{2}{c|}{($M,b$,Cost,$l$,$\pi$) for extreme circuits} & $\!\!\!\!\!$Min circuit costs\\
$\!\!\!\!\!$$n$$\!\!\!\!\!$ & Min & Max & Avg & Min & Max & Avg & $\!\!\!\!\!$Min cost &$\!\!\!\!\!$ Max cost & $\!\!\!\!\!$(4-LUT, Mod-mult, Mux)\\
\hline
$\!\!\!\!\!$7$\!\!\!\!\!$	&36	  &150	&97.7& 3 & 6 & 4.8 & $\!\!\!\!\!$(85,2,36,3,8) & $\!\!\!\!\!$(115,2,150,6,44) & $\!\!\!\!\!$(36,0,0) [2, 4, 16]\\
$\!\!\!\!\!$8$\!\!\!\!\!$	&99	  &326	&192.4& 5 & 7 & 5.9 & $\!\!\!\!\!$(217,5,39,3,6) & $\!\!\!\!\!$(209,3,655,7,90) & $\!\!\!\!\!$(39,0,0) [5, 25, 191]\\
\hline
\hline
$\!\!\!\!\!$9$\!\!\!\!\!$	&61	  &631	&375.9&  4 & 8 & 6.8 & $\!\!\!\!\!$(511,3,54,4,12) & $\!\!\!\!\!$(497,3,1191,8,210)& $\!\!\!\!\!$(54,0,0) [3, 9, 81, 429]\\
$\!\!\!\!\!$10$\!\!\!\!\!$	&121  &1099	&689.4&  5 & 9 & 7.8 & $\!\!\!\!\!$(635,2,121,5,28) & $\!\!\!\!\!$(713,3,1747,9,330)& $\!\!\!\!\!$(58,43,20)[4, 16, 256, 131]\\
$\!\!\!\!\!$11$\!\!\!\!\!$	&68	  &1691	&992.3&  4 & 10& 8.3 & $\!\!\!\!\!$(1285,2,68,4,16) & $\!\!\!\!\!$(1841,3,2584,10,786)& $\!\!\!\!\!$(68,0,0)[2, 4, 16, 256]\\
$\!\!\!\!\!$12$\!\!\!\!\!$	&146  &2511	&1624.5& 5 & 11 & 9.4 & $\!\!\!\!\!$(4069,5,46,3,8) & $\!\!\!\!\!$(3817,5,3601,11,1730)& $\!\!\!\!\!$(46,0,0)[5, 25, 625]\\
$\!\!\!\!\!$13$\!\!\!\!\!$	&75	  &3463	&2332.6& 4 & 12 & 10.3 & $\!\!\!\!\!$(5461,2,75,4,14) & $\!\!\!\!\!$(8153,3,4876,12,3930)&$\!\!\!\!\!$(75,0,0)[2, 4, 16, 256]\\
$\!\!\!\!\!$14$\!\!\!\!\!$  &179  &4680	&3224.9& 5 & 13 & 11.2& $\!\!\!\!\!$(10261,2,179,5,30) & $\!\!\!\!\!$(14849,3,6282,13,7170)& $\!\!\!\!\!$(88,63,28)[4, 16, 256, 3970]\\
\hline
\hline
\multicolumn{10}{|l|}{Trend lines for $b=2$ ~~~~~~~ Max($n$)=$3.861n^3 - 40.61n^2 + 187.3n - 578.9$ ~~~~~~~ Avg($n$)=$ 1.979n^3 + 12.32n^2 - 512.9n + 2563.0$}\\
\hline
\multicolumn{10}{l}{Extrapolated values for $b=2$ ~~~~~~~~ $n$: (max,avg)} \\
\hline
\multicolumn{10}{|l|}{20: (17811,13065); 50: (389886,255093); 100: (3473051,2053473); 200: (29300481,16224783); 300: (100647711,54390493)}\\
\hline
\end{tabular}
}
\end{table}

\section{Comparison with prior art} \label {sec:comp}
Prior work on circuits for modular multiplication and modular exponentiation typically describes circuit sizes by a closed-form expression in terms of the number of input qubits. Those circuits typically take on the order of $n^2$ gates for modular multiplication and $n^3$ for modular exponentiation.\footnote{QFT-based circuits exhibit slower asymptotic growth, but are viewed impractical for $<1000$ qubits or less.} The best cases almost always exhibit the same asymptotic growth. In contrast, our circuits for modular multiplication by 2, 3 and 5 (as well as their inverses) require only a linear number of gates. In the more general case, our optimization is algorithmic in nature, therefore a closed-form expression cannot be given {\em a priori} and comparisons require software implementations of our proposed algorithms. To compare the asymptotic number of gates in the proposed mod-mult and mod-exp circuits, we use the trend lines for maximum and average gate counts.

\subsection{Modular multiplication}
In \cite{Beckman1996}, circuits for $n$-qubit modular multiplication uses $n$ conditional mod $M$ additions. The addition mod $M$ is constructed by a multiplexed adder and a comparison operator where the former is based on multiplexed full and half adders. Considering one enable bit in \cite[Formulae 5.12 \& 5.17]{Beckman1996} for multiplexed full and half adders leads to $[2,2,2,1]$\footnote{Following \cite{Beckman1996}, $[c_0, c_1, c_2, c_3]$ indicates a circuit with $c_0$ NOT, $c_1$ CNOT, $c_2$ C$^2$NOT, and $c_3$ C$^3$NOT gates.} and $[2,2,1,0]$ gates in the worst case. Hence, worst-case gate counts for mod-mult in \cite{Beckman1996} are given by Formula \ref{eq:Beckman1996worst} \cite[Formula 6.4]{Beckman1996} leading to $[16n^2-16n,8n^2 + 16n - 18, 24n^2 - 56n + 24, 4n^2-8n+4]$ gates.

\begin{small}
\begin{equation}
\label{eq:Beckman1996worst}
\begin{array}{l}
4(n-1)^2 [2,2,2,1] + 4(n-1)[2,2,1,0] + 8(n-1)[n, 2, 2n - 3, 0] + \\
4(n-1)[0, 0, 1, 0] + 2(n-1)[0, 1, 0, 0] + 2[0, n, 0, 0] + 2[0, 2n, 0, 0]
 \end{array}
\end{equation}
\end{small}

Following \cite[Formula 6.4]{Beckman1996} leads to Formula \ref{eq:Beckman1996avg} for the average gate count in mod-mult. Similar to the worst case, $2n+1$ ancillae are used and cleared at the end of computation.

\begin{small}
\begin{equation}
\label{eq:Beckman1996avg}
\begin{array}{l}
4(n-1)^2 [1/2, 3/2, 3/2, 1/2] + 4(n-1)[1/2, 5/4, 1/2, 0] + \\
8(n-1)[n-1/2, 3/2, 3/2n-5/2, 0]+4(n-1)[0, 0, 1, 0] + 2(n-1)[0, 1, 0, 0] + \\
2[0, 1/2n, 0, 0] + 2[0, n, 0, 0]
 \end{array}
\end{equation}
\end{small}

To account for the number of \T ~and \CC ~gates, one can apply the cost model in \cite{MaslovS2011}\footnote{This cost model evaluates circuit implementation via estimating the number of two-qubit gates required to implement it. Inverters are ignored because they may be merged into 2-qubit gates. In this generic model, an $n$-qubit Toffoli gate (either with positive or negative controls) can be decomposed into $2n-5$ \T gates.} ~which leads to $8n^2 + 16n - 18$ \CC ~and $36n^2 - 80n + 36$ \T ~gates in the worst case and $6n^2-16n+13$ \CC ~and $24n^2 -50n + 26$ \T ~in the average case with $2n+1$ ancillae --- (486 \CC, 1240 \T), (622 \CC,1700 \T), (774 \CC,2232 \T), (942 \CC,2836 \T), (1126 \CC, 3512 \T), (1326 \CC, 4260 \T) for  $n=7, 8 \cdots 12$ in the worst case and (195 \CC, 852 \T), (269 \CC, 1162 \T), (355 \CC, 1520 \T), (453 \CC, 1926 \T), (563 \CC, 2380 \T), (685 \CC, 2882 \T) for  $n=7, 8 \ldots 12$ in the average case. More recent work optimizes ancillae~\cite{Takahashi2010} and circuit depth \cite{VanMeter2005}, resulting in larger circuits. The trend lines for \T-cost in the proposed modular multiplcation circuits are
$5.309n^2 - 11.59n + 4.5$, and $3.351n^2 + 7.127n - 78.57$ in the worst and average cases, respectively (Table \ref{tab:max_avg}).

\subsection{Modular exponentiation}
In \cite{Beckman1996}, $n$-qubit modular exponentiation is constructed from $\approx 2n$ conditional modular multiplications. For a modular multiplication with an enable bit, $n$ conditional mod $M$ additions are chained. Hence, each mod $M$ addition has a pair of enable bits. The average CNOT and Toffoli gate counts for $n$-qubit modular exponentiation are $14n^3 + 5n^2 -18n + 13$, and $46n^3 - 107n^2 + 92n - 25$, respectively \cite{Beckman1996}. In this configuration, $2n+3$ ancillae are used and cleared at the end of computation.

In \cite{VedralEtAl1995}, modular exponentiation is performed by setting Register 2 to $|1\rangle$ and applying $n$ conditional mod-mult $b^{2^i}\%M$ modules followed by a controlled multiplication network $b^{-2^i} \%M$ that clears the ancillae: $7n+1$ ancillae in total. Overall the algorithm needs $20n^2-5n$ adders with $4n-3$ \CC ~and $4n-4$ \T ~gates which leads to $96n^3 - 84n^2 + 15n$ \CC ~and $80n^3 - 100n^2 + 20n$ \T ~gates.
The adder structure of \cite{VedralEtAl1995} was improved in \cite{VanMeter2005} to include $2n-3$ \CC  ~and $3n-3$ \T ~gates leading to $40n^3 - 70n^2 + 15^n$ \CC ~and $60n^3 - 75n^2 + 15n$ \T ~gates for modular exponentiation.

Modular multiplications in our proposed structure are unconditional (Section \ref{sec:control}). To consider the effect of structural and algorithmic optimizations for modular exponentiation without considering the effect of ideas proposed for modular multiplication, here we use the same structure as in \cite{Beckman1996} except that each mod $M$ addition has a single enable bit whereas two enable bits were used in \cite{Beckman1996}. The average numbers of CNOT and Toffoli gates in modular multiplication are $6n^2-16n+13$ and $24n^2 -50n + 26$, respectively.

The first and last controlled SWAPs in a 2-to-2 multiplexer needs $n$ Toffoli and $n$ CNOT gates. Other controlled SWAP gates need two additional CNOTs. Finally, note that gate count of a $(4,n)$-LUT are $O(n)$. Precisely, $1/2 \times (8n)$ C$^4$NOT gates are required for $(4,n)$-LUT on average with one ancilla. With two zero-initialized ancillae, each C$^4$NOT gate can be decomposed into 5 Toffoli gates. Overall, the average number of Toffoli gates is $20n$.

Combining the above calculations with our proposed structure for modular exponentiation shown in Formula \ref{eq:mod-exp} leads to $6n^3 - 39n^2 +76n -62$ CNOT, and $24n^3 - 145n^2 + 243 n - 104$ Toffoli gates. As for the number of ancillae, aside from $2n+1$ for mod-mult, we need $2n$ ancillae for Register 1 and Register 2 in Shor's algorithm, and $n$ ancillae for swap register --- $5n+1$ in total. Applying the proposed mod-mult circuits in mod-exp instead of \cite{Beckman1996} reduces the leading orders in \CC ~and \T ~gates from $6n^3$ and $24n^3$ to $5.309n^3$ and $3.351n^3$, respectively.

\begin{small}
\begin{equation}
\label{eq:mod-exp}
\begin{array}{l}
20n(0 ~\mathcal C,1 ~\mathcal T) + (n-4)(6n^2-16n+13 ~\mathcal C, 24n^2 -50n + 26 ~\mathcal T) + \\
2 (n ~\mathcal C,n ~\mathcal T) + (n-5)(n+2 ~\mathcal C,n ~\mathcal T)
\end{array}
\end{equation}
\end{small}

In \cite{VanMeter2005}, depth-optimized circuits for modular exponentiation were constructed by parallelizing modular multiplications and using depth-optimized adders. With arbitrary-distance interaction between qubits, the authors reduced the asymptotic depth of modular exponentiation to $O(n \log^2 n)$. However, their circuits need $\sim100n$ ancillae, use a large number of gates, and assume unbounded gate parallelism, which can make them impractical with current technologies. For $n=128$, the latency (circuit depth) of the best technique in \cite[Algorithm E, Table II]{VanMeter2005} is $1.96 \times 10^4$ CNOT, and $1.71 \times 10^5$ Toffoli gates with 12657 ancillae. For $n=128$ our mod-exp circuits need $1.1 \times 10^7$ CNOT, and $7.0 \times 10^6$ Toffoli gates with 641 ancillae. If \cite[algorithm G, Table II]{VanMeter2005} with 660 ancillae is used for comparison, the latency is $2.48 \times 10^5$ \CC, and $1.50 \times 10^7$ \T ~gates. Even though our circuits are not optimized for depth, the actual number of gates seems comparable to the depth of depth-optimized circuits in \cite{VanMeter2005}.

\section{Conclusions and future research} \label{sec:conc}

  In this paper, we proposed linear-size circuits for several special cases in modular multiplication
  and used them to develop a shortest-path formalism for finding compact generic mod-mult circuits.
  Our results can be viewed as the first illustration of automated logic synthesis and optimization
  for modular multiplication circuits with superior results compared to mathematical circuit constructions.
  Our circuits are also the first not to require Bennett's technique, and this produces significant savings.
  The above results are directly applicable to modular exponentiation circuits, for which we propose
  several additional improvements.


 While previous techniques for reversible logic optimization operate at the bit level \cite{SaeediM2011}, our research used {\em register-transfer level (RTL) primitives} to optimize reversible circuits. This higher-level perspective facilitates much greater scalability than for previous algorithms.
  The RTL primitives we proposed
  in Table \ref{tab:ops} are good candidates for direct implementations in terms of specific
  quantum technologies. Such implementations may be faster and less error-prone than the decompositions
  into elementary gates that we have shown. They can also support a higher level of programming of quantum computers, where sequences of operators demonstrated in Tables \ref{tab:65} and \ref{tab:max_avg}
  can be issued directly to the quantum computer without intermediate levels of software translation.

  Despite concrete evidence of smaller circuits for mod-exp, our research leaves a number of
  open challenges. In particular, the algorithms for synthesizing mod-exp circuits that we have implemented do not scale easily beyond $15$-bit $M$ values. Our follow-up method \cite{MarkovPRA} constructs near-optimal circuits from execution traces of a GCD algorithm and reports circuits for up to $512$-bit $M$ values generated in less than half an hour.
  Departing from register-level structure of our current mod-mult circuits, bit-level local optimization \cite{SaeediM2011} may further reduce gate counts. Follow-up methods in \cite{ShafaeiDATE,ShafaeiJETC} optimize circuits in reversible Look-up Tables (LUTs) that we have identified. Further reductions may be achievable by leaving the Boolean domain \cite{MaslovS2011}.



\end{document}